\documentclass[aps,prd,superscriptaddress,showpacs,showkeys,a4paper,superscriptaddress,10pt,nofootinbib,notitlepage]{revtex4-1}

\usepackage[utf8]{inputenc}
\usepackage[english]{babel}
\usepackage{amsmath}
\usepackage{amssymb}
\usepackage{amsfonts}
\usepackage{graphicx}
\usepackage{bm}
\usepackage{cancel}
\usepackage{upgreek}
\usepackage{mathtools}
\usepackage[T1]{fontenc}

\usepackage{xcolor}
\definecolor{link_blue}{RGB}{52,46,157}

\usepackage[colorlinks,citecolor=link_blue,linkcolor=link_blue,urlcolor=link_blue]{hyperref}
\usepackage[tiny,center,uppercase]{titlesec}
\usepackage[a4paper,textwidth=380pt,textheight=592pt,top=107pt,left=80pt]{geometry}

\renewcommand{\vec}{\boldsymbol}
\newcommand{\la}{\langle}
\newcommand{\ra}{\rangle}

\newcommand{\thp}[2]{\vec #1\cdot\vec #2}
\newcommand{\opA}[1]{\boldsymbol{\mathsf{#1}}}
\newcommand{\opa}[1]{\mathsf{#1}}
\newcommand\ri{\mathrm{i}}

\DeclareMathOperator{\Div}{div}
\allowdisplaybreaks[2]

\begin{document}
	
\title{Soliton-like solution in quantum electrodynamics}

\author{O.\ D.\ Skoromnik}
\email[Corresponding author: ]{olegskor@gmail.com}
\affiliation{Max Planck Institute for Nuclear Physics, Saupfercheckweg 1, 69117 Heidelberg, Germany}
\author{I.\ D.\ Feranchuk}
\email[Corresponding author: ]{ilya.feranchuk@tdt.edu.vn}
\affiliation{Atomic Molecular and Optical Physics Research Group, Ton Duc Thang University, 19 Nguyen Huu Tho Str., Tan Phong Ward, District 7, Ho Chi Minh City, Vietnam}
\affiliation{Faculty of Applied Sciences, Ton Duc Thang University, 19 Nguyen Huu Tho Str., Tan Phong Ward, District 7, Ho Chi Minh City, Vietnam}
\affiliation{Belarusian State University, 4 Nezavisimosty Ave., 220030, Minsk,   Belarus}
\author{C. H. Keitel}
\affiliation{Max Planck Institute for Nuclear Physics, Saupfercheckweg 1, 69117 Heidelberg, Germany}

\begin{abstract}
  A novel soliton-like solution in quantum electrodynamics is obtained via a self-consistent field method. By writing the Hamiltonian of quantum electrodynamics in the Coulomb gauge, we separate out a classical component in the density operator of the electron-positron field. Then, by modeling the state vector in analogy with the theory of superconductivity, we minimize the functional for the energy of the system. This results in the equations of the self-consistent field, where the solutions are associated with the collective excitation of the electron-positron field---the soliton-like solution. In addition, the canonical transformation of the variables allowed us to separate out the total momentum of the system and, consequently, to find the relativistic energy dispersion relation for the moving soliton.
\end{abstract}

\maketitle

\section{Introduction}
\label{sec:introduction}
Solitons or solitary waves are the solutions of nonlinear equations of motion, which describe a localized field state and possess a nondispersive energy density \cite{RajaramanB1982solitons}. Initially obtained in hydrodynamics \cite{RayleighA1876waves,KortewegA1895change} and solid state physics \cite{PhysRev.108.1175,BlattB1971theory,AbrikosovB1975methods} they quickly spread in different areas of physics and nowadays they also play an important role in quantum field theory, high energy physics and cosmology \cite{WeinbergB2012classical}.

Among the plethora of soliton solutions which have been found, a large amount is related to model systems and in the general case it is not clear, which physical object corresponds to this soliton-like solution. For example, the existence of the Dirac monopole \cite{DiracA1931quantized} can immediately explain the charge quantization condition, however, none of the monopoles has been experimentally observed. Despite of this, it has been proven that monopoles necessarily arise as soliton solutions in certain gauge field theories \cite{PolyakovA1974particle,HOOFT1974276,Bolognesi2002337}.

At the same time, some situations exist in quantum field theory when the soliton-like solutions can be experimentally observed \cite{PhysRevLett.117.055301,PhysRevLett.117.055302,Chen:2015rt}. One of the most known examples is the polar model of a metal---the polaron problem \cite{Froehlich1954,Mitra198791,RevModPhys.63.63,Spohn1987278,0022-3719-17-24-012,PhysRev.97.660}. In this model an electron is confined to a potential well, which is created due to the interaction with phonons of a crystal resulting in a localized state with a renormalized mass, which is substantially different from the one of the ``bare'' electron. This model correctly predicts the observable characteristics of the charge carriers in a crystal.

Soliton solutions are significant for the nonperturbative description of states in quantum field theories and some results can not be obtained via a perturbative basis. For example, in the above mentioned polaron problem in the strong coupling regime the perturbation theory does not lead to the desired solution and the modeling of a state vector as a localized state in a self-consistent potential formed by the classical component of a quantum field is required. A similar situation arises for strong interactions, where the modeling of the state vector as a localized state has led to some success in the description of a hadron, the so called SLAC ``bag'' model of a quark \cite{PhysRevD.11.1094},\cite{Bolognesi2005150,Bolognesi200693}. Moreover, it was recently demonstrated that a nonperturbative treatment, in which a soliton solution is separated out in the zeroth-order approximation, leads to the regularization of the perturbation-theory series in the problem of a particle interacting with a scalar quantum field \cite{PhysRevD.92.125019}.

In the present work we are interested in seeking a soliton-like solution in the physically important theory, which describes one of the four fundamental interactions, namely quantum electrodynamics (QED). It is well known that the two constants contained in the QED Hamiltonian, i.e. $e_{0}$ and $m_{0}$---the ``bare'' electron charge and the ``bare'' mass---are not known. These two constants depend on the momentum cutoff \cite{bjorken1965,LandauQED,AkhiezerB1965quantum,PeskinB1995introduction} and are excluded from the theory through the renormalization procedure \cite{CollinsB1984renormalization,SalmhoferB1999renormalization,SchwartzB2014quantum,AkhiezerB1965quantum}, introducing the physical values of the electron charge and the mass in the Hamiltonian. However, $e_{0}$ and $m_{0}$ remain unknown and, therefore, we can consider them as free parameters of the theory. Next, we assume that the soliton-like solution in quantum electrodynamics is mainly formed by the self-interaction of the electron-positron field and neglect the contribution of the transverse electromagnetic field. As a result, we model the state vector of the electron-positron system in analogy with Ref.~\cite{PhysRevD.11.1094,FeranchukA2007sigma}. In addition, by exploiting the self-consistent field method \cite{HartreeA1928wave-1,HartreeA1928wave-2,FockA1930naeherungs} we obtain a system of equations which describes in a self-consistent way the collective excitation of the electron-positron field, i.e. yielding evidence that our initial assumptions are reasonable. The solution of this system of equations is associated with the soliton-like solution in QED.

With respect to this, we would also like to mention that a related approach was exploited in a series of works \cite{ACIKGOZ1995126,PhysRevA.45.7740,BARUT19921469,PhysRevA.39.2796,PhysRevA.32.3187,Barut1983,PhysRevD.16.161}, where the total electromagnetic field was separated into two parts, namely the external field and the electron self-field. This second part is generated by the nonquantized electron current and thus excluded from the action via the equations of motion. As a result of this procedure, the effects such as the spontaneous emission \cite{PhysRevA.32.3187}, the vacuum polarization \cite{ACIKGOZ1995126}, the Lamb shift \cite{PhysRevA.45.7740,Barut1983} in the absence of the external field, as well as $g-2$ \cite{PhysRevA.39.2796} in the presence of the external field were evaluated.

The article is organized in the following way. In Sec.~\ref{sec:equation-of-motion}, starting from the QED Hamiltonian written in the Coulomb gauge in the Schr\"{o}dinger representation based on the self-consistent field method, we derive the system of Dirac equations with the self-consistent field for the quasi-particle collective excitation of the electron-positron system at rest. Proceeding to Sec.~\ref{sec:separation-variables} we discuss the separation of variables, which lead to the equations for the radial part of the Dirac bispinors. Furthermore, we calculate the integral characteristics, e.g. the total energy of the collective excitation of the electron-positron field. In Sec.~\ref{sec:solut-second-kind} we discuss the solution of the second kind, which possesses the opposite sign of energy. Next, in Sec.~\ref{sec:solit-like-solut} we demonstrate that similar two kinds of solutions with the opposite sign of charge exist. With this we conclude the formulation for the soliton-like solution at rest and transfer to Sec.~\ref{sec:states-with-nonzero}, in which we discuss the moving solitons. In that Sec.~\ref{sec:states-with-nonzero} we perform the canonical transformation of the variables and separate the total momentum of the system. After this we calculate the energy of the moving soliton and show that its energy dispersion relation, i.e. the dependence of the energy on the total momentum is given through the well known relativistic energy-momentum relation. At last, the summary of the paper, the discussion of the obtained results and an intuitive, simple quasi-classical estimation are presented in Sec.~\ref{sec:discussion}. Finally, the details of all relevant calculations can be found in Appendices~\ref{sec:exp_value} to \ref{sec:eval-matr-elem}.

\section{Equations of the self-consistent field}
\label{sec:equation-of-motion}

Let us start from writing the QED Hamiltonian in the Coulomb gauge \cite{bjorken1965,HeitlerB1954quantum} in the Schr\"{o}dinger representation
\begin{align}
  \label{eq:1}
  \opa H_{\mathrm{QED}} &= \sum_{\vec k \lambda}\omega_{\vec k}\opa c^\dag_{\vec k \lambda}\opa c_{\vec k \lambda} + \int d\vec x :\uppsi^\dag(\vec x)(\vec \alpha\cdot\opA p +\beta m_0)\uppsi(\vec x): \nonumber
  \\
                        &+ \frac{e_0^2}{8\pi}\int \frac{d\vec x d\vec y}{|\vec x - \vec y|}:\uprho(\vec x)::\uprho(\vec y): + \int d\vec x:\uppsi^\dag(\vec x)\vec \alpha\cdot (-e_0 \opA A(\vec x))\uppsi(\vec x):
\end{align}
and introducing the notations.

Through the text we use natural units $\hbar = c = 1$ and all operators are written in straight font. In Eq.~(\ref{eq:1}) $\opA p = -\ri\nabla$ is the momentum operator, the two colon symbol $::$ describes the normal ordering of operators, $e_{0}$, $m_{0}$ are the charge and the mass of the bare electron, $\vec \alpha$ and $\beta$ the Dirac matrices and $\uppsi(\vec x)$, $\uppsi^{\dag}(\vec x)$ the operators at the position $\vec x$ of the electron-positron field in the secondary-quantized representation
\begin{align}
  \uppsi(\vec x) &= \sum_{\vec ps} \frac{1}{\sqrt{V}}\sqrt{\frac{m_{0}}{\epsilon_{\vec p}}}(\opa a_{\vec ps}u_{\vec ps}e^{\ri\thp{p}{x}}+\opa b_{\vec ps}^\dag v_{\vec ps}e^{-\ri\thp{p}{x}}), \label{eq:2}
	\\
	\uppsi^\dag(\vec x) &= \sum_{\vec ps} \frac{1}{\sqrt{V}}\sqrt{\frac{m_{0}}{\epsilon_{\vec p}}}(\opa a^\dag_{\vec ps}u^\dag_{\vec ps}e^{-\ri\thp{p}{x}}+\opa b_{\vec ps} v^\dag_{\vec ps}e^{\ri\thp{p}{x}}). \label{eq:3}
\end{align}

In Eqs.~(\ref{eq:2}) and (\ref{eq:3}) $\opa a^\dag_{\vec ps}$, $\opa a_{\vec ps}$ are the creation and annihilation operators of the bare electron field with the corresponding bispinor $u_{\vec ps}$ and $(\hat p - m_{0})u_{\vec ps} = 0$. A hat on the top of quantities is defined as the contraction of the Dirac gamma matrices with the four vectors $\hat f = \sum_{\mu}\gamma^{\mu}f_{\mu}$. $\opa b^\dag_{\vec ps}$, $\opa b_{\vec ps}$ are the creation and annihilation operators of the bare positron field with the corresponding bispinor $v_{\vec ps}$ and $(\hat p + m_{0})v_{\vec ps} = 0$. $\vec p$ and $s$ are the momentum and the helicity of the electron (positron) field, respectively. The operators of the electron-positron field anti-commute, with the only two nonzero anti-commutators $\{\opa a_{\vec ps},\opa a^{\dag}_{\vec p^{\prime}s^{\prime}}\} = \delta_{ss^{\prime}} \delta_{\vec p \vec p^{\prime}}$ and $\{\opa b_{\vec ps},\opa b^{\dag}_{\vec p^{\prime}s^{\prime}}\} = \delta_{ss^{\prime}} \delta_{\vec p \vec p^{\prime}}$ and commute with the operators of the photon field. $\uprho(\vec x) = \uppsi^{\dag}(\vec x)\uppsi(\vec x)$ is the density of the electron-positron field
\begin{align}
  \label{eq:4}
  :\uprho: = \frac{1}{V}\sum_{\mathclap{\vec ps,\vec p^\prime s^\prime}}\sqrt{\frac{m_{0}}{\epsilon_{\vec p}}}\sqrt{\frac{m_{0}}{\epsilon_{\vec p^\prime}}}&\Bigg(\opa a_{\vec ps}^\dag \opa a_{\vec p^\prime s^\prime} u^\dag_{\vec ps}u_{\vec p^\prime s^\prime}e^{-\ri(\vec p - \vec p^\prime)\cdot \vec x}+\opa a_{\vec ps}^\dag \opa b_{\vec p^\prime s^\prime}^\dag u^\dag_{\vec ps}v_{\vec p^\prime s^\prime}e^{-\ri(\vec p + \vec p^\prime)\cdot \vec x} \nonumber
	\\
	&+\opa b_{\vec ps} \opa a_{\vec p^\prime s^\prime} v^\dag_{\vec ps}u_{\vec p^\prime s^\prime}e^{\ri(\vec p + \vec p^\prime)\cdot \vec x}-\opa b_{\vec p^\prime s^\prime}^\dag \opa b_{\vec p s} v^\dag_{\vec p s}v_{\vec p^\prime s^\prime}e^{\ri(\vec p - \vec p^\prime)\cdot \vec x}\Bigg).
\end{align}

$\opA A(\vec x)$ is the vector potential of the transverse electromagnetic field
\begin{align}
  \label{eq:5}
  \opA A(\vec x) = \sum_{\vec k \lambda}\frac{\vec e_{\vec k \lambda}}{\sqrt{2V\omega_{\vec k}}}\left(\opa c_{\vec k \lambda}e^{\ri \thp{k}{x}} + \opa c^{\dag}_{\vec k\lambda}e^{-\ri\thp{k}{x}}\right),
\end{align}
with $\opa c^{\dag}_{\vec k \lambda}$, $\opa c_{\vec k \lambda}$ being the creation and annihilation operators of the photon with the wave vector $\vec k$, the frequency $\omega_{\vec k} = |\vec k|$ and the polarization $\lambda$. The operators of the photon field commute with the operators of the electron-positron field with the only nonvanishing commutator $[\opa c_{\vec k\lambda}, \opa c^{\dag}_{\vec k^{\prime}\lambda^{\prime}}] = \delta_{\lambda\lambda^{\prime}}\delta_{\vec k \vec k^{\prime}}$. We also denote through $\opA P = \int d\vec x \uppsi(\vec x)\opA p \uppsi(\vec x)$ the total momentum of the electron-positron system.

The QED Hamiltonian (\ref{eq:1}) consists of four terms. The first two terms describe the energies of the free electromagnetic and the electron-positron fields. The third term, being quadratic in the density $\uprho(\vec x)$, is the so-called instantaneous interaction between charges, while the fourth one represents the interaction between the transverse electromagnetic field $\Div\opA A(\vec x) = 0$ and the current of the electron-positron field $\opA j = \uppsi^{\dag}(\vec x)\vec \alpha \uppsi(\vec x)$.

As was described in the Introduction we are seeking for the soliton-like solution, corresponding to the case of the vacuum for the transverse electromagnetic field $\opA A(\vec x)$. Consequently, we consider the vacuum average with respect to the state vectors of the latter. For this reason we introduce the reduced Hamiltonian operator
\begin{align}
  \label{eq:6}
  \opa H^{\prime}_{\mathrm{QED}} = \langle0_{\mathrm{ph}}|\opa H_{\mathrm{QED}}|0_{\mathrm{ph}}\rangle &= \int d\vec x :\uppsi^\dag(\vec x)(\vec \alpha\cdot\opA p +\beta m_0)\uppsi(\vec x): \nonumber
  \\
  &\mspace{90mu}+ \frac{e_0^2}{8\pi}\int \frac{d\vec x d\vec y}{|\vec x - \vec y|}:\uprho(\vec x)::\uprho(\vec y):,
\end{align}
which becomes the starting point for all the subsequent analysis.

It is well known that in quantum field theory the solitons or solitary waves can not be obtained by treating the interaction terms on the perturbative basis \cite{RajaramanB1982solitons,RevModPhys.61.763,RevModPhys.79.1139}. For example, in order to obtain a soliton-like solution in the simplest case of a scalar field \cite{PhysRevD.11.1094} with the Hamiltonian
\begin{align}
  \label{eq:7}
  \opa H_{\mathrm{S}} =  \int d x \left\{\frac{1}{2}\uppi^2(x) + \frac{1}{2}\left(\frac{\partial \upphi (x)}{\partial x}\right)^2 + \frac{g}{4}\left(\upphi^{2}(x)-\frac{m^{2}}{g}\right)^{2}\right\},
\end{align}
one displaces the classical component $\phi(x)$ from the field operators $\upphi(x)$ and minimizes the functional
\begin{align}
  \label{eq:8}
  \mathfrak{I}[\phi(x)] &= \int d x \left\{\frac{1}{2}\left(\frac{\partial \phi (x)}{\partial x}\right)^2 + \frac{g}{4}\left(\phi^{2}(x)-\frac{m^{2}}{g}\right)^{2}\right\},
  \\
  \frac{\delta\mathfrak{I}}{\delta \phi(x)} &= 0. \label{eq:9}
\end{align}

The transition from Eq.~(\ref{eq:7}) to Eq.~(\ref{eq:8}) is based on the application of the variational principle when the trial state vector $|\Psi_{\mathrm{S}}\{\phi(x)\}\rangle$ of the initial quantum system is chosen as a coherent state, with a coherent state parameter $\phi(x)$
\begin{align}
  \label{eq:10}
  \mathfrak{I} = \langle\Psi_{\mathrm{S}}\{\phi(x)\}|\opa H_{\mathrm{S}}|\Psi_{\mathrm{S}}\{\phi(x)\}\rangle.
\end{align}

As a result an approximate substitution of the linear Schr\"{o}dinger equation for the determination of the state vector $|\Psi_{\mathrm{S}}\rangle$ with the nonlinear equation for the classical component $\phi(x)$ is performed.

However, the fermionic nature of the operators of the electron-positron field in the QED Hamiltonian (\ref{eq:6}) does not allow one to substitute them with the corresponding classical functions (the corresponding expression will contain the Grassman variables \cite{RajaramanB1982solitons}). However, the nonlinear part still can be separated out in this case. For this we refer to the well know Hartree method of the self-consistent field in the description of an atom \cite{LandauQM}. There, the interaction term has the same structure as the interaction term in Eq.~(\ref{eq:6}) and exactly this term leads to a nonlinearity. Then, starting from the variational method the equations of the self-consistent field are derived, which are indeed the nonlinear equations.

Consequently, in order to obtain the soliton-like solution in quantum electrodynamics we will not split the reduced Hamiltonian $\opa H^{\prime}_{\mathrm{QED}}$ into the ``bare'' and the interaction parts. Instead we will try to apply the method of the self-consistent field \cite{PhysRevD.11.1094,LandauQM,RevModPhys.54.913,PhysRevD.89.025008,PhysRevLett.111.121602,PhysRevD.78.065043}. For this reason, we neglect the quantum fluctuations in the density of the electron-positron field. This results in the replacement of the exact density operator $:\uprho(\vec x):$ through the mean density
\begin{align}
  \label{eq:11}
  :\uprho(\vec x):\, \eqsim \langle:\uprho(\vec x):\rangle \equiv \langle\psi_{0}|:\uprho(\vec x):|\psi_{0}\rangle,
\end{align}
where the expectation value is calculated with some trial state vector $|\psi_{0}\rangle$. Proceeding further as in the Hartree method, i.e. by calculating the functional for the energy one obtains
\begin{align}
  \label{eq:12}
  \mathfrak{J}[|\psi_{0}\rangle] &= \int d\vec x \langle\psi_{0}|:\uppsi^\dag(\vec x)(\vec \alpha\cdot\opA p +\beta m_0)\uppsi(\vec x):|\psi_{0}\rangle \nonumber
  \\
  &\mspace{90mu}+ \frac{e_0^2}{8\pi}\int \frac{d\vec x d\vec y}{|\vec x - \vec y|}\langle\psi_{0}|:\uprho(\vec x):|\psi_{0}\rangle\langle\psi_{0}|:\uprho(\vec y):|\psi_{0}\rangle
\end{align}

As a result, our task is to adequately choose the state vector $|\psi_{0}\rangle$. For this, we employ an analogy with the theory of superconductivity \cite{PhysRev.108.1175,BlattB1971theory,AbrikosovB1975methods}, where the trial state vector takes into account the coupling between electrons with different momenta, which form the Cooper pair. Consequently, in the case of QED we suppose that the soliton-like state can be formed by the whole spectrum of the single electron and positron states  with all possible momenta and spins. Since the states $\opa a^{\dag}_{\vec qs}|0\rangle$ and $\opa b^{\dag}_{\vec qs}|0\rangle$ span the single-particle subspace they provide a complete basis for expanding the trial function. For this reason, the most general trial state vector of the collective excitation of the electron-positron field can be chosen as a linear combination of these single-particle states
\begin{align}
  \label{eq:13}
  |\psi_0\rangle = \sum_{\vec q r}(U_{\vec q r}\opa a^{\dag}_{\vec q r}+V_{\vec q r}\opa b^{\dag}_{\vec q r})|0\ra,
\end{align}
with arbitrary unknown mixing coefficients $U_{\vec qr}$ and $V_{\vec qr}$, describing the population of the various single-particle states of the electron-positron field. Therefore, if the soliton solution exists, it will be described by these coefficients.

We mention here, that this nonperturbative approach based on a modeling of a state vector, which we call the operator method \cite{FeranchukB2015nonperturbative} was successfully applied in a large amount of quantum mechanical problems. For example, the modelling of the initial state vector for the most pictorial case of the anharmonic oscillator leads to an approximation of the energy levels, which is uniformly convergent to the exact numerical results in the whole range of variation of the coupling constant. In addition we have recently demonstrated the effectiveness of this approach in a nonperturbative description of the interaction between a particle and a scalar quantum field \cite{PhysRevD.92.125019}. Moreover, a different point of view on the similar Nambu-Jona-Lasinio problem can be obtained with the use of the path integral formalism in QFT \cite{ALKOFER1996139,CHRISTOV199691}.

Coming back to the state vector $|\psi_{0}\rangle$, we require it to be normalized, which leads to the condition on the coefficients
\begin{align}
  \label{eq:14}
  \langle\psi_{0}|\psi_{0}\rangle = \sum_{\vec qr}(|U_{\vec qr}|^2+|V_{\vec qr}|^2) = 1.
\end{align}

Another condition on these coefficients is associated with the fact that the charge operator $\opa Q = e_{0} \sum_{\vec qs}(\opa a^{\dag}_{\vec q s}\opa a_{\vec qs} - \opa b^{\dag}_{\vec q s}\opa b_{\vec qs})$ commutes with the QED Hamiltonian $\opa H_{\mathrm{QED}}$ and consequently any collective excitation of the electron-positron system should possess some charge $e$
\begin{align}
  \label{eq:15}
  e = \langle\psi_{0}|\opa Q|\psi_{0}\rangle = e_0\sum_{\vec qr}\left(|U_{\vec q r}|^2 - |V_{\vec q r}|^2\right),
\end{align}
which in the general case is different from the ``bare'' electron charge $e_{0}$.

Therefore, we can immediately conclude from Eq.~(\ref{eq:15}) that the unknown coefficients $U_{\vec qr}$, $V_{\vec q r}$ can not be equally normalized. For this reason, we introduce the quantity $C$, which describes the relative population of the electron field with respect to the positron one. As a result, if we normalize $U_{\vec qr}$, $V_{\vec q r}$ independently
\begin{align}
  \label{eq:16}
  \sum_{\vec qs}|U_{\vec qs}|^2 = \frac{1}{1+C}, \quad \sum_{\vec qs}|V_{\vec qs}|^2 = \frac{C}{1+C},
\end{align}
the normalization condition Eq.~(\ref{eq:14}) for the state vector $|\psi_{0}\rangle$ will be automatically fulfilled for an arbitrary value of $C$. We want to stress here that we are seeking for the nontrivial solution, when the coefficient functions $U_{\vec qr}$ and $V_{\vec qr}$ are differently normalized and consequently the observed charge is nonvanishing.

We continue with the calculation of the functional defined via Eq.~(\ref{eq:12}), which is discussed in detail in Appendix~\ref{sec:exp_value}. This yields for the functional $\mathfrak{J}$
\begin{align}
  \label{eq:17}
  \mathfrak{J}[\Psi(\vec  x),\Psi^c(\vec x)] &= \int d\vec x \Bigg\{\Psi^\dag (\vec x) \left[(\vec \alpha\cdot\opA p + \beta m_0) + \frac{1}{2}e_0 \varphi (\vec x) \right] \Psi (\vec x)  \nonumber
  \\
                            &-\Psi^{c\dag} (\vec x) \left[(\vec \alpha\cdot \opA p + \beta m_0)+\frac{1}{2}e_0 \varphi (\vec x) \right] \Psi^c (\vec x)\Bigg\},
\end{align}
where we introduced the potential of the self-consistent field $\varphi(\vec x)$
\begin{align}
  \label{eq:18}
  \varphi (\vec x) = \frac{e_0}{4 \pi} \int \frac {d \vec{y}} {|\vec x - \vec{y}|} \left[\Psi^{\dag}(\vec y) \Psi (\vec y) - \Psi^{c\dag} (\vec y) \Psi^{c} (\vec y)\right]
\end{align}
and the inverse Fourier transforms
\begin{align}
  \Psi(\vec x) &= \sum_{\vec qr}\frac{1}{\sqrt{V}}\sqrt{\frac{m_{0}}{\epsilon_{\vec q}}}U_{\vec q r}u_{\vec q r}e^{i\thp{q}{x}}, \label{eq:19}
  \\
  \Psi^c(\vec x) &= \sum_{\vec qr}\frac{1}{\sqrt{V}}\sqrt{\frac{m_{0}}{\epsilon_{\vec q}}}V^*_{\vec q r}v_{\vec q r}e^{-i\thp{q}{x}}. \label{eq:20}
\end{align}
of the unknown coefficients $U_{\vec qr}$, $V^{*}_{\vec q r}$ respectively. In addition, according to their definition $\Psi(\vec x)$ and $\Psi^{c}(\vec x)$ are the Dirac bispinor wave functions.

The wave functions $\Psi(\vec x)$ and $\Psi^{c}(\vec x)$ satisfy a normalization condition, which follows from Eq.~(\ref{eq:16}), i.e.
\begin{align}
  \mathfrak{N}[\Psi(\vec  x)] &= \int {d \vec{x}} \Psi^{\dag} (\vec x) \Psi (\vec x) = \frac{1}{1+C}, \label{eq:21}
  \\
  \mathfrak{N}_{1}[\Psi^c(\vec x)] &= \int {d \vec{x}} \Psi^{c\dag} (\vec x) \Psi^c (\vec x) = \frac{C}{1+C}. \label{eq:22}
\end{align}

One can also determine the asymptotic behavior of the self-consistent potential for the large values of $|\vec x|$. Indeed, if we suppose that the functions $\Psi(\vec y)$ and $\Psi^{c}(\vec y)$ tend to zero when $|\vec y|\to\infty$, then the expansion of the denominator in Eq.~(\ref{eq:18}) for the large values of $|\vec x|$ yields
\begin{align}
  \label{eq:23}
  \varphi(|\vec x| \to \infty) &\approx \frac{e_{0}}{4\pi|\vec x|}\int d\vec y\left[\Psi^{\dag}(\vec y) \Psi (\vec y) - \Psi^{c\dag} (\vec y) \Psi^{c} (\vec y)\right] \nonumber
  \\
  &=\frac{e_{0}}{4\pi|\vec x|}\frac{1-C}{1+C}.
\end{align}

Before proceeding, we want to discuss the difference of QED with respect to the hadronic models \cite{PhysRevD.10.4130,PhysRevD.11.1094} regarding the change of the vacuum energy for the vacuum state and the single-charge state. For example, in Ref.~\cite{PhysRevD.10.4130} the authors considered a bosonic field as in Eq.~(\ref{eq:7}) and found that, if the first quantum correction $\upeta(x)$ to the classical component of the bosonic field $\phi(x)$ is taken into account, i.e. $\upphi(x) = \phi(x) + \upeta(x)$, then the vacuum expectation value of terms quadratic in $\upeta(x)$ is not completely cancelled with the vacuum energy and is of the same order of magnitude as $\phi(x)$. Consequently, this contribution should be taken into account. Contrary to this case, in QED it is well known \cite{jauch2012theory} that the vacuum diagrams, i.e. the diagrams with no external lines, do not contribute into any observable values. For this reason, it can be demonstrated (see Appendix~\ref{sec:change-vacuum-energy}) that the vacuum energy in the single-charge state is identical to the one in the vacuum state, and therefore, there is no change in vacuum energies.

Let us come back to the functional for the energy of the system, defined by Eq.~(\ref{eq:17}). As we already mentioned above, our starting point was the linear Schr\"{o}dinger equation for the system state vector. However, the replacement of the density operator through its classical value and the corresponding modeling of the state vector brought us to the functional which has terms of the fourth order with respect to the variational functions $\Psi(\vec  x)$ and $\Psi^c(\vec x) $. These wave functions can be considered as the ones of the electron and positron components of the unknown soliton-like solution. Moreover, we would like to stress here that the soliton-like solution is described via the pair of coefficient functions $\{U_{\vec qr},V^{*}_{\vec qr}\}$ or equivalently via their inverse Fourier transforms $\{\Psi(\vec x), \Psi^{c}(\vec x)\}$ and these functions should be always considered in pairs and never independently of each other.

In order to determine the equations for the wave functions $\Psi(\vec x)$ and $\Psi^{c}(\vec x)$ or equivalently for their Fourier transforms $U_{\vec qr}$, $V^{*}_{\vec q r}$ we proceed as in the usual variational method \cite{MessiahB1981quantum}, in accordance with the SLAC bag model of the quark \cite{PhysRevD.11.1094}, namely we introduce two Lagrange multipliers $\Lambda$ and $\Lambda_{c}$ in order to satisfy the two additional constrains of Eqs.~(\ref{eq:21})-(\ref{eq:22}) and find the minimum of the functional
\begin{align}
  \label{eq:24}
  \mathfrak{T}[\Psi(\vec  x),\Psi^c(\vec x)] = \mathfrak{J}[\Psi(\vec  x),\Psi^c(\vec x)] - \Lambda \mathfrak{N}[\Psi(\vec  x)] - \Lambda_{c} \mathfrak{N}_{1}[\Psi^c(\vec x)]
\end{align}
with respect to the wave functions $\Psi(\vec x)$ and $\Psi^{c}(\vec x)$. The variation of this functional is described in Appendix~\ref{sec:var1}. Consequently, this yields the two nonlinear equations
\begin{align}
  &[(-\ri\vec \alpha\cdot\nabla + \beta m_0) + e_0 \varphi (\vec x)]\Psi (\vec x) =  \Lambda \Psi (\vec x), \label{eq:25}
  \\
  &[(-\ri\vec \alpha\cdot\nabla + \beta m_0) + e_0 \varphi (\vec x)]\Psi^c (\vec x) =  -\Lambda_{c} \Psi^c(\vec x), \label{eq:26}
  \\
  &\varphi (\vec x) = \frac{e_0}{4 \pi} \int \frac {d \vec{y}} {|\vec x - \vec{y}|} \left[\Psi^{\dag}(\vec y) \Psi (\vec y) - \Psi^{c\dag} (\vec y) \Psi^{c} (\vec y)\right], \nonumber
  \\
  &\int {d \vec{x}} \Psi^{\dag} (\vec x) \Psi (\vec x) = \frac{1}{1+C},\quad \int {d \vec{x}} \Psi^{c\dag} (\vec x) \Psi^c (\vec x) = \frac{C}{1+C}. \nonumber
\end{align}

\section{Separation of variables. The method of solution. Integral characteristics}
\label{sec:separation-variables}

In the previous section based on the self-consistent field method, we constructed a nonlinear system of integro-differential equations, which describes the collective excitation of the electron-positron system in the absence of the photon field. In general, the solution of this system of equations is a nontrivial mathematical problem, as the separation of variables is impossible to perform due to the nonlinearity. Consequently, we separate the variables by employing an ansatz for the wave functions, which we describe in what follows. We also note here that the analogous approach is used in the polaron problem \cite{Mitra198791,RevModPhys.63.63,Spohn1987278,0022-3719-17-24-012,PhysRev.97.660} and in the ``bag'' model of a quark \cite{PhysRevD.11.1094,PhysRevD.10.4114,PhysRevD.10.4130,PhysRevD.10.4138}.

In order to proceed we first of all consider the case when the total momentum of the electron-positron system is equal to zero, i.e. $\langle\psi_{0}|\opA P|\psi_{0}\rangle = 0$ (later we will discuss the case $\langle{\opA P}\rangle \neq 0$). In this case the system does not possess any preferable vectors defining some direction. Oppositely to this situation, when $\langle\opA P\rangle \neq 0$ the direction of motion is preferable. As the second step we note that the self-consistent potential $\varphi(\vec x)$ has a boundary condition as a spherically symmetric function Eq.~(\ref{eq:23}). Furthermore, it is well known that the Dirac equation allows the separation of variables in a spherical coordinate system \cite{AkhiezerB1965quantum,LandauQED,FlueggeB1994practical}. Consequently, we introduce the spherically symmetric ansatz for the wave functions
\begin{align}\label{eq:27}
  \Psi_{jlM} =\begin{pmatrix}
    g(r) \Omega_{jlM}\\
    \ri f(r) \Omega_{jl^\prime M}
  \end{pmatrix}, \quad \Psi^c_{jlM} = \begin{pmatrix}
    g_1(r) \Omega_{jlM}\\
    \ri f_1(r) \Omega_{jl^\prime M}
  \end{pmatrix},
\end{align}
where $j$ is the quantum number of the total angular momentum operator $\opA J$, $l$ the quantum number of the orbital angular momentum operator $\opA L$, $l^{\prime} = 2j - l$, $M$ the quantum number of $\opa J_{z}$ and $\Omega_{jlM}$ the spherical spinors \cite{AkhiezerB1965quantum,LandauQED,FlueggeB1994practical}. The properties of $\Omega_{j l M}$ are briefly presented in Appendix~\ref{sec:deriv-radi-equat}. The self-consistent potential is calculated through the density $\rho(\vec x)$, which in turn is calculated through the wave functions themselves. Consequently, the ansatz (\ref{eq:27}) results in the spherically symmetric self-consistent potential and therefore the variables in the Dirac equation can be separated in the spherical coordinates in a self-consistent way.

Further simplification is associated with the fact that we are trying to seek the state with the lowest nonzero energy. Consequently, as the large eigenvalues $j$ of the total angular momentum operator $\opA J$ correspond to the larger energy it is quite natural to assume that our state possesses the minimal possible eigenvalue, namely $j = 1/2$. As a result, the two values of the eigenvalue $l$ are possible, i.e. either $l = 0$ and $l^{\prime} = 1$ or $l = 1$ and $l^{\prime} = 0$. In the following, we will demonstrate that the solutions for both these situations exist. Until then we fix the values for $l$ and $l^{\prime}$ as $l = 0$ and $l^{\prime} = 1$.

The last remark reflects the situation that $M = \pm 1/2$ is a two-fold degenerate eigenvalue. For this reason, the most general linear combination of the wave functions (\ref{eq:27}) can be written as
\begin{equation}
  \begin{aligned}
    \Psi(\vec x) &= A_{\frac{1}{2}} \Psi_{\frac{1}{2}, 0, \frac{1}{2}}+A_{-\frac{1}{2}}\Psi_{\frac{1}{2},0,-\frac{1}{2}} = \begin{pmatrix}
      g(r)\chi_{0}
      \\
      \ri f(r)\chi_{1}
    \end{pmatrix},
    \\
    \Psi^c(\vec x) &= A_{\frac{1}{2}}^{c} \Psi^c_{\frac{1}{2}, 0, \frac{1}{2}}+A_{-\frac{1}{2}}^{c}\Psi^c_{\frac{1}{2},0,-\frac{1}{2}} = \begin{pmatrix}
      g_1(r)\chi^{c}_{0}
      \\
      \ri f_1(r)\chi^{c}_{1}
    \end{pmatrix},
  \end{aligned}\label{eq:28}
\end{equation}
where the spinors $\chi_{0}$, $\chi_{1}$, $\chi_{0}^{c}$ and $\chi_{1}^{c}$ are defined in Appendix~\ref{sec:deriv-radi-equat}. In addition, the coefficients of these linear combinations satisfy $|A_{\frac{1}{2}}|^{2}+|A_{-\frac{1}{2}}|^{2} = 1$, $|A^{c}_{\frac{1}{2}}|^{2}+|A^{c}_{-\frac{1}{2}}|^{2} = 1$ and $A^{*}_{\frac{1}{2}}A^{c}_{\frac{1}{2}}+A^{*}_{-\frac{1}{2}}A^{c}_{-\frac{1}{2}}=0$, such that the wave functions $\Psi(\vec x)$ and $\Psi^{c}(\vec x)$ satisfy the orthogonality relation $\int d\vec x \Psi^{\dag}(\vec x)\Psi^{c}(\vec x) = 0$ and the normalization relations (\ref{eq:21}), (\ref{eq:22}) (see Appendix~\ref{sec:deriv-radi-equat}).

Let us come directly to the separation of variables in Eqs.~(\ref{eq:25})--(\ref{eq:26}). Due to the choice of the wave functions (\ref{eq:28}) the angular dependence is fully determined with the spherical spinors $\chi_{0}$, $\chi_{1}$, $\chi_{0}^{c}$ and $\chi_{1}^{c}$. Consequently we need to find the remaining radial functions $g(r)$, $f(r)$, $g_{1}(r)$ and $f_{1}(r)$. This is performed by plugging the wave functions (\ref{eq:28}) into the two Dirac Eqs.~(\ref{eq:25})--(\ref{eq:26}) as well as in the definition of the self-consistent potential Eq.~(\ref{eq:18}). This procedure is described in great detail in Appendix~\ref{sec:deriv-radi-equat}. The final result written in the dimensionless variables Eq.~(\ref{eq:deriv-radi-equat:15}) (see Appendix~\ref{sec:deriv-radi-equat}) reads
\begin{align}
  \left\{\begin{aligned}
      &u^\prime(x) - \frac{u(x)}{x} - (1-\phi(x) ) v(x) = \lambda v(x),
      \\
      &v^\prime(x) + \frac{v(x)}{x} - (1+\phi(x)) u(x) = -\lambda u(x),
      \\
      &u_c^\prime(x) - \frac{u_c(x)}{x} - (1-\phi(x) ) v_c(x) = -\lambda_c v_c(x),
      \\
      &v_c^\prime(x) + \frac{v_c(x)}{x} - (1+\phi(x)) u_c(x) = \lambda_c u_c(x),
      \\
      &\phi(x) = \frac{\alpha_0}{x}\int_0^x \rho(y)dy + \alpha_0 \int_x^\infty \frac{\rho(y)}{y}dy,
      \\
      &\rho(x) = u^2(x) + v^2(x) - u_c^2(x) - v_c^2(x)
      \\
      &\int_0^\infty dx[u^2(x) + v^2(x)] = \frac{1}{1+C}, \quad \int_{0}^{\infty}dx[u_c^2(x) + v_c^2(x)] = \frac{C}{1+C}.
    \end{aligned}\right. \label{eq:29}
\end{align}

This system of equations should be complemented with the boundary conditions resulting from the asymptotic behavior of the functions $u(x)$, $v(x)$, $u_{c}(x)$ and $v_{c}(x)$ near $x=0$ and $x\to\infty$ respectively:
\begin{equation}
  \begin{aligned}
    &u(x) \sim F_{0}x\left(1+\frac{1-(\lambda - \phi(0))^2}{6}x^2\right),& &v(x) \sim F_{0}\frac{1-(\lambda - \phi(0))}{3}x^2,& &x\rightarrow0,
    \\
    &u(x) \sim F_{\infty} e^{-\sqrt{1- \lambda^2}x},& &v(x) \sim -F_{\infty}\sqrt{\frac{1-\lambda}{1+\lambda}} e^{-\sqrt{1-\lambda^2}x},& &x \rightarrow \infty,
  \end{aligned}\label{eq:30}
\end{equation}
with the corresponding equations for $u_{c}(x)$ and $v_{c}(x)$. Here $F_{0}$  and $F_{\infty}$ are arbitrary constants of integration.

The eigenvalue problem for the nonlinear integro-differential Eqs.~(\ref{eq:29}) with the corresponding boundary conditions (\ref{eq:30}) was obtained from the linear Schr\"{o}dinger equation with the help of the self-consistent field method and describes the collective excitation of the electron-positron system in quantum electrodynamics in the absence of the photon field. The self-consistent potential $\phi(x)$ is calculated with the help of functions $u(x)$, $v(x)$, $u_{c}(x)$ and $v_{c}(x)$. This system of equations is an analog of the Hartree equations in the description of an atom. Consequently, its solution can be found only numerically. For this reason in what follows we present the numerical algorithm for the solution of this system of equations.

First of all, we note that the functions $u(x)$, $v(x)$, $u_{c}(x)$ and $v_{c}(x)$ are differently normalized. We, however, would like to replace them with functions that are equally normalized. Therefore, we introduce the new normalized wave functions $u_{N}(x)$, $v_{N}(x)$, $u_{cN}(x)$ and $v_{cN}(x)$ as
\begin{equation}
  \label{eq:31}
  \begin{aligned}
    &u(x) = \sqrt{\frac{1}{1+C}}u_N(x), \quad  v(x) = \sqrt{\frac{1}{1+C}}v_N(x),
    \\
    &u_{c}(x) = \sqrt{\frac{C}{1+C}}u_{cN}(x), \quad  v_{c}(x) = \sqrt{\frac{C}{1+C}}v_{cN}(x),
    \\
    &\int_0^\infty dx [u_{N}^2(x) + v_{N}^2(x)] = \int_0^\infty dx [u_{cN}^2(x) + v_{cN}^2(x)] = 1.
  \end{aligned}
\end{equation}

As a result the density $\rho(x)$ in the self-consistent potential becomes a function of $C$
\begin{equation}
  \label{eq:32}
  \rho_{N}(x) = \frac{1}{1+C}(u_{N}^2(x) + v_{N}^2(x)) - \frac{C}{1+C}(u_{cN}^2(x) + v_{cN}^2(x)),
\end{equation}
and the system of Eqs.~(\ref{eq:29}) transforms as
\begin{equation}
  \label{eq:33}
    \left\{\begin{aligned}
      &u_{N}^\prime(x) - \frac{u_N(x)}{x} - (1-\phi(x) ) v_N(x) = \lambda v_N(x),
      \\
      &v_N^\prime(x) + \frac{v_N(x)}{x} - (1+\phi(x)) u_N(x) = -\lambda u_N(x),
      \\
      &u_{cN}^\prime(x) - \frac{u_{cN}(x)}{x} - (1-\phi(x) ) v_{cN}(x) = -\lambda_c v_{cN}(x),
      \\
      &v_{cN}^\prime(x) + \frac{v_{cN}(x)}{x} - (1+\phi(x)) u_{cN}(x) = \lambda_c u_{cN}(x),
      \\
      &\phi(x) = \frac{\alpha_0}{x}\int_0^x \rho_{N}(y)dy + \alpha_0 \int_x^\infty \frac{\rho_{N}(y)}{y}dy,
      \\
      &\rho(y) = \frac{1}{1+C}(u_{N}^2(y) + v_{N}^2(y)) - \frac{C}{1+C}(u_{cN}^2(y) + v_{cN}^2(y)),
      \\
      &\int_0^\infty dx (u_{N}^2(x) + v_{N}^2(x)) = \int_0^\infty dx (u_{cN}^2(x) + v_{cN}^2(x)) = 1.
    \end{aligned}\right.
\end{equation}

The boundary conditions, however, are not changed (up to notations of constants $F_{0}$ and $F_{\infty}$).

The solution of this system of equations have been performed numerically with the use of the continuous analog of Newton method \cite{KOMAROV1978153,GAREEV1977116,AirapetyanA1999continuous,ERMAKOV1981235,GavurinA1958nonlinear,PuzyninA1999generalized,ZhidkovA1973continuous}, which is described in detail in Appendix~\ref{sec:cont-anal-newt}. During the solution we first fixed the values of the parameters $C$ and $\alpha_{0}$. Then the system of equations has been solved as described in Appendix~\ref{sec:cont-anal-newt} and the two eigenvalues $\lambda$ and $\lambda_{c}$ were determined. It was found that for all values of $C$ and $\alpha_{0}$ for which the solution exists the eigenvalues $\lambda$ and $-\lambda_{c}$ coincide with each other, i.e. $\lambda = -\lambda_{c}$. This very important circumstance allows us to simplify the system of Eqs.~(\ref{eq:33}) significantly as the sets of functions $u_{N}(x)$, $v_{N}(x)$ and $u_{cN}(x)$, $v_{cN}(x)$ coincide with each other. As the immediate consequence, the system of four Eqs.~(\ref{eq:33}) transforms into the system of two equations.

For the following it is convenient to change notations. We introduce the new wave functions $u_{0}(x)$, $v_{0}(x)$, which are normalized to unity and the parameter
\begin{equation}
  \label{eq:34}
  q = \alpha_{0}\frac{1 - C}{1 + C}.
\end{equation}

Consequently, the system of Eqs.~(\ref{eq:33}) transforms into the form
\begin{equation}
  \label{eq:35}
  \left\{\begin{aligned}
      &u_0^\prime(x) - \frac{u_0(x)}{x} - (1-\phi(x) ) v_0(x) = \lambda v_0,
      \\
      &v_0^\prime(x) + \frac{v_0(x)}{x} - (1+\phi(x)) u_0(x) = -\lambda u_0,
      \\
      &\phi(x) = q \phi_0(x), \quad q = \alpha_0 \frac{1-C}{1+C},
      \\
      &\phi_0(x) = \frac{1}{x}\int_0^x \rho_0(y)dy + \int_x^\infty \frac{\rho_0(y)}{y}dy,
      \\
      &\rho_0(x) = u_0^2(x)+v_0^2(x),
      \\
      &\int_0^\infty (u_0^2 + v_0^2)dx = 1.
    \end{aligned}\right.
\end{equation}
which becomes the starting point of all subsequent considerations.

We stress the importance of the parameter $q$, which is associated with the self-similarity in our system, manifesting in the equality of the radial functions $g(r) = g_{1}(r)$ and $f(r) = f_{1}(r)$.

Another important conclusion comes from the fact that the spin part of the wave functions is determined up to linear combinations of the spherical spinors $\Omega_{\frac{1}{2},l,\pm\frac{1}{2}}$ in  $\chi_{0}$, $\chi_{1}$ and $\chi_{0}^{c}$, $\chi_{1}^{c}$ in
\begin{equation}
  \label{eq:36}
  \Psi(\vec x) = \begin{pmatrix}
    g_0(r) \chi_0\\
    \ri f_0(r) \chi_1
  \end{pmatrix}, \quad \Psi^{c}(\vec x) = \begin{pmatrix}
    g_0(r) \chi_0^{c}\\
    \ri f_0(r) \chi_1^{c}
  \end{pmatrix}.
\end{equation}
This reflects the arbitrariness in the choice of the spin quantization direction, which is quite natural when the total momentum of the system $\langle\opA P\rangle$ is equal to zero.

As the last step we need to determine the integral characteristics of the system. For this, we first mention that in the case of nonlinear equations the total energy of the system is not equal to the sum of the corresponding eigenvalues. For example, in the Hartree method the total energy of the system is not equal to the sum of the Lagrange multipliers $\epsilon_{i}$ introduced for the solution of the corresponding Schr\"{o}dinger equations. Instead, the mean energy of the interaction should be subtracted from this sum \cite{LandauQM}. Consequently, in our case the total energy of the system is not equal to the sum of $\lambda$ and $\lambda_{c}$. As a result, we can derive two integral characteristics. The first integral characteristic results from the equations of motion, while the second one is an outcome of the direct calculation of the functional Eq.~(\ref{eq:17}).

In order to find the first integral characteristic we multiply the first of Eqs.~(\ref{eq:35}) by $v_{0}(x)$, the second one by $u_{0}(x)$ and subtract the first result from the second one. This yields
\begin{align}
  T &+ q\Pi = \lambda, \label{eq:37}
  \\
  T &= \int_{0}^{\infty} dx \left([u_0^\prime(x) v_0(x) - v_0^\prime(x) u_0(x)] - \frac{2u_0(x) v_0(x)}{x} + [u_0^2(x) - v_0^2(x)]\right) \nonumber
  \\
  &\equiv B + D,\label{eq:38}
  \\
  \Pi &= \int_{0}^{\infty}dx \phi_0(x) [u_0^2(x) + v_0^2(x)]. \label{eq:39}
  \\
  B &= \int_{0}^{\infty}dx\left\{[u_{0}^{\prime}(x) v_{0}(x) - v^{\prime}_{0}(x) u_{0}(x)] - \frac{2u_{0}(x)v_{0}(x)}{x}\right\}, \label{eq:40}
  \\
  D &= \int_{0}^{\infty}dx[u_{0}^{2}(x) - v_{0}^{2}(x)]. \label{eq:41}
\end{align}

The second integral characteristic, namely the total energy of the system can be easily obtained from the functional (\ref{eq:17}). In that equation we add and subtract in the square brackets $\frac{1}{2} e_{0}\varphi(\vec x)$ and then employ the equations of the self-consistent field (\ref{eq:25}), (\ref{eq:26}). Consequently one obtains
\begin{align}
  \label{eq:42}
  E_{0} &= \int d\vec x \Psi^\dag(\vec x) \left(\Lambda - \frac{1}{2}e_0 \varphi(\vec x)\right)\Psi(\vec x) - \int d\vec x \Psi^{c\dag}(\vec x) \left(\Lambda - \frac{1}{2}e_0 \varphi(\vec x)\right)\Psi^c(\vec x) \nonumber
  \\
        &=\frac{1-C}{1+C}\left(\Lambda - \frac{1}{2} \int dr e_0 \varphi(r) \left[(rg_0)^2 + (rf_0)^2\right]\right) \nonumber
  \\
        &= \frac{m_{0}}{\alpha_{0}}\left(q \lambda - \frac{q^{2}}{2}\Pi\right) = \frac{m_{0}}{\alpha_{0}}\left(qT + \frac{q^{2}}{2}\Pi\right),
\end{align}
where on the last step we expressed $(1-C)/(1+C)$ through the parameter $q$ (\ref{eq:34}) and introduced the dimensionless variables (\ref{eq:deriv-radi-equat:15}). We should note here that the direct calculation from the functional yields, of course, the same result.

At last we want to point out that the radial equations (\ref{eq:35}) can be obtained by varying the functional for the energy Eq.~(\ref{eq:42}) with respect to $u_{0}(x)$ and $v_{0}(x)$.

\section{Solution of the second kind}
\label{sec:solut-second-kind}

In the previous section we have determined the state vector $|\psi_{0}\rangle$ and the integral characteristics, which describe the collective excitation of the electron-positron system. Before proceeding with the analysis of the solutions we need to demonstrate that the solution of the second kind exists
\begin{align}
  \label{eq:43}
  &|\psi_{0}^{\prime}\rangle = \sum_{\vec qr}(U^{\prime}_{\vec qr}\opa a^{\dag}_{\vec qr} + V^{\prime}_{\vec qr}\opa b^{\dag}_{\vec qr})|0\rangle,
\end{align}
which is orthogonal to the solution of the first kind and satisfies the normalization condition
\begin{align}
  \label{eq:44}
  \langle\psi^{\prime}_{0}|\psi^{\prime}_{0}\rangle = 1, \quad \langle\psi_{0}|\psi_{0}^{\prime}\rangle = 0.
\end{align}

The ansatz for the inverse Fourier transforms $\Psi^{\prime}(\vec x)$, $\Psi^{\prime c}(\vec x)$ of the mixing coefficients $U^{\prime}_{\vec qr}$, $V^{\prime*}_{\vec qr}$ in the state vector $|\psi_{0}^{\prime}\rangle$, which satisfies the conditions (\ref{eq:44}) can be chosen as
\begin{align}
  \label{eq:45}
  &U^{\prime}_{\vec qr} \rightarrow \Psi^{\prime}(\vec x) = (\vec \alpha\cdot\vec \nu)\Psi(\vec x), \quad V^{\prime*}_{\vec qr} \rightarrow \Psi^{\prime c}(\vec x) = (\vec \alpha\cdot\vec \nu)\Psi^{c}(\vec x),
  \\
  &\int {d \vec{x}} \Psi^{\prime\dag} (\vec x) \Psi^{\prime} (\vec x) = \frac{1}{1+C},\quad \int {d \vec{x}} \Psi^{\prime c\dag} (\vec x) \Psi^{\prime c} (\vec x) = \frac{C}{1+C}. \label{eq:46}
\end{align}
Here $\vec \nu$ is an arbitrary unit vector $(\vec \nu\cdot\vec \nu) = 1$, along which the quantization axis of the spherical spinors is directed. We want to emphasize that since the vector $\vec \nu$ is arbitrary, the direction of the quantization axis is also arbitrary. Consequently, the orthogonality relations
\begin{equation}
  \label{eq:47}
  \begin{aligned}
      \int d\vec x \Psi^{\prime\dag}(\vec x) \Psi(\vec x) &= \int d\vec x \Psi^{\dag}(\vec x)(\vec \alpha\cdot\vec \nu) \Psi(\vec x) = 0,
      \\
      \int d\vec x \Psi^{\prime c\dag}(\vec x) \Psi(\vec x) &= \int d\vec x \Psi^{c\dag}(\vec x)(\vec \alpha\cdot\vec \nu) \Psi^{c}(\vec x) = 0
  \end{aligned}
\end{equation}
are automatically fulfilled, due to the orthogonality of the spherical harmonics $Y_{00}$ and $Y_{1m}$, $m=\{-1,0,1\}$.

Further, we proceed exactly as in Sec.~\ref{sec:equation-of-motion} and calculate the functional for the energy. For this we note, that since the square of the Dirac matrix $\vec \alpha^{2} = 1$ and $\vec \nu$ is the unit vector, the term which is quadratic in the density does not change as $(\vec\alpha\cdot\vec \nu)^{2} = 1$, and consequently, we obtain
\begin{align}
  \label{eq:48}
  \mathfrak{J}^{\prime}[\Psi^{\prime}(\vec  x),\Psi^{\prime c}(\vec x)] &= \int d\vec x \Bigg[\Psi^{\dag}(\vec x)(\vec \alpha\cdot\vec \nu)(\vec \alpha\cdot\opA{p} +\beta m_0) (\vec \alpha\cdot\vec \nu)\Psi(\vec x)
  \\
  &\mspace{120mu}- \Psi^{c\dag}(\vec x)(\vec \alpha\cdot\vec \nu)(\vec \alpha\cdot\opA{p} +\beta m_0)(\vec \alpha\cdot\vec \nu)\Psi(\vec x)\Bigg] \nonumber
  \\
  &+ \frac{e_{0}^{2}}{8\pi}\int \frac{d\vec x d\vec y}{|\vec x - \vec y|}\left[\Psi^{\dag}(\vec x)\Psi(\vec x) - \Psi^{c\dag}(\vec x)\Psi^{c}(\vec x)\right]\left[\Psi^{\dag}(\vec y)\Psi(\vec y) - \Psi^{c\dag}(\vec y)\Psi^{c}(\vec y)\right]. \nonumber
\end{align}

The reduction of this functional from the full three dimensional form into the one dimensional one, containing only the integral characteristics (\ref{eq:38})--(\ref{eq:42}) and the radial functions $u_{0}(x)$, $v_{0}(x)$ is presented in Appendix \ref{sec:calc-expect-kinet}. The result yields
\begin{align}
  \label{eq:49}
  \mathfrak{J}^{\prime}[\Psi^{\prime}(\vec  x),\Psi^{\prime c}(\vec x)] &= - E_{0} 
  \\
  &+ \underbrace{\frac{m_{0}}{\alpha_{0}}q\left[ \int_{0}^{\infty}dx \left(\frac{2}{3}(u_{0}^{\prime}(x)v_{0}(x) - v_{0}^{\prime}(x)u_{0}(x))-\frac{4}{3}\frac{u_{0}(x)v_{0}(x)}{x}\right) + q\Pi\right]}_{X}.\nonumber
\end{align}

From this equation we observe that the functional $\mathfrak{J}^{\prime}$ is actually different from the functional $\mathfrak{J}$ and consequently, the functions $u_{0}(x)$ and $v_{0}(x)$ are different from the analogous functions of the solution of the first kind.

However, we can assume that the solutions of these two kinds are analogous to the positive and negative energy solutions of the free Dirac equation for a single particle. As a result we use the same functions $u_{0}(x)$ and $v_{0}(x)$ in both solutions and require that the energy of the system for the solution of the second kind is exactly equal to $-E_{0}$, i.e.
\begin{align}
  \label{eq:50}
  E_{0}^{\prime} = - E_{0}.
\end{align}

For this reason we would like to require that the quantity $X$ in Eq.~(\ref{eq:49}) vanishes. Consequently, in order to satisfy this condition we introduce the second Lagrange multiplier $\mu$ in the functional $\mathfrak{J}$. This will lead, of course, to the equations for the radial functions $u_{0}(x)$ and $v_{0}(x)$, which are different from Eqs.~(\ref{eq:35}) of the previous section. Therefore, our task is to establish weather the solution of these modified equations exists.

As was mentioned in the last paragraph of the previous section, the equations for the radial functions can be obtained from the variation of the functional Eq.~(\ref{eq:42}). Therefore, we will start from this functional $\mathfrak{J}$ and modify it, in order to incorporate the additional condition $X = 0$. This new functional $\mathfrak{I}$ reads as
\begin{align}
  \mathfrak{I} &= \frac{m_0}{\alpha_{0}} q \int dx \Bigg([u_0^\prime(x) v_0(x) - v_0^\prime(x) u_0(x)] - \frac{2u_0(x) v_0(x)}{x} \nonumber
  \\
               &\mspace{120mu}+ [u_0^2(x)- v_0^2(x)] +\frac{q}{2} \phi_0(x) [u_0^2(x) + v_0^2(x)] \Bigg) \nonumber
  \\
                 &- \frac{m_{0}}{\alpha_{0}}q \lambda\int_{0}^{\infty}dx[u_{0}^{2}(x) + v_{0}^{2}(x)] \nonumber
  \\
               &- \frac{m_{0}}{\alpha_{0}}q \mu \int_{0}^{\infty}dx\Bigg[\left(\frac{2}{3}[u_{0}^{\prime}(x)v_{0}(x) - v_{0}^{\prime}(x)u_{0}(x)] -\frac{4}{3} \frac{u_{0}(x)v_{0}(x)}{x}\right) \nonumber
  \\
                 &\mspace{180mu}+ q\phi_{0}(x)[u_{0}^{2}(x) + v_{0}^{2}(x)]\Bigg]. \label{eq:51}
\end{align}

The Lagrange multipliers $\lambda$ and $\mu$ require the two additional conditions to be fulfilled, namely $X=0$ and the normalization condition of the functions $u_{0}(x)$ and $v_{0}(x)$, respectively. The variation of this new functional is calculated in Appendix~\ref{sec:vari-funct-refeq}, the resulting modified equations are
\begin{equation}
  \label{eq:52}
  \begin{aligned}
      &\left(1 - \frac{2}{3}\mu\right)u_{0}^{\prime}(x) - \left(1 - \frac{2}{3}\mu\right)\frac{u_{0}(x)}{x} - [1 + \lambda - q(1 - 2\mu) \phi_{0}(x)]v_{0}(x) = 0,
      \\
      &\left(1 - \frac{2}{3}\mu\right)v_{0}^{\prime}(x) + \left(1 - \frac{2}{3}\mu\right)\frac{v_{0}(x)}{x} - [1 - \lambda + q (1 - 2\mu) \phi_{0}(x)]u_{0}(x) = 0,
      \\
      &\phi_{0}(x) = \frac{1}{x}\int_{0}^{x}dy [u_{0}^{2}(y) + v_{0}^{2}(y)] + \int_{x}^{\infty}dy\frac{[u_{0}^{2}(y) + v_{0}^{2}(y)]}{y}.
  \end{aligned}
\end{equation}

It is a remarkable fact that this new system of equations can be reduced to exactly the same form as the initial system of Eqs.~(\ref{eq:35}). The differences consist in the redefinition of the parameter $q$ and the rescaling of the radial variable $x$. Let us demonstrate this.

We start from the change of variable $x\to z$
\begin{equation}
  \label{eq:53}
  x = \left(1 - \frac{2}{3}\mu\right)z, \quad \frac{d}{dx} = \frac{1}{1 - \frac{2}{3}\mu}\frac{d}{dz}
\end{equation}
and introduce the new radial functions $\bar u_{0}(z)$ and $\bar v_{0}(z)$, which are normalized to unity
\begin{align}
  \label{eq:54}
  u_{0}(x) &= a \bar{u}_{0}(z), \quad v_{0}(x) = a \bar{v}_{0}(z),
  \\
  \int_{0}^{\infty} dx (u_{0}^{2}(x) + v_{0}^{2}(x)) &= \left(1 - \frac{2}{3}\mu\right)a^{2}\int_{0}^{\infty} dz (\bar{u}_{0}^{2}(z) + \bar{v}_{0}^{2}(z)) = 1 \quad \Rightarrow a = \frac{1}{\sqrt{1 - \frac{2}{3}\mu}}. \nonumber
\end{align}

Furthermore, as demonstrated in Appendix~\ref{sec:change-vari-self} this change of variables leads to the change of the amplitude of the self-consistent potential
\begin{align}
  \label{eq:55}
  \phi_{0}\left(\left(1 - \frac{2}{3}\mu\right)z\right) = \frac{1}{\left(1 - \frac{2}{3}\mu\right)}\bar{\phi}_{0}(z).
\end{align}

The consequence of this is that the system of Eqs.~(\ref{eq:52}) transforms into
\begin{equation}
  \begin{aligned}
     &\bar{u}_{0}^{\prime}(z) - \frac{\bar{u}_{0}(z)}{z} - \left[1 + \lambda - q\frac{(1 - 2\mu)}{\left(1 - \frac{2}{3}\mu\right)} \bar{\phi}_{0}(z)\right]\bar{v}_{0}(z) = 0,
      \\
     &\bar{v}_{0}^{\prime}(z) + \frac{\bar{v}_{0}(z)}{z} - \left[1 - \lambda + q\frac{(1 - 2\mu)}{\left(1 - \frac{2}{3}\mu\right)} \bar{\phi}_{0}(z)\right]\bar{u}_{0}(z) = 0,
      \\
      &\bar{\phi}_{0}(z) = \frac{1}{z}\int_{0}^{z}dy [\bar{u}_{0}^{2}(y) + \bar{v}_{0}^{2}(y)] + \int_{z}^{\infty}dy\frac{[\bar{u}_{0}^{2}(y) + \bar{v}_{0}^{2}(y)]}{y}.
  \end{aligned}\label{eq:56}
\end{equation}

We conclude that it is identical to the system of Eqs.~(\ref{eq:35}) up to the renormalization for the magnitude $q$ of the self-consistent potential. For this reason, the bar on the top of the radial functions and the self-consistent potential will be omitted below.

For the determination of the Lagrange multiplier $\mu$ we need to find the expression for $X$ in the new variables. As follows from Appendix~\ref{sec:change-vari-self} this relation is given through
\begin{align}
  \label{eq:57}
  X &= a^{2}\int_{0}^{\infty}dz\left\{\frac{2}{3}[u_{0}^{\prime}(z) v_{0}(z) - v^{\prime}_{0}(z) u_{0}(z)] - \frac{4}{3}\frac{u_{0}(z)v_{0}(z)}{z} + q \phi_{0}(z)[u_{0}^{2}(z) + v_{0}^{2}(z)]\right\}\nonumber
  \\
  &= a^{2}\left(\frac{2}{3}B + q \Pi\right).
\end{align}

The amplitude of the self-consistent potential in the radial Eqs.~(\ref{eq:56}) is now different from the one in Eqs.~(\ref{eq:35}). However, during the numerical solution we specify the total magnitude, which we call $\bar q$, and which is related to $q$ as
\begin{equation}
  \label{eq:58}
  \bar q = q\frac{(1 - 2\mu)}{\left(1 - \frac{2}{3}\mu\right)}.
\end{equation}

For this reason, in order to determine the Lagrange multiplier $\mu$ from Eq.~(\ref{eq:57}), we express $q$ through $\bar q$
\begin{equation}
  \label{eq:59}
  q = \bar{q}\frac{1-\frac{2}{3}\mu}{1 - 2 \mu}
\end{equation}
and plug in it into the definition of $X$, Eq.~(\ref{eq:57}). This yields
\begin{equation}
  \frac{2}{3}B + \bar{q}\frac{1-\frac{2}{3}\mu}{1 - 2 \mu} \Pi = 0 \Rightarrow \quad  \mu = \frac{B + \frac{3}{2}\bar{q}\Pi}{2B + \bar{q}\Pi}.\label{eq:60}
\end{equation}

As a result, we have achieved the goal, namely the new set of the radial functions together with the Lagrange multipliers $\lambda$ and $\mu$ has been determined, which leads to the energy $E_{0}^{\prime} = -E_{0}$ for the solution of the second kind.

However, in contrast to the integral characteristic Eq.~(\ref{eq:37}) yielding from the equations of motion, the expression for the energy $E_{0}$ is now different from the one defined via Eq.~(\ref{eq:42}). This is related to the fact, that the actual parameter, which defines the solution is not equal to $q$, but to $\bar q$. Consequently, the new value for the energy of the solution of the first kind should be expressed through $\bar q$. As demonstrated in Appendix~\ref{sec:change-vari-self} this new value is equal to
\begin{equation}
  \label{eq:61}
  E_{0} =\frac{m_{0}}{\alpha_{0}} \frac{\bar{q}}{1 - 2\mu} \left(B + \left(1 - \frac{2}{3}\mu\right) D + \frac{\left(1 - \frac{2}{3}\mu\right)}{1 - 2\mu}\frac{\bar{q}}{2}\Pi\right).
\end{equation}

\begin{figure}[t]
  \centering
  \includegraphics[width=0.6\textwidth]{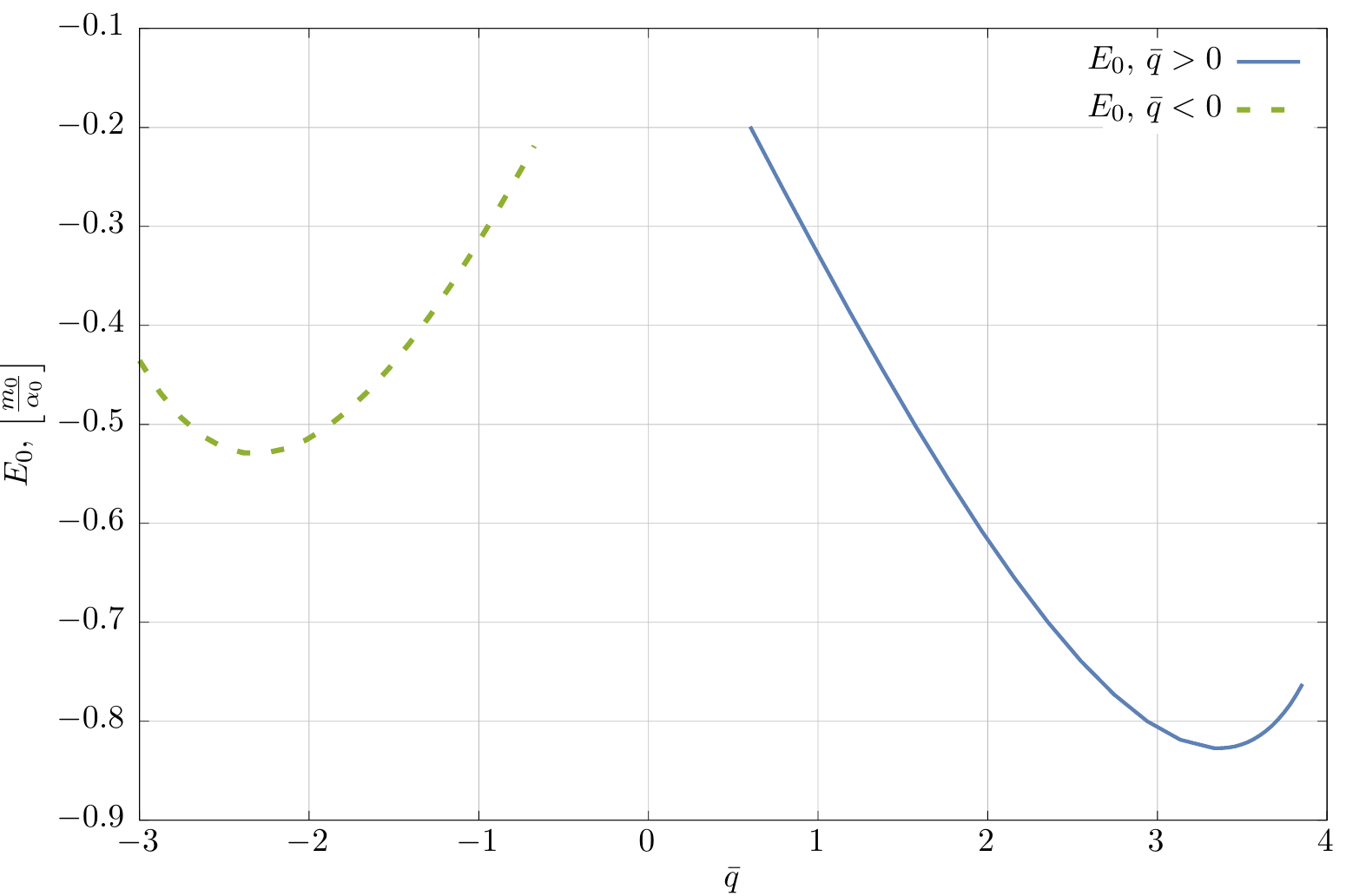}
  \\
  \includegraphics[width=0.48\textwidth]{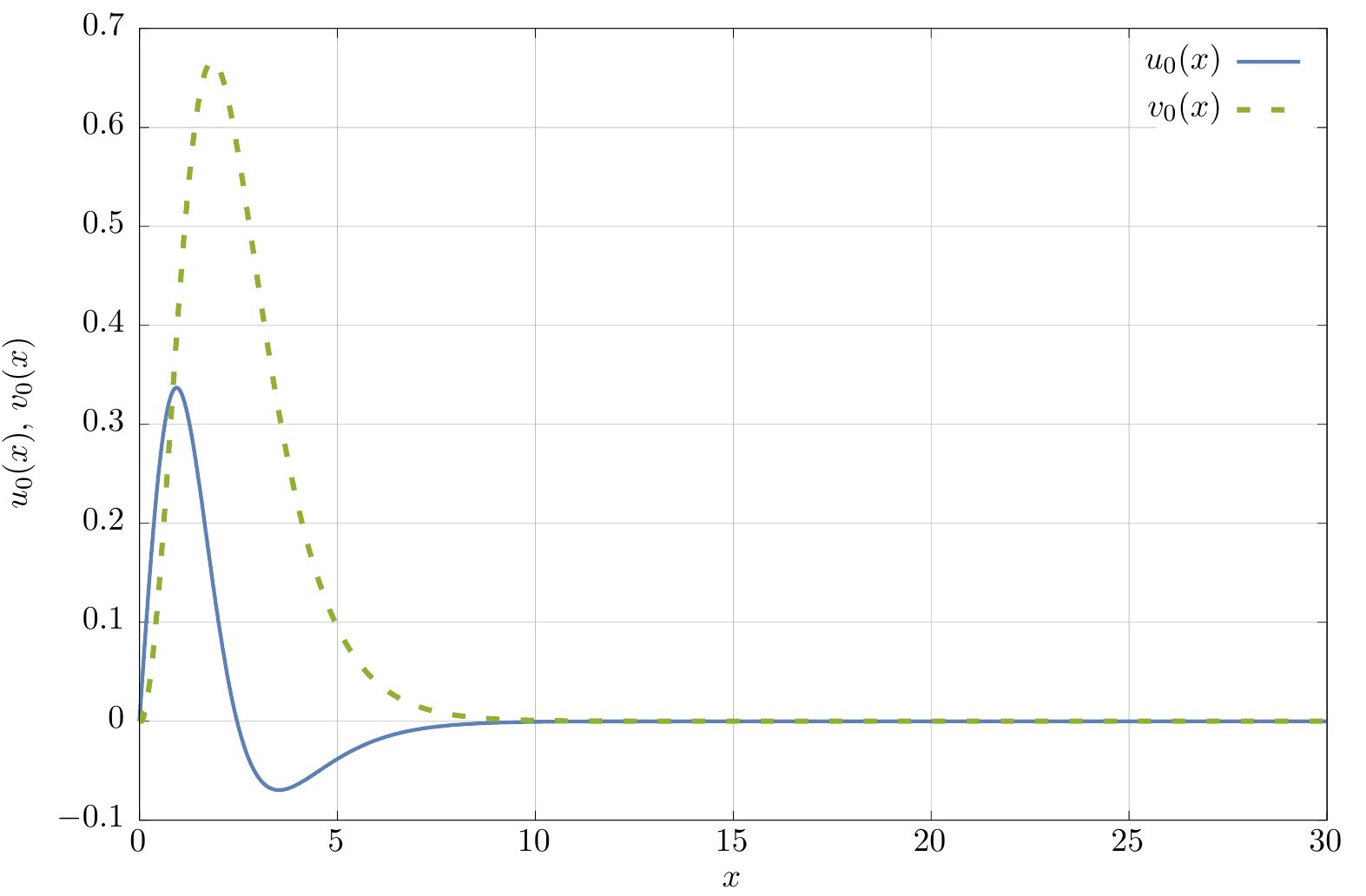}
  \includegraphics[width=0.48\textwidth]{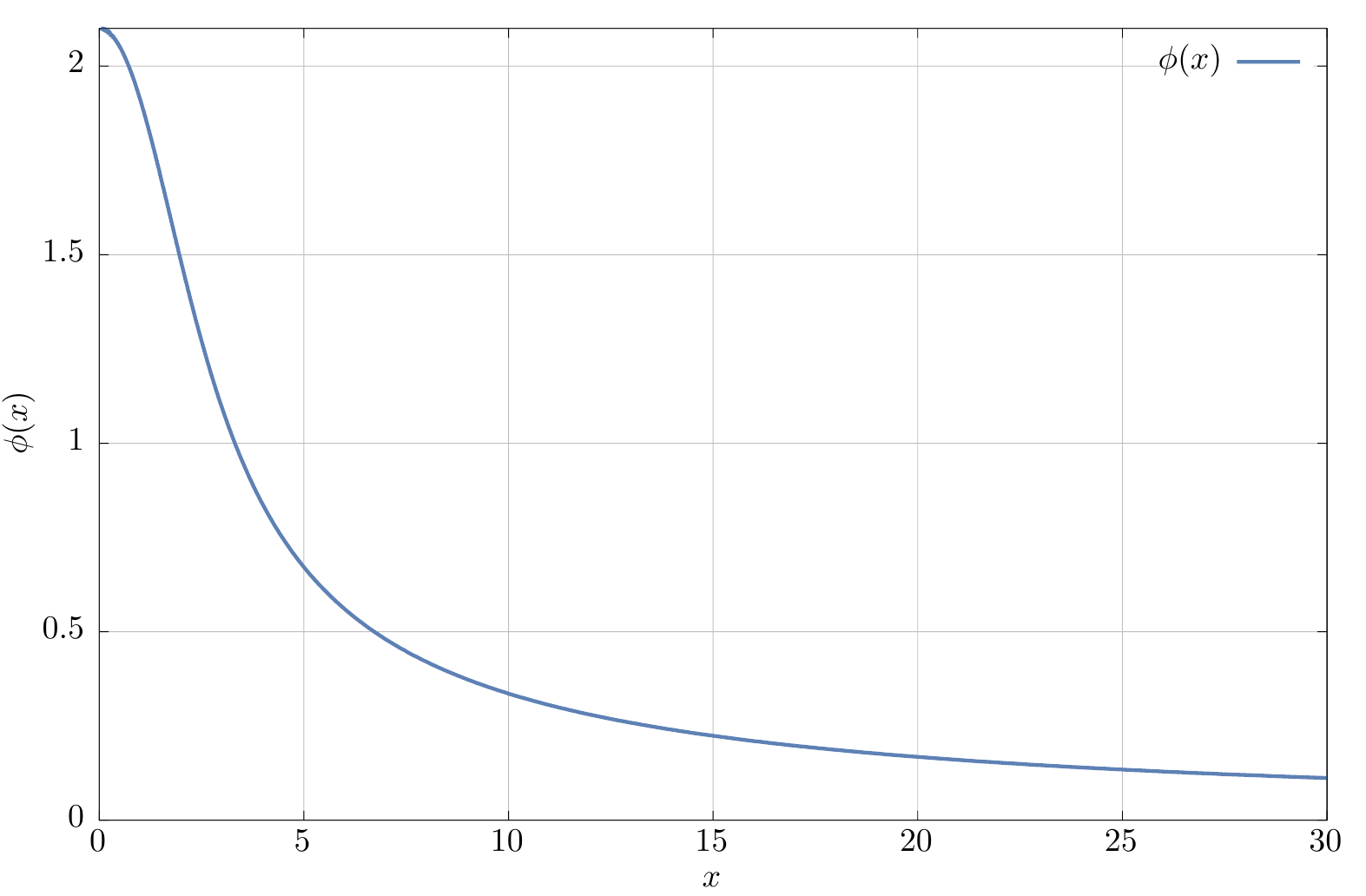}
  \caption{\label{fig:1} (color online) Top pane: The dependence of the energy $E_{0}$, (\ref{eq:61}) of the collective excitation of the electron-positron system on the parameter $\bar q$. The dashed green line represents $\bar q<0$ and the blue solid line $\bar q>0$. Bottom panes: The dependence of the universal functions $u_{0}(x)$ and $v_{0}(x)$ as well as the self-consistent potential $\phi(x)$ on the dimensionless coordinate $x = m_{0}r$.}
\end{figure}

As described above we are seeking for the solutions with the lowest nonzero energy $E_{0}$. Consequently, we investigated how the energy $E_{0}$ in Eq.~(\ref{eq:61}) depends on the parameter $\bar q$, namely whether the minimum of this function exists. The result of the evaluation is presented in Fig.~\ref{fig:1}. We have identified the two regions of the parameter $\bar q$, for which the solution exist, namely $\bar q<0$ and $\bar q>0$. However, the absolute value of the energy $E_{0}$ for $\bar q>0$ is larger, than for the case $\bar q<0$. For this reason, since we are looking for the state with the lowest nonzero energy, we discarded the value $\bar q < 0$ and determined the radial functions, the self-consistent potential, the values of $E_{0}$, $\lambda$ and $\mu$ in the point $\bar q_{0}$ of the minimum of the energy of the system, for the region $\bar q>0$
\begin{equation}
  \label{eq:62}
  \begin{aligned}
    E_{0}\left[\frac{m_{0}}{\alpha_{0}}\right] &= -0.82720 \pm 10^{-5},& &\bar q_{0} = 3.360 \pm 0.001,& &\lambda = 0.007 \pm 0.002,
    \\
    \mu &= -478.0 \pm 0.1,& &T = -1.480 \pm 0.001,& &\bar q_{0}\Pi = 1.486 \pm 0.002,
  \end{aligned}
\end{equation}
where the error boundaries are defined by the accuracy of the numerical calculations.

The radial functions $u_{0}(x)$, $v_{0}(x)$ and the self-consistent potential for the above value of $\bar q_{0}$ are presented in Fig.~\ref{fig:1}.

From these numerical results we can conclude that the parameter $\mu$ is a large value. This allows one to simplify the expression for the energy $E_{0}$ of the system, since the largest contribution is given through the quantity $D$, which describes the integral difference between the densities $u_{0}^{2}(x)$ and $v_{0}^{2}(x)$, respectively. Consequently,
\begin{align}
  \label{eq:63}
  E_{0} \approx \frac{m_{0}}{\alpha_{0}}\frac{\bar q_{0}}{3}D, \quad D = -0.735 \pm 10^{-4}.
\end{align}

\section{Soliton-like solution with a different sign of charge}
\label{sec:solit-like-solut}
In the previous section we have constructed two types of solutions, which possess the same charge $e$
\begin{align}
  \label{eq:64}
  e = e_{0}\frac{1-C}{1+C}
\end{align}
and two different signs of the energy $\pm E_{0}$. However, it appears that exactly the same two solutions can be constructed with the opposite sign of charge $e$. Indeed, let us choose a different ansatz in comparison with Eq.~(\ref{eq:27}), for the inverse Fourier transforms of the new mixing coefficients $\tilde U_{\vec qr}$ and $\tilde V_{\vec qr}$. We will denote this new type of solutions with the tilde symbol
\begin{align}
  \label{eq:65}
  &|\tilde{\psi}_{0}\rangle = \sum_{\vec qr}\left(\tilde{U}_{\vec qr}\opa a^{\dag}_{\vec qr} + \tilde{V}_{\vec qr}\opa b_{\vec qr}^{\dag}\right)|0\rangle,
  \\
  \tilde{\Psi}(\vec x) &= \sum_{\vec qr}\frac{1}{\sqrt{V}}\sqrt{\frac{m_{0}}{\epsilon_{\vec q}}}\tilde{U}_{\vec q r}u_{\vec q r}e^{i\thp{q}{x}}, \quad  \tilde{\Psi}^c(\vec x) = \sum_{\vec qr}\frac{1}{\sqrt{V}}\sqrt{\frac{m_{0}}{\epsilon_{\vec q}}}\tilde{V}^*_{\vec q r}v_{\vec q r}e^{-i\thp{q}{x}}, \nonumber
  \\
  &\int {d \vec{x}} \tilde\Psi^{\dag} (\vec x) \tilde\Psi (\vec x) = \frac{\tilde{C}}{1+\tilde{C}},\quad \int {d \vec{x}} \tilde\Psi^{c\dag} (\vec x) \tilde\Psi^{c} (\vec x) = \frac{1}{1+\tilde{C}},\label{eq:66}
\end{align}
where the inverse Fourier transforms are defined as
\begin{align}
  \label{eq:67}
  \tilde\Psi(\vec x) = \sqrt{\frac{\tilde{C}}{1+\tilde{C}}}\begin{pmatrix}
    -\ri \tilde{f}(r) \tilde{\chi}_1
    \\
    \tilde{g}(r) \tilde{\chi}_0
  \end{pmatrix}, \quad \tilde\Psi^{c}(\vec x) = \sqrt{\frac{1}{1+\tilde{C}}}\begin{pmatrix}
    -\ri \tilde{f}(r) \tilde{\chi}_1^{c}
    \\
    \tilde{g}(r) \tilde{\chi}_0^{c}
  \end{pmatrix}.
\end{align}

Here we flipped the two component spinors in the Dirac bispinors and changed the normalization of the corresponding wave functions. This is equivalent to fix $l = 1$ and $l^{\prime} = 0$.

One can calculate the expectation value of the charge operator with the help of the state vector (\ref{eq:65})
\begin{equation}
  \label{eq:68}
  \tilde{e} = e_{0}\sum_{\vec qr}\left(|\tilde{U}_{\vec q r}|^2 - |\tilde{V}_{\vec q r}|^2\right) = e_0\frac{\tilde{C}-1}{\tilde{C}+1}.
\end{equation}

Moreover, by direct substitution, it can be shown that the radial functions $\tilde{g}(r)$ and $\tilde{f}(r)$ in $\tilde\Psi(\vec x)$ and $\tilde\Psi^{c}(\vec x)$ satisfy exactly the same equations as the radial functions $g(r)$ and $f(r)$ in $\Psi(\vec x)$ and $\Psi^{c}(\vec x)$. As a result, the radial functions with the tilde symbol $\tilde{g}(r)$ and $\tilde{f}(r)$ are equal to the radial functions $g(r)$ and $f(r)$. This fixes as in the previous section the numerical value for $\bar{q}_{0} > 0$ and, since $\alpha_{0}>0$, the quantity $C > 1$.

Therefore, the quantity $\tilde{C} = C$ and, consequently, this new soliton-like solution possesses the opposite sign of charge
\begin{equation}
  \label{eq:69}
  \tilde{e} = -e.
\end{equation}

In addition we pay attention to the fact that since the soliton state of the first kind is described via a pair of wave functions $\{\Psi(\vec x),\Psi^{c}(\vec x)\}$, one might expect that the soliton solution with the opposite charge sign can be obtained by applying the charge conjugation operator $\opa C$ to $\{\Psi(\vec x),\Psi^{c}(\vec x)\}$. By observing the structure of the wave functions $\tilde{\Psi}(\vec x)$ and $\tilde{\Psi}^{c}(\vec x)$ and by exploiting the definition of the charge conjugation operator \cite{LandauQED} we indeed find that $\{\tilde{\Psi}(\vec x),\tilde{\Psi}^{c}(\vec x)\} = \{\beta\alpha_{2}\Psi^{*}(\vec x),\beta\alpha_{2}\Psi^{c*}(\vec x)\}$.

Furthermore, it can be demonstrated that the total energy of this solution is equal to the energy of the solution of the first kind.

\section{States with nonzero total momentum}
\label{sec:states-with-nonzero}

In Secs.~\ref{sec:separation-variables} and \ref{sec:solut-second-kind} we have determined the solutions of two types which have two different signs of $E_{0}$, for the case when the total momentum of the system $\langle\opA P\rangle$ is equal to zero. However, it is well known that the QED Hamiltonian is Lorentz invariant \cite{bjorken1965,LandauQED,AkhiezerB1965quantum} and consequently, we need to generalize the results of the previous sections for the case when $\langle\opA P\rangle\neq0$.

In the following we will employ the method of canonical transformations of the variables, which was introduced in Refs.~\cite{GROSS19761,TyablikovA1951adyabatic,BogolyubovA1950about}. In this method one introduces the collective field coordinate $\vec R$, which is canonically conjugated to the momentum $\opA P$. By construction the momentum $\opA P$ coincides with the total momentum of the system. We want to note here, that a similar problem appears in the context of the Nambu-Jona-Lasinio model \cite{0954-3899-18-9-008}.

Let us demonstrate this on the example of the separation of the center of mass variable in the system of $N$ particles with the coordinates $\vec r_{a}$.

Firstly, we introduce the center of mass coordinate $\vec R_{1}$ and the relative coordinates $\vec x_{a}$ in the usual way
\begin{equation}
  \label{eq:70}
  \vec R_1 = \frac{1}{N}\sum_{a=1}^{N}\vec r_a. \quad \vec r_a = \vec R_1 + \vec{x}_a, \quad \sum_{a=1}^{N} \vec{x}_a = 0.
\end{equation}
As can be seen from Eq.~(\ref{eq:70}) the introduction of $\vec R_{1}$ is compensated by the condition imposed on the relative coordinates $\vec x_{a}$.

Secondly, we calculate the operators of the new momenta of the system according to their definition
\begin{equation}
  \label{eq:71}
  \begin{aligned}
    \opA p_a = - \ri \nabla_a = - \frac{\ri}{N}\nabla_{\vec R_1} + \opA p^\prime_a, \quad \opA P_{1} = -\ri\nabla_{\vec R_1} \quad \opA p^\prime_a =- \ri \nabla_{\vec x_a} + \frac{\ri}{N}\sum_{b=1}^{N}\nabla_{\vec x_b}, \quad \sum_{a=1}^{N} \opA p^\prime_a = 0.
  \end{aligned}
\end{equation}
Consequently, we may conclude from Eq.~(\ref{eq:71}) that the momentum $\opA P_{1}$ describes the collective motion of the system.

In what follows\footnote{This discussion is based on § 13.1 of the Ref.~\cite{bjorken1965}, §64-65 of Ref.~\cite{LandauQM} and §6.8 of Ref.~\cite{FeynmanB1998statistical}} we would like to apply a similar procedure for the reduced Hamiltonian of QED, namely to separate out the center of mass of the collective excitation of the electron-positron system. For this we note that the secondary-quantized representation in quantum mechanics is based on the equality of the matrix elements, which are calculated in two different representations for the wave function of the system\cite{LandauQM,FeynmanB1998statistical,bjorken1965}. In the first representation the system is described via the wave function, which depends on the coordinates of the individual particles while in the second one the system is described by the distribution of the occupation numbers of particles over different states. Since the ``total'' operator of the whole system, e.g. the total energy or the total momentum, is represented as a sum of the single-particle operators, i.e. the linear relation, the matrix elements of this ``total'' operator are equal to each other in these two different representations. That is, the reduced Hamiltonian of QED can be written in the completely equivalent coordinate representation as
\begin{align}
  \opa H^{\prime}_{\mathrm{QED}} = \lim_{N\rightarrow\infty}\sum_{a=1}^N \left\{\vec \alpha_a \cdot \opA p_a + \beta_a m_0 + \frac{e_0^2}{8\pi}\sum_{b=1}^N \frac{1}{|\vec r_a - \vec r_b|} \right\}. \label{eq:72}
\end{align}

Consequently, in this representation we can apply the relations (\ref{eq:70}) and (\ref{eq:71}) for the reduced Hamiltonian (\ref{eq:72}). This yields
\begin{align}
  \label{eq:73}
  \opa H^{\prime}_{\mathrm{QED}} &= \lim_{N\rightarrow\infty}\sum_{a=1}^N \Bigg\{\frac{\vec \alpha_{a}\cdot\opA P_{1}}{N} + \vec \alpha_a \cdot (-\ri\nabla_{\vec x_{a}}) + \beta_a m_0 \nonumber
  \\
  &\mspace{173mu}+ \sum_{b=1}^N\left(\frac{e_0^2}{8\pi} \frac{1}{|\vec x_a - \vec x_b|} + \frac{\ri}{N}\nabla_{\vec x_{b}}\right) \Bigg\}.
\end{align}

Furthermore, as demonstrated in Appendix~\ref{sec:calc-jacob-determ} the absolute value of the Jacobian determinant of the variable transformations Eqs.~(\ref{eq:70}), (\ref{eq:71}) is equal to $N^{3}$. Therefore, during the calculation of the matrix elements of an arbitrary operator $\opa M$ for the system containing $N$ particles we will change the variables as
\begin{align}
  \la \Phi_1|\opa M|\Phi_2\ra &= \int d\vec r_1 \cdots d\vec r_N \Phi_1^*(\{\vec r_i\})\opa M \Phi_2(\{\vec r_i\}) \nonumber
  \\
  &= \int N^{3} d\vec R_1 d\vec x_1 \cdots d\vec x_{N-1} \Phi_1^*(\vec R_1,\{\vec x_i\})\opa M \Phi_2(\vec R_1,\{\vec x_i\}), \label{eq:74}
\end{align}
As a result it is natural to introduce the new variables
\begin{equation}
  \begin{aligned}
    N \vec R_1 &= \vec R,
    \\
    N \opA P &= \opA P_{1},
  \end{aligned}\label{eq:75}
\end{equation}
and consequently Eqs.~(\ref{eq:74}) and (\ref{eq:73}) transform as follows
\begin{align}
  \opa H^{\prime}_{\mathrm{QED}} &= \lim_{N\rightarrow\infty}\sum_{a=1}^N \Bigg\{\vec \alpha_{a}\cdot\opA P + \vec \alpha_a \cdot (-\ri\nabla_{\vec x_{a}}) + \beta_a m_0 \nonumber
  \\                                 &\mspace{160mu}+\sum_{b=1}^N\left(\frac{e_0^2}{8\pi} \frac{1}{|\vec x_a - \vec x_b|} + \frac{\ri}{N}\nabla_{\vec x_{b}}\right) \Bigg\}  \label{eq:76}
  \\
  \la \Phi_1|\opa M|\Phi_2\ra &= \int d\vec R d\vec x_1 \cdots d\vec x_{N-1} \Phi_1^*(\vec R,\{\vec x_i\})\opa M \Phi_2(\vec R,\{\vec x_i\}).  \label{eq:77}
\end{align}

As the last step of the derivation we need to return into the secondary-quantized representation, for which we investigate the single-particle Hamiltonians $\opa H_{a}$
\begin{align}
  \opa H_{a} = \vec \alpha_a \cdot \opA p_a+\vec\alpha_a\cdot\opA P + \beta_a m_0. \label{eq:78}
\end{align}
From this equation we can immediately conclude that the coordinate $\vec R$ is a cyclic one. Therefore, the solution of the Dirac equation with the Hamiltonian (\ref{eq:78}) is easily found and reads
\begin{equation}
  \label{eq:79}
  \psi_a = \frac{1}{\sqrt{V}}\sqrt{\frac{m_0}{\epsilon_{\vec P + \vec p_a}}}u(\vec P + \vec p_a,s)e^{\ri\thp{P}{R}+\ri\thp{p_a}{x_a}},
\end{equation}
where $u(\vec P + \vec p_a,s)$ are the bispinors of the free Dirac equation with the momentum $\vec P + \vec p_a$. Consequently, one can deduce from the single-particle solutions Eq.~(\ref{eq:79}) that in order to return into the secondary-quantized representation the following substitution for the field operators and bispinors is required
\begin{align*}
  &\opa a_{\vec p s} \rightarrow \opa a_{\vec P+\vec p, s}e^{\ri\thp{P}{R}},\quad \opa b_{\vec p s} \rightarrow \opa b_{\vec P+\vec p, s}e^{\ri\thp{P}{R}}
  \\
  &u_{\vec ps}\rightarrow u(\vec P + \vec p,s), \quad v_{\vec ps}\rightarrow v(\vec P + \vec p,s),
  \\
  &\epsilon_{\vec p}\rightarrow \epsilon_{\vec P+\vec p}.
\end{align*}
As a result, the secondary quantized-wave functions read
\begin{equation}
  \begin{aligned}
    \uppsi(\vec x, \vec R, \vec P) &= \sum_{\vec ps} \frac{1}{\sqrt{V}}\sqrt{\frac{m_{0}}{\epsilon_{\vec P+\vec p}}}\left(\opa a_{\vec P +\vec p,s}u(\vec P + \vec p,s)e^{\ri \thp{p}{x}+\ri\thp{P}{R}}+\opa b_{\vec P +\vec p,s}^\dag v(\vec P + \vec p,s)e^{-\ri \thp{p}{x}-\ri\thp{P}{R}}\right),
    \\
    \uppsi^\dag(\vec x, \vec R, \vec P) &= \sum_{\vec ps} \frac{1}{\sqrt{V}}\sqrt{\frac{m_{0}}{\epsilon_{\vec P+\vec p}}}\left(\opa a^\dag_{\vec P +\vec p,s}u^\dag(\vec P + \vec p,s)e^{-\ri \thp{p}{x}-\ri\thp{P}{R}}+\opa b_{\vec P +\vec p,s} v^\dag(\vec P + \vec p,s)e^{\ri \thp{p}{x}+\ri\thp{P}{R}}\right).
  \end{aligned}\label{eq:80}
\end{equation}

By exploiting this expression we can write down the reduced Hamiltonian of QED
\begin{align*}
  \opa H^{\prime}_{\mathrm{QED}} &= \int d\vec x :\uppsi^\dag(\vec x,\vec R,\vec P)(\vec\alpha\cdot \opA p + \vec\alpha\cdot \opA P +\beta m_0)\uppsi(\vec x,\vec R,\vec P): \nonumber
  \\
  &\mspace{180mu}+ \frac{e_0^2}{8\pi}\int \frac{d\vec x d\vec y}{|\vec x - \vec y|}:\uprho(\vec x,\vec R,\vec P)::\uprho(\vec y,\vec R,\vec P): \nonumber
	\\
	&- \lim_{N\rightarrow\infty}\frac{1}{2N}\int d\vec x d\vec y:\uppsi^\dag(\vec x,\vec R,\vec P)\vec\alpha\uppsi(\vec x,\vec R,\vec P)\cdot\uppsi^\dag(\vec y,\vec R,\vec P)(-\ri\nabla_{\vec y})\uppsi(\vec y,\vec R,\vec P):.
\end{align*}

The last term vanishes when $N$ tends to infinity and we finally obtain
\begin{align}
  \label{eq:81}
  \opa H^{\prime}_{\mathrm{QED}} &= \int d\vec x :\uppsi^\dag(\vec x,\vec R,\vec P)(\vec\alpha\cdot \opA p + \vec\alpha\cdot \opA P +\beta m_0)\uppsi(\vec x,\vec R,\vec P): \nonumber
  \\
  &+ \frac{e_0^2}{8\pi}\int \frac{d\vec x d\vec y}{|\vec x - \vec y|}:\uprho(\vec x,\vec R,\vec P)::\uprho(\vec y,\vec R,\vec P):.
\end{align}

Concluding, we have achieved the goal and have separated the total momentum of the system in the reduced QED Hamiltonian.

Proceeding with the construction of the soliton-like solutions for a nonzero total momentum, we require the collective excitation of the electron-positron system to be Lorentz invariant, i.e. its relativistic energy dispersion law must be fulfilled for an arbitrary total momentum of the system
\begin{equation}
  \label{eq:82}
  E^{2}(\vec P) = \vec P^{2} + E_{0}^{2}.
\end{equation}

It is well known that in the general case of a strong coupling theory, e.g. the polaron problem \cite{Mitra198791,RevModPhys.63.63,PhysRev.97.660,0022-3719-17-24-012,Spohn1987278} or in the problem of a particle interacting with a scalar field \cite{PhysRevD.92.125019} the dispersion law $E_{\mathrm{p}}(\vec P)$ (the index p stands for polaron) is a very complicated function of the total momentum of the system. Moreover, the energy dispersion relation is associated directly with the mass renormalization. For this reason, the complicated dispersion law is expanded in a power series for small momenta, resulting in the expression $E_{\mathrm{p}}(\vec P) \approx E_{0} + \vec P^{2}/(2m^{*})$, where $m^{*}$ is the renormalized mass. This happens due to the nonlinear interaction between the particle and the field or equivalently because the momentum operator of the particle does not commute with the interaction part.

However, in our problem the situation is different, which can be concluded from the upcoming fact. It follows from the reduced QED Hamiltonian Eq.~(\ref{eq:76}) that the interaction part does not depend on the coordinate of the center of mass $\vec R$. For this reason, the total momentum operator $\opA P$ commutes with the reduced Hamiltonian of QED and, therefore, the self-consistent potential does not change when the translation of the system is performed. Moreover, by observing Eq.~(\ref{eq:81}) we can conclude that the coordinate $\vec R$, which is conjugated to the total momentum $\opA P$ is a cyclic one. Furthermore, the total momentum is coupled only to the spin degrees of freedom. In addition, it is well known that the relativistic motion can be considered as the transformations in the spinor space. For example, in Ref. \cite{bjorkenB1964} the solution of the free Dirac equation has been found firstly for the case of the particle at rest. Then it was demonstrated that the transformations in the spinor space lead to the solution of the Dirac equation for an arbitrary momentum.

By exploiting this analogy, we introduce the state vector $|\psi_{\vec P}\rangle$, which is normalized to unity and describes the collective excitation of the electron-positron system with nonzero total momentum $\vec P$ and try to represent this state vector as a linear combination of the obtained above resting solutions. Furthermore, we assume that the dependence on the total momentum is completely absorbed in the coefficients of the linear combination. In other words, we try to solve the Schr\"{o}dinger equation with the help of the basis consisting of a finite number of the known state vectors.

The state $|\psi_{0}^{\prime}\rangle$ contains an arbitrary vector $\vec \nu$ Eq.~(\ref{eq:45}), which we direct along the momentum $\vec P$, i.e. $\vec P = P \vec \nu$. Proceeding, we form a linear combination of the solutions of the first and the second kinds
\begin{equation}
  \label{eq:83}
  |\psi_{\vec P}\rangle = K(\vec P) |\psi_{0}\rangle + L(\vec P) |\psi_{0}^{\prime}\rangle.
\end{equation}

The state vectors $|\psi_{0}\rangle$ and $|\psi_{0}^{\prime}\rangle$ are normalized to unity and orthogonal to each other, i.e
\begin{align}
  \label{eq:84}
  \langle\psi_{0}|\psi_{0}\rangle = 1, \quad \langle\psi^{\prime}_{0}|\psi^{\prime}_{0}\rangle = 1, \quad \langle\psi_{0}|\psi_{0}^{\prime}\rangle = 0.
\end{align}

We require the state vector $|\psi_{\vec P}\rangle$ to be normalized, which yields the condition on the coefficients of the linear combination
\begin{align}
  \label{eq:85}
  \langle\psi_{\vec P}|\psi_{\vec P}\rangle = 1 \Rightarrow \quad |K(\vec P)|^{2} + |L(\vec P)|^{2} = 1,
\end{align}

As was mentioned above we need to solve the Schr\"{o}dinger equation
\begin{align}
  \label{eq:86}
  \opa H_{\mathrm{QED}}^{\prime}(\vec P) |\psi_{\vec P}\rangle = E(\vec P)|\psi_{\vec P}\rangle,
\end{align}
or by plugging the definition of $|\psi_{\vec P}\rangle$ from Eq.~(\ref{eq:83})
\begin{align}
  \label{eq:87}
  \opa H_{\mathrm{QED}}^{\prime}(\vec P)\left[K(\vec P) |\psi_{0}\rangle + L(\vec P) |\psi_{0}^{\prime}\rangle\right] = E(\vec P)\left[K(\vec P) |\psi_{0}\rangle + L(\vec P) |\psi_{0}^{\prime}\rangle\right].
\end{align}

Let us project this expression on $|\psi_{0}\rangle$ and $|\psi_{0}^{\prime}\rangle$. With the help of Eq.~(\ref{eq:84}) this yields
\begin{equation}
  \label{eq:88}
  \begin{aligned}
    K(\vec P) \langle\psi_{0}|\opa H_{\mathrm{QED}}^{\prime}(\vec P)|\psi_{0}\rangle + L(\vec P) \langle\psi_{0}|\opa H_{\mathrm{QED}}^{\prime}(\vec P)|\psi_{0}^{\prime}\rangle &= E(\vec P)K(\vec P),
    \\
    K(\vec P) \langle\psi_{0}^{\prime}|\opa H_{\mathrm{QED}}^{\prime}(\vec P)|\psi_{0}\rangle + L(\vec P) \langle\psi_{0}^{\prime}|\opa H_{\mathrm{QED}}^{\prime}(\vec P)|\psi_{0}^{\prime}\rangle &= E(\vec P)L(\vec P).
  \end{aligned}
\end{equation}

According to Secs.~\ref{sec:separation-variables}--\ref{sec:solut-second-kind} the matrix elements
\begin{align}
  \label{eq:89}
  \langle\psi_{0}|\opa H_{\mathrm{QED}}^{\prime}(\vec P)|\psi_{0}\rangle = E_{0}, \quad \langle\psi_{0}^{\prime}|\opa H_{\mathrm{QED}}^{\prime}(\vec P)|\psi_{0}^{\prime}\rangle = -E_{0},
\end{align}
since the expectation value
\begin{align}
  \label{eq:90}
  &\langle\psi_{0}|\int d\vec x :\uppsi^\dag(\vec x,\vec R,\vec P)(\vec\alpha\cdot \opA P)\uppsi(\vec x,\vec R,\vec P):|\psi_{0}\rangle \nonumber
  \\
  &\mspace{120mu}= \int d\vec x \Bigg(\Psi^{\dag}(\vec x) \vec \alpha\cdot\vec P \Psi(\vec x) - \Psi^{c\dag}(\vec x)(-\vec \alpha\cdot\vec P) \Psi^{c}(\vec x)\Bigg) = 0,
\end{align}
due to the orthogonality of the spherical harmonics $Y_{00}$ and $Y_{1m}$, $m=\{-1,0,1\}$. In addition, $\exp(\pm\ri\vec P\cdot\vec R)$ in the density operator $\uprho(\vec x,\vec R,\vec P)$ does not contribute, as the only nonzero matrix elements are proportional to $\opa a^{\dag}_{\vec ps}\opa a_{\vec ps}$ and $\opa b^{\dag}_{\vec ps}\opa b_{\vec ps}$. Moreover, the same calculation leads to the second equation in Eq.~(\ref{eq:89}) for $\Psi^{\prime}(\vec x)$, $\Psi^{c\prime}(\vec x)$. Therefore, the system of Eqs.~(\ref{eq:88}) transforms into the following form
\begin{equation}
  \label{eq:91}
   \begin{aligned}
    K(\vec P) E_{0} + L(\vec P) \langle\psi_{0}|\opa H_{\mathrm{QED}}^{\prime}(\vec P)|\psi_{0}^{\prime}\rangle &= E(\vec P)K(\vec P),
    \\
    K(\vec P) \langle\psi_{0}^{\prime}|\opa H_{\mathrm{QED}}^{\prime}(\vec P)|\psi_{0}\rangle - L(\vec P) E_{0} &= E(\vec P)L(\vec P).
  \end{aligned}
\end{equation}

The calculation of the two remaining matrix elements is presented in Appendix~\ref{sec:eval-matr-elem} and the result reads
\begin{align}
  \label{eq:92}
  \langle\psi_{0}^{\prime}|\opa H_{\mathrm{QED}}^{\prime}(\vec P)|\psi_{0}\rangle = \langle\psi_{0}|\opa H_{\mathrm{QED}}^{\prime}(\vec P)|\psi_{0}^{\prime}\rangle = P.
\end{align}

This allows us to rewrite the system of Eqs.~(\ref{eq:91}) for the determination of the energy $E(\vec P)$ as
\begin{equation}
  \label{eq:93}
  \begin{aligned}
    K(\vec P) E_{0} + L(\vec P) P &= E(\vec P)K(\vec P),
    \\
    K(\vec P) P - L(\vec P) E_{0} &= E(\vec P)L(\vec P).
  \end{aligned}
\end{equation}

This is a system of linear equations and in order to obtain nontrivial solutions its determinant should be equal to zero. Therefore,
\begin{equation}
  \label{eq:94}
  \begin{vmatrix}
    E_{0} - E(\vec P) && P
    \\
    P && - [E_{0} + E(\vec P)]
  \end{vmatrix} = 0,
\end{equation}
or by expanding the determinant we obtain the dispersion law
\begin{equation}
  E_{0}^{2} + P^{2} = E^{2}(\vec P).\label{eq:95}
\end{equation}

Consequently, we have achieved the goal and have found the state vector $|\psi_{\vec P}\rangle$, with the corresponding eigenvalue $E(\vec P)$, which yields the relativistic dispersion law for the soliton-like solution. This result can be understood in a way, that due to the motion, the operator $\vec \alpha\cdot\vec P$ mixes the two different states $|\psi_{0}\rangle$, $|\psi_{0}^{\prime}\rangle$ with the energies $\pm E_{0}$, such that the relativistic dispersion law holds. This result could have been only achieved as the interaction part in the reduced QED Hamiltonian is invariant under the translation of the system as a whole under the vector $\vec R$.

\section{Quasi-classical picture and survey of the results}
\label{sec:discussion}
In this work we presented a novel soliton-like solution in quantum electrodynamics, which was obtained by modeling the state vector of the system in analogy with the theory of superconductivity, by separating out the classical component in the density operator (\ref{eq:11}) and by variation of the functional for the total energy of the system. This leads to the equations of the self-consistent field (\ref{eq:25})--(\ref{eq:26}). We based our derivations on the assumptions that the parameters of the initial QED Hamiltonian, which are the ``bare'' electron charge and the mass are unknown values.

Next by exploiting the spherically symmetric ansatz for the Dirac wave functions, which resulted in the spherically symmetric self-consistent potential we separated the variables in the Dirac equation. Due to the commutation of the charge operator with the QED Hamiltonian the normalization condition for the electron component in the state vector $|\psi_{0}\rangle$, i.e. $U_{\vec qr}$, $\Psi(\vec x)$ can not be equal to the normalization of the corresponding positron component $V_{\vec qr}^{*}$, $\Psi^{c}(\vec x)$. Consequently, we introduced the quantity $C$, which describes the population of the electron component with respect to that of the positron and, which determines the sign of charge
\begin{align}
  e = e_{0}\frac{1-C}{1+C} \label{eq:96}
\end{align}
of the quasi-particle collective excitation of the electron-positron system. In addition the soliton-like solution is described via the pair of functions $\{\Psi(\vec x),\Psi^{c}(\vec x)\}$ or equivalently via their Fourier transforms $\{U_{\vec qr},V^{*}_{\vec qr}\}$.

After separation of variables we obtained a system of equations (\ref{eq:29}) for the radial functions $g(r)$, $f(r)$ and $g_{1}(r)$, $f_{1}(r)$, which determine the density of the self-consistent potential. This system of integro-differential equations is a nonlinear eigenvalue problem. In order to provide the solution we employed the continuous analog of Newton method. During the solution we have found that the radial functions $g(r)$, $f(r)$ and $g_{1}(r)$, $f_{1}(r)$ are equal to each other, which exhibits the self-similarity. This allowed us to determine the parameter
\begin{align}
  q = \alpha_{0}\frac{1-C}{1+C},\label{eq:97}
\end{align}
which defines the magnitude of the self-consistent potential.

According to the uncertainty principle, the localization of the electronic and positronic components of the charge density in a finite volume of space leads to the corresponding uncertainty in their momentum. Due to the Coulomb attraction between charges, the positive kinetic energy of the fluctuations compensates for the negative potential energy and the system equilibrates. This can be viewed as the physical reason for the self-consistent solution. In order to clarify this statement we will provide a simple qualitative quasi-classical estimation below.

Let us introduce a characteristic parameter $a$ of the localization region in space for both components of the charge density. The uncertainty of the momentum is then defined by the parameter $u \sim 1/a$. The integral densities of the electron and positron components we specify as $\rho_{-} = 1/(1+C)$ and $\rho_{+} = C / (1+C)$ correspondingly. In addition, we consider that the state vector is normalized according to Eq.~(\ref{eq:14}) such that $\rho_{-} + \rho_{+} = 1$. If the charge is localized in the spacial region $a < m_{0}^{-1}$, then the momentum uncertainty $u > m_{0}$ and the relativistic description for the kinetic energy is required. Consequently, we can write down the quasi-classical estimation for the energy of the system
\begin{align}
  E \approx \rho_{-}\sqrt{u^{2}+m_{0}^{2}} - \rho_{+}\sqrt{u^{2}+m_{0}^{2}} + \rho_{-} e_{0}\varphi_{\mathrm{C}} - \rho_{+}e_{0}\varphi_{\mathrm{C}},\label{eq:98}
\end{align}
where $\varphi_{\mathrm{C}}$ is the estimation for the self-consistent potential, which is created by the electronic and positronic components of the charge density
\begin{align}
  \varphi_{\mathrm{C}} = \frac{\rho_{-}e_{0}}{4\pi a} - \frac{\rho_{+}e_{0}}{4\pi a} = \frac{e_{0}}{4\pi}\frac{1-C}{1+C} u.\label{eq:99}
\end{align}

Consequently, we can write down the estimation for the energy of the system
\begin{align}
  E\approx \frac{q}{\alpha_{0}}\sqrt{u^{2} + m_{0}^{2}} + \frac{q^{2}}{\alpha_{0}} u.\label{eq:100}
\end{align}

The parameter $q$ is related to the soliton charge, such that in accordance with Eq.~(\ref{eq:96}) $e = 4\pi q / e_{0}$. The soliton charge is in turn an integral of motion and consequently defines the stability of the soliton state with minimal energy. The value $q = 0$ corresponds to the state with vanishing charge and energy. However, the energy of the system, Eq.~(\ref{eq:100}), possesses a nontrivial minimum for $q\neq 0$. Indeed, the variation of $E$ with respect to $q$ and $u$ leads to the equations
\begin{align}
  \frac{u}{\sqrt{u^2 + m_0^2}} + q = 0, \quad \sqrt{u^2 + m_0^2} + 2 q u = 0. \label{eq:101}
\end{align}

Since $u>0$ according to its definition, a nontrivial solution exists when $q < 0$ or equivalently $C>1$ and reads
\begin{align}
  u = m_{0}, \quad q = -\frac{1}{\sqrt{2}}, \quad E = -\frac{m_{0}}{2\alpha_{0}}. \label{eq:102}
\end{align}

Concluding, even a rough quasi-classical estimation demonstrates that the soliton-like solution is energetically more preferable than the solution with vanishing energy. We also observe that the numerical value of the coefficient in the energy Eq.~(\ref{eq:102}) is close to the exact quantum mechanical result Eq.~(\ref{eq:63}) (compare $-0.5$ versus $\bar{q}_{0} D / 3 \approx -0.823$). 

Returning back to the exact formulation, we introduced the solution of the second kind, with the state vector $|\psi_{0}^{\prime}\rangle$, which is orthogonal to the state vector of the solution of the first kind and is also normalized to unity. We have demonstrated that the energy $E_{0}^{\prime}$ of this second solution has the opposite sign with regard to the energy of the solution of the first kind, i.e. $E_{0}^{\prime} = -E_{0}$. This condition is manifested by the parameter $\mu$ which is associated with the renormalization of the magnitude of the self-consisted potential
\begin{align}
  q = \bar q \frac{1-\frac{2}{3}\mu}{1 - 2\mu} \approx 1.118,\quad \bar q \approx 3.360, \quad \mu \approx 478. \label{eq:103}
\end{align}

Finally, by concluding the formulation for the resting soliton-like solutions we determined two analogous kinds of solutions with opposite sign of charge.

Next we accomplished the transition to the moving soliton and performed the canonical transformation of the field variables. This allowed us to separate out the center of mass coordinate, with the canonically conjugated total momentum of the system. Our results are based on the equivalence of the two different representations for the reduced QED Hamiltonian and the fact that the interaction part is invariant under translations of the system.

By forming a linear combination of the obtained solutions of the first and the second kinds we have found the dependence of the energy of the moving soliton on the total momentum of the system. This has removed the arbitrariness in the orientation of the quantization axis of the spinor part of the wave function in analogy with the motion of a free electron and has lead to the well known relativistic energy-momentum relation
\begin{align}
  E_{0}^{2} + P^{2} = E^{2}(\vec P). \label{eq:104}
\end{align}

At last, we want to discuss the stability of the soliton-like solution with respect to its decay with the emission of a photon, as it happens during the annihilation of the bound state of electron and positron --- positronium.

First, we notice that the state vector of positronium is bilinear in the creation operators
\begin{align}
  |\Psi_{\mathrm{Ps}}\rangle \sim \opa a^{\dag}_{\vec p}\opa b^{\dag}_{\vec q}|0\rangle,\label{eq:105}
\end{align}
which corresponds to a two-particle excitation with vanishing charge. Consequently, the transition matrix element into the state with one photon $|\gamma_{\vec k}\rangle = \opa c^{\dag}_{\vec k}|0\rangle$ from the interaction Hamiltonian of QED is not equal to zero, i.e.,
\begin{align}
  M^{\mathrm{Ps}}_{if} \sim \langle\gamma_{\vec k}|\int d\vec x: \uppsi^{\dag}(\vec x)\vec \alpha\cdot(-e_{0}\opa A(\vec x))\uppsi(\vec x):|\Psi_{\mathrm{Ps}}\rangle \neq 0.\label{eq:106}
\end{align}

At the same time, the state vector $|\psi_{0}\rangle$ Eq.~(\ref{eq:13}) is a linear combination of single-particle excitations. For this reason, the transition matrix element $M^{\mathrm{S}}_{if}$ is identically equal to zero, which corresponds to the conservation of charge and implies the stability of the soliton-like solution.

Finally, we want to briefly address the physical meaning of the obtained soliton-like solution in QED. We expect that this solution can describe the observable characteristics of the ``physical'' electron. However, this assumption requires an additional, comprehensive analysis, which we envisage to perform in subsequent works.

\begin{acknowledgments}
  The authors are grateful to S. I. Feranchuk, S. Cavaletto and S. Bragin for valuable discussions.
\end{acknowledgments}

\appendix
\section{Calculation of the expectation value $\langle\psi_{0}|\opa H^{\prime}_{\mathrm{QED}}|\psi_{0}\rangle$}
\label{sec:exp_value}

In order to calculate the expectation value $\langle\psi_{0}|\opa H^{\prime}_{\mathrm{QED}}|\psi_{0}\rangle$ we note that the vacuum average of the product of the creation and annihilation operators is not equal to zero only for the equal number of the former and the latter. As we calculate the expectation value in the mean field theory all operators in the Hamiltonian can be written in the general form as
\begin{align}
  \label{eq:exp_value1}
  :\uppsi^{\dag}(\vec x)\opa F \uppsi(\vec x): = \frac{1}{V}\sum_{\mathclap{\vec ps,\vec p^\prime s^\prime}}\sqrt{\frac{m_{0}}{\epsilon_{\vec p}}}\sqrt{\frac{m_{0}}{\epsilon_{p^\prime}}}&\Bigg(\opa a_{\vec ps}^\dag \opa a_{\vec p^\prime s^\prime}  u^\dag_{\vec ps}e^{-\ri\thp{p}{x}}\opa F u_{\vec p^\prime s^\prime}e^{\ri\vec p^\prime\cdot \vec x} \nonumber
  \\
  &\mspace{20mu}+\opa a_{\vec ps}^\dag \opa b_{\vec p^\prime s^\prime}^\dag u^\dag_{\vec ps}e^{-\ri\thp{p}{x}} \opa F v_{\vec p^\prime s^\prime}e^{-\ri\vec p^\prime\cdot \vec x} \nonumber
  \\
  &\mspace{20mu}+\opa b_{\vec ps} \opa a_{\vec p^\prime s^\prime} v^\dag_{\vec ps}e^{\ri\vec p\cdot \vec x}\opa F u_{\vec p^\prime s^\prime}e^{\ri\vec p^\prime\cdot \vec x}\nonumber
  \\
  &\mspace{20mu}-\opa b_{\vec p^\prime s^\prime}^\dag \opa b_{\vec p s} v^\dag_{\vec p s}e^{\ri\vec p\cdot \vec x}\opa F v_{\vec p^\prime s^\prime}e^{-\ri\vec p^\prime\cdot \vec x}\Bigg),
\end{align}
where the operator $\opa F$ in the case of the kinetic energy is equal to $\vec \alpha\cdot\opA p +\beta m_0$ and in the case of the density to $1$.

Consequently, after the counting of the number of creation and annihilation operators in the corresponding matrix elements $\langle\psi_{0}|:\uppsi^{\dag}(\vec x)\opa F \uppsi(\vec x):|\psi_{0}\rangle$ and the omission of the vanishing terms one obtains
\begin{align}
  \label{eq:exp_value2}
  \int d\vec x\langle\psi_{0}|:\uppsi^{\dag}(\vec x)\opa F \uppsi(\vec x):|\psi_{0}\rangle &= \int d\vec x \frac{1}{V}\sum_{\mathclap{\vec ps,\vec p^\prime s^\prime}}\sqrt{\frac{m_{0}}{\epsilon_{\vec p}}}\sqrt{\frac{m_{0}}{\epsilon_{p^\prime}}}\nonumber
  \\
  &\times\Bigg(U^{*}_{\vec ps}  u^\dag_{\vec ps}e^{-\ri\thp{p}{x}}\opa F U_{\vec p^{\prime}s^{\prime}}u_{\vec p^\prime s^\prime}e^{\ri\vec p^\prime\cdot \vec x}\nonumber
  \\
    &\mspace{190mu}- V_{\vec p s}v^\dag_{\vec p s}e^{\ri\vec p\cdot \vec x}\opa F V^{*}_{\vec p^{\prime} s^{\prime}}v_{\vec p^\prime s^\prime}e^{-\ri\vec p^\prime\cdot \vec x}\Bigg) \nonumber
  \\
  &=\int d\vec x \left(\Psi^{\dag}(\vec x)\opa F \Psi(\vec x) - \Psi^{c\dag}(\vec x)\opa F \Psi^{c}(\vec x)\right),
\end{align}
where we have introduced the inverse Fourier transforms of the coefficients
\begin{align}
  \Psi(\vec x) &= \sum_{\vec qr}\frac{1}{\sqrt{V}}\sqrt{\frac{m_{0}}{\epsilon_{\vec q}}}U_{\vec q r}u_{\vec q r}e^{i\thp{q}{x}}, \label{eq:exp_value3}
  \\
  \Psi^c(\vec x) &= \sum_{\vec qr}\frac{1}{\sqrt{V}}\sqrt{\frac{m_{0}}{\epsilon_{\vec q}}}V^*_{\vec q r}v_{\vec q r}e^{-i\thp{q}{x}}. \label{eq:exp_value4}
\end{align}
Consequently, the expectation value of the various terms in the reduced QED Hamiltonian looks like
\begin{align}
  \langle\psi_{0}|&:\uppsi^\dag(\vec x)(\vec \alpha\cdot\opA p +\beta m_0)\uppsi(\vec x):|\psi_{0}\rangle \nonumber
  \\
                  &= \left(\Psi^{\dag}(\vec x)(\vec \alpha\cdot\opA p +\beta m_0) \Psi(\vec x) - \Psi^{c\dag}(\vec x)(\vec \alpha\cdot\opA p +\beta m_0) \Psi^{c}(\vec x)\right), \label{eq:exp_value5}
  \\
  \langle\psi_{0}|&:\uprho(\vec x):|\psi_{0}\rangle \nonumber
  \\
  &=\left(\Psi^{\dag}(\vec x)\Psi(\vec x) - \Psi^{c\dag}(\vec x)\Psi^{c}(\vec x)\right) \label{eq:exp_value6}
\end{align}

\section{Change of the vacuum energy of the single-charge state with respect to the vaccum state}
\label{sec:change-vacuum-energy}

We would like to demonstrate that the normal ordering of operators in the reduced QED Hamiltonian is equivalent to counting all energies from the vacuum energy. In other words we want to demonstrate that the difference of the expectation value of the single-charge state $|\psi_{0}\rangle$ with respect to the vacuum state $|0\rangle$ yields the functional (\ref{eq:17})
\begin{align}
  \label{eq:change-vacuum-energy:1}
  \langle\psi_{0}|\opa H'_{\mathrm{QED}}|\psi_{0}\rangle - \langle0|\opa H'_{\mathrm{QED}}|0\rangle = \mathfrak{J}[\Psi(\vec x),\Psi^{c}(\vec x)],
\end{align}
if the normal ordering of operators is not used.

We would like to separate out the classical component in the density operator. For this reason, let us rewrite the quadratic in density term as
\begin{align}
  \label{eq:change-vacuum-energy:2}
  \frac{e_{0}^{2}}{8\pi}\int d\vec x d\vec y \frac{\uprho(\vec x)\uprho(\vec y)}{|\vec x - \vec y|} = \frac{e_{0}^{2}}{8\pi}\int d\vec x d\vec y \frac{\uprho(\vec x)\mathbb{I}\uprho(\vec y)}{|\vec x - \vec y|},
\end{align}
where we have introduced the identity operator $\mathbb{I}$ between the densities. This identity operator is equal to
\begin{align}
    \label{eq:change-vacuum-energy:3}
  \mathbb{I} = |0\rangle\langle0|+ |\psi_{0}\rangle\langle\psi_{0}| + |\psi_{1}\rangle\langle\psi_{1}| + |\psi_{2}\rangle\langle\psi_{2}| + \cdots.
\end{align}

The states $|\psi_{i}\rangle$ form a complete set and the state $|\psi_{2}\rangle$, for example, is equal to
\begin{align}
  \label{eq:change-vacuum-energy:4}
  |\psi_2\rangle = \sum_{\vec qs}(U_{\vec qs}\opa a^{\dag}_{\vec qs}+V_{\vec qs}\opa b^{\dag}_{\vec qs})(\opa a^\dag_{\vec k l} \opa b^\dag_{\vec k^\prime l^\prime}+\opa a^\dag_{\vec k l} \opa a^\dag_{\vec k^\prime l^\prime} + \opa b^\dag_{\vec k l} \opa b^\dag_{\vec k^\prime l^\prime})|0\ra.
\end{align}

In addition, the states $|\psi_{1}\rangle, |\psi_{2}\rangle, \ldots$ represent transitions into intermediate states with a higher number of electron-positron pairs and consequently correspond to the diagrams with a higher number of vertices. Consequently, in the zeroth-order approximation we drop the terms $|\psi_{1}\rangle\langle\psi_{1}|, |\psi_{2}\rangle\langle\psi_{2}|, \ldots$ in the projector and consider them as higher order corrections. This is a similar approximation to Ref.\cite{PhysRevD.10.4130}, where the authors kept only quadratic terms in $\upeta(\vec x)$ in the zeroth-order approximation of the expansion of the bosonic field $\upphi(x) = \phi(x) + \upeta(x)$.

Proceeding, we firstly calculate the vacuum expectation value. For this we evaluate
\begin{align}
  \label{eq:change-vacuum-energy:5}
  (|0\rangle\langle0| + |\psi_{0}\rangle\langle\psi_{0}|)\uprho(\vec x)|0\rangle = |0\rangle\langle0|\uprho(\vec x)|0\rangle,
\end{align}
since the matrix element $\langle\psi_{0}|\uprho(\vec x)|0\rangle$ vanishes. Consequently, the vacuum expectation value from the reduced QED Hamiltonian is equal to
\begin{align}
  \label{eq:change-vacuum-energy:6}
  \langle0|\opa H'_{\mathrm{QED}}|0\rangle = -2\sum_{\vec p}\sqrt{|\vec p|^{2} + m_{0}^{2}}+\frac{e_{0}^{2}}{8\pi}\int \frac{d\vec x d\vec y}{|\vec x - \vec y|}\langle0|\uprho(\vec x)|0\rangle \langle0|\uprho(\vec y)|0\rangle.
\end{align}

In a similar fashion, when we calculate the expectation value with a single-charge state we obtain
\begin{align}
  \label{eq:change-vacuum-energy:7}
  (|0\rangle\langle0| + |\psi_{0}\rangle\langle\psi_{0}|)\uprho(\vec x)|\psi_{0}\rangle = |\psi_{0}\rangle\langle\psi_{0}|\uprho(\vec x)|\psi_{0}\rangle.
\end{align}

Taking into account the anticommutation relation between the positronic operators $\{\opa b_{\vec ps},\opa b^{\dag}_{\vec ps}\} = \delta(\vec p - \vec p')\delta_{ss'}$ and the normalization of the single-charge state vector $\langle\psi_{0}|\psi_{0}\rangle = 1$ one easily finds that
\begin{align}
  \label{eq:change-vacuum-energy:8}
  \langle\psi_{0}|\opa H'_{\mathrm{QED}}|\psi_{0}\rangle = \langle\psi_{0}|:\opa H'_{\mathrm{QED}}:|\psi_{0}\rangle + \langle0|\opa H'_{\mathrm{QED}}|0\rangle
\end{align}
and consequently Eq.~(\ref{eq:change-vacuum-energy:1}) holds.

\section{Variation of the functional $\mathfrak{T}$}
\label{sec:var1}

In this Appendix we describe the variation of the functional defined by Eq.~(\ref{eq:24}). Here, we need to take into account that the self-consistent potential $\varphi(\vec x)$ is a function of $\Psi(\vec  x)$ and $\Psi^c(\vec x)$. For this reason, the variation of the corresponding term with the potential in the functional is performed as
\begin{align}
  \label{eq:var11}
  \delta_{\Psi^{\dag}}\left(\Psi^{\dag}(\vec x)\frac{1}{2}e_{0}\varphi(\vec x)\Psi(\vec x)\right) &= \delta_{\Psi^{\dag}} \Bigg[\frac{e_0^{2}}{4\pi} \int \frac {d\vec x d \vec{y}} {|\vec x - \vec{y}|} \nonumber
  \\
  &\times\Bigg(\Psi^\dag(\vec x)\Psi(\vec x)\left[\Psi^{\dag}(\vec y) \Psi (\vec y) - \Psi^{c\dag} (\vec y) \Psi^{c} (\vec y)\right] \nonumber
  \\
  &\mspace{90mu}-\Psi^{c\dag}(\vec x)\Psi^c(\vec x)\left[\Psi^{\dag}(\vec y) \Psi (\vec y) - \Psi^{c\dag} (\vec y) \Psi^{c} (\vec y)\right]\Bigg)\Bigg] \nonumber
  \\
  &\mspace{-80mu}=\frac{e_{0}^{2}}{4\pi}\int \frac {d\vec x d \vec{y}} {|\vec x - \vec{y}|}\Bigg[\delta_{\Psi^{\dag}}[ \Psi^\dag(\vec x)]\rho(\vec y)\Psi(\vec x)+\Psi^\dag(\vec x)\Psi(\vec x) \delta_{\Psi^{\dag}}[\Psi^\dag(\vec y)]\Psi(\vec y)\nonumber
  \\
  &\mspace{150mu}-\Psi^{c\dag}(\vec x)\Psi^c(\vec x)\delta_{\Psi^{\dag}}[\Psi^\dag(\vec y)]\Psi(\vec y)\Bigg],
\end{align}
or interchanging $\vec x \leftrightarrow \vec y$ in the last terms of the previous equation we conclude that
\begin{align}
  \label{eq:var12}
  \delta_{\Psi^{\dag}}\left(\Psi^{\dag}(\vec x)\frac{1}{2}e_{0}\varphi(\vec x)\Psi(\vec x)\right) = \int \delta_{\Psi^{\dag}}\Psi^\dag(\vec x) e_0\varphi(\vec x)\Psi(\vec x) d\vec x.
\end{align}

The variations with respect to $\Psi(\vec x)$, $\Psi^{c}(\vec x)$ and $\Psi^{c\dag}(\vec x)$ are performed in an analogous way.

Consequently, the variation of the functional $\mathfrak{T}$ looks like
\begin{align}
  \label{eq:var13}
  \delta\mathfrak{T} = \int d\vec x \Bigg\{&\delta(\Psi^\dag (\vec x)) \left[(\vec \alpha\cdot\opA p + \beta m_0) + e_0 \varphi (\vec x) -\Lambda\right] \Psi (\vec x)  \nonumber
  \\
                     &-\delta(\Psi^{c\dag} (\vec x)) \left[(\vec \alpha\cdot \opA p + \beta m_0)+e_0 \varphi (\vec x) + \Lambda_{c}\right] \Psi^c (\vec x) \nonumber
  \\
                     &+\Psi^\dag (\vec x) \left[(\vec \alpha\cdot\opA p + \beta m_0) + e_0 \varphi (\vec x) -\Lambda\right] \delta(\Psi (\vec x))  \nonumber
  \\
                     &-\Psi^{c\dag} (\vec x) \left[(\vec \alpha\cdot \opA p + \beta m_0)+e_0 \varphi (\vec x) +\Lambda_{c}\right] \delta(\Psi^c (\vec x))\Bigg\}.
\end{align}

\section{Derivation of the radial equations}
\label{sec:deriv-radi-equat}

Before proceeding with the derivation of the radial equations of motion we introduce the spherical spinors $\Omega_{jlM}$, which can be obtained, through the addition of the angular momentum $\opA J$ and the spin operators $\opA S$ \cite{AkhiezerB1965quantum}:
\begin{align}
  \label{eq:deriv-radi-equat:1}
  \Omega_{jlM} = \left(\begin{aligned}
      C^{jM}_{l,M-\frac{1}{2},\frac{1}{2},\frac{1}{2}}Y_{l,M-\frac{1}{2}}(\vec n)
      \\
      C^{jM}_{l,M+\frac{1}{2},\frac{1}{2},-\frac{1}{2}}Y_{l,M+\frac{1}{2}}(\vec n)
    \end{aligned}\right),
\end{align}
where the Clebsch–Gordan coefficients $C^{jM}_{l,M-\frac{1}{2},\frac{1}{2},\pm\frac{1}{2}}$ are defined as \cite{AkhiezerB1965quantum}
\begin{equation}
  \begin{aligned}
    C^{jM}_{l,M-\frac{1}{2},\frac{1}{2},\frac{1}{2}} &= \sqrt{\frac{l+M+\frac{1}{2}}{2l+1}}, &j = l+\frac{1}{2},& &C^{jM}_{l,M+\frac{1}{2},\frac{1}{2},-\frac{1}{2}} = \sqrt{\frac{l-M+\frac{1}{2}}{2l+1}}, \quad j = l+\frac{1}{2},
    \\
    C^{jM}_{l,M-\frac{1}{2},\frac{1}{2},\frac{1}{2}} &= -\sqrt{\frac{l-M+\frac{1}{2}}{2l+1}}, &j = l-\frac{1}{2},& &C^{jM}_{l,M+\frac{1}{2},\frac{1}{2},-\frac{1}{2}} = \sqrt{\frac{l+M+\frac{1}{2}}{2l+1}}, \quad j = l-\frac{1}{2}
  \end{aligned}\label{eq:deriv-radi-equat:2}
\end{equation}
and $Y_{l,M\pm1/2}$ are the ordinary spherical harmonics \cite{LandauQM,AkhiezerB1965quantum}.

In the following we will need the spherical spinors $\Omega_{jlM}$ for $j = 1/2$, $l=\{0,1\}$ and $M = \pm1/2$. Consequently, with the help of Eqs.~(\ref{eq:deriv-radi-equat:1}) - (\ref{eq:deriv-radi-equat:2}) one obtains
\begin{equation}
  \begin{aligned}
    \Omega_{\frac{1}{2},0,\frac{1}{2}} &= \begin{pmatrix}
      Y_{00}(\vec n)
      \\
      0
    \end{pmatrix}, &\Omega_{\frac{1}{2},0,-\frac{1}{2}}& = \begin{pmatrix}
      0
      \\
      Y_{00}(\vec n)
    \end{pmatrix},
    \\
    \Omega_{\frac{1}{2},1,\frac{1}{2}} &= \begin{pmatrix}
      -\sqrt{\frac{1}{3}} Y_{10}(\vec n)
      \\
      \sqrt{\frac{2}{3}}Y_{11}(\vec n)
    \end{pmatrix}, &\Omega_{\frac{1}{2},1,-\frac{1}{2}}& = \begin{pmatrix}
      -\sqrt{\frac{2}{3}}Y_{1-1}(\vec n)
      \\
      \sqrt{\frac{1}{3}} Y_{10}(\vec n)
    \end{pmatrix},
  \end{aligned}\label{eq:deriv-radi-equat:3}
\end{equation}
where
\begin{equation}
  \begin{aligned}
    &Y_{00}(\vec n) = \sqrt{\frac{1}{4\pi}},& &Y_{10}(\vec n) = \sqrt{\frac{3}{4\pi}} \cos{\theta},&
    \\
    &Y_{11}(\vec n) = -\sqrt{\frac{3}{8\pi}}e^{\ri \varphi}\sin{\theta},& &Y_{1-1}(\vec n) = \sqrt{\frac{3}{8\pi}}e^{-\ri \varphi}\sin{\theta}.&
  \end{aligned}\label{eq:deriv-radi-equat:4}
\end{equation}

The following useful properties \cite{LandauQED,AkhiezerB1965quantum} of the spherical spinors will be used below
\begin{equation}
  \begin{aligned}
    &\vec \sigma\cdot\vec n\ \Omega_{jlM} = - \Omega_{jl^\prime M},& &\int do \Omega_{jlM}^{\dag}\Omega_{j^{\prime}l^{\prime}M^{\prime}} = \delta_{jj^{\prime}}\delta_{ll^{\prime}}\delta_{MM^{\prime}}
    \\
    &\vec \sigma\cdot\opA p\ \Omega_{jlM} = \ri \frac{1+\varkappa}{r}\Omega_{jl^\prime M},& &\vec \sigma\cdot\opA p\ \Omega_{jl^\prime M} = \ri \frac{1-\varkappa}{r}\Omega_{jlM},
    \\
    &\varkappa = l(l+1) - j(j+1) - \frac{1}{4},& &\vec n = \vec x / r.
  \end{aligned}\label{eq:deriv-radi-equat:5}
\end{equation}
Here $\vec \sigma$ denotes the vector of the Pauli matrices and $\int do$ represents the integration over the angular variables in a spherical coordinate system $do = \sin\theta d\theta d\varphi$.

Let us also introduce the abbreviations
\begin{equation}
  \begin{aligned}
    \chi_{l} &= A_{\frac{1}{2}}\Omega_{\frac{1}{2}, l, \frac{1}{2}} + A_{-\frac{1}{2}}\Omega_{\frac{1}{2}, l, -\frac{1}{2}}, \quad \chi_{l}^{c} = A^{c}_{\frac{1}{2}}\Omega_{\frac{1}{2}, l, \frac{1}{2}} + A^{c}_{-\frac{1}{2}}\Omega_{\frac{1}{2}, l, -\frac{1}{2}},
  \end{aligned}\label{eq:deriv-radi-equat:6}
\end{equation}
where the coefficients $A_{\pm\frac{1}{2}}$ and $A^{c}_{\pm\frac{1}{2}}$ satisfy the normalization condition
\begin{align}
  |A_{\frac{1}{2}}|^2 + |A_{-\frac{1}{2}}|^2 = |A^{c}_{\frac{1}{2}}|^2 + |A^{c}_{-\frac{1}{2}}|^2 = 1.\label{eq:deriv-radi-equat:7}
\end{align}

Moreover, with the help of Eqs.~(\ref{eq:deriv-radi-equat:5})-(\ref{eq:deriv-radi-equat:7}) one can write
\begin{equation}
  \begin{aligned}
    \chi_{l}^\dag \chi_{l} &= \frac{1}{4\pi},\quad \int d o \chi_{l}^\dag \chi_{l^\prime} = \delta_{l l^\prime};
    \\
    \vec \sigma\cdot\opA p \chi_{0} &= 0, \quad \vec \sigma\cdot\opA p \chi_{1} = \frac{2\ri}{r}\chi_{0},
    \\
    \vec \sigma\cdot\vec n \chi_{0} &= -\chi_{1}, \quad \vec \sigma\cdot\vec n \chi_{1} = -\chi_{0},
  \end{aligned}\label{eq:deriv-radi-equat:8}
\end{equation}
with the analogous expressions for $\chi^{c}_{l}$.

Since all the relevant quantities have been defined we can start the calculation of the density $\rho(\vec x) = \Psi^{\dag}(\vec x)\Psi(\vec x) - \Psi^{c\dag}(\vec x)\Psi^{c}(\vec x)$. For this purpose, we note that the most general linear combination of the wave functions, which leads to the spherically symmetric density can be written as
\begin{equation}
  \begin{aligned}
    \Psi(\vec x) &= A_{\frac{1}{2}} \Psi_{\frac{1}{2}, 0, \frac{1}{2}}+A_{-\frac{1}{2}}\Psi_{\frac{1}{2},0,-\frac{1}{2}} = \begin{pmatrix}
      g(r)\chi_{0}
      \\
      \ri f(r)\chi_{1}\end{pmatrix},
    \\
    \Psi^c(\vec x) &= A^{c}_{\frac{1}{2}} \Psi^c_{\frac{1}{2}, 0, \frac{1}{2}}+A^{c}_{-\frac{1}{2}}\Psi^c_{\frac{1}{2},0,-\frac{1}{2}} = \begin{pmatrix}
      g_1(r)\chi^{c}_{0}
      \\
      \ri f_1(r)\chi_{1}^{c}\end{pmatrix}.
  \end{aligned}\label{eq:deriv-radi-equat:9}
\end{equation}

In addition, the functions $\Psi(\vec x)$ and $\Psi^{c}(\vec x)$ should satisfy the orthogonality relation $\int d\vec x \Psi^{\dag}(\vec x)\Psi^{c}(\vec x) = 0$. This is achieved with a suitable choice of the coefficients $A_{\pm1/2}$ and $A^{c}_{\pm1/2}$, i.e. $A^{*}_{\frac{1}{2}}A^{c}_{\frac{1}{2}}+A^{*}_{-\frac{1}{2}}A^{c}_{-\frac{1}{2}}=0$. One can easily demonstrate that the choice of $\chi_{l} = \frac{1}{\sqrt{2}}(\Omega_{1/2,l,1/2}+\Omega_{1/2,l,-1/2})$ and $\chi_{l}^{c} = \frac{1}{\sqrt{2}}(\Omega_{1/2,l,1/2}-\Omega_{1/2,l,-1/2})$, $l=0,1$ leads to the orthogonality of $\Psi(\vec x)$ and $\Psi^{c}(\vec x)$.

By calculating the density $\rho(\vec x)$ with these wave functions and by using the properties of the spherical spinors $\chi$ of Eq.~(\ref{eq:deriv-radi-equat:8}) one obtains
\begin{align}
  \rho(r) = \frac{|g|^{2} + |f|^{2}}{4\pi} - \frac{|g_{1}|^{2} - |f_{1}|^{2}}{4\pi}.\label{eq:deriv-radi-equat:10}
\end{align}

Now we are ready to continue the derivation of the radial equations of motion. For this we firstly note that Eqs.~(\ref{eq:25}) and (\ref{eq:26}) are identical up to the notations indicated by the index $c$. For this reason, we will derive the radial equation of motion only for the wave function $\Psi(\vec x)$ as the final result for $\Psi^{c}(\vec x)$ can be simply obtained by adding the index $c$.

First of all we rewrite the Dirac Eq.~(\ref{eq:25}) in matrix form for the two-component spinors
\begin{equation}
  \begin{aligned}
    \begin{pmatrix}
      e_0 \varphi + m_0 & \vec \sigma \cdot \opA p
      \\
      \vec \sigma \cdot \opA p & e_0 \varphi - m_0
    \end{pmatrix}\begin{pmatrix}
      g(r)\chi_{0}
      \\
      \ri f(r)\chi_{1}
    \end{pmatrix} &= \Lambda \begin{pmatrix}
      g(r)\chi_{0}
      \\
      \ri f(r)\chi_{1}
    \end{pmatrix}.
  \end{aligned}\label{eq:deriv-radi-equat:11}
\end{equation}

With the help of Eqs.~(\ref{eq:deriv-radi-equat:5}) we determine how the operator $\vec\sigma\cdot\opA p$ acts on the two-component wave function
\begin{equation}
  \begin{aligned}
    \vec \sigma \cdot \opA p (g(r) \chi_0) &= -\ri g^\prime(r)\thp{\sigma}{n}\chi_0 + g(r)\vec \sigma \cdot \opA p \chi_0 = \ri g^\prime(r) \chi_1,
    \\
    \vec \sigma \cdot \opA p (\ri f(r) \chi_1) &=  f^\prime(r)\thp{\sigma}{n}\chi_1 + \ri f(r)\vec \sigma \cdot \opA p \chi_1 = - f^\prime(r) \chi_0 - \frac{2 f(r)}{r}\chi_0.
  \end{aligned}\label{eq:deriv-radi-equat:12}
\end{equation}
Consequently, by plugging Eq.~(\ref{eq:deriv-radi-equat:12}) into Eq.~(\ref{eq:deriv-radi-equat:11}) one obtains
\begin{equation}
  \begin{aligned}
    &f^\prime(r) + \frac{2f(r)}{r} - (e_0 \varphi(r) + m_0)g(r) = -\Lambda g(r),
    \\
    &g^\prime(r) + (e_0\varphi(r) - m_0)f(r) = \Lambda f(r),
  \end{aligned}\label{eq:deriv-radi-equat:13}
\end{equation}
which is cast into the final form through the use of the identity $f^\prime(r) = ((rf(r))^\prime - f(r))/r$
\begin{equation}
  \begin{aligned}
    &(rg(r))^\prime - \frac{rg(r)}{r} - (m_0 - e_0\varphi(r))rf(r) = \Lambda f(r),
    \\
    &(rf(r))^\prime + \frac{rf(r)}{r} - (m_0 + e_0 \varphi(r) )rg(r) = -\Lambda g(r).
  \end{aligned}\label{eq:deriv-radi-equat:14}
\end{equation}

We introduce the dimensionless variables as follows
\begin{equation}
  \begin{aligned}
    &x = r m_0, \quad \Lambda = m_0 \lambda, \quad e_0 \varphi(r) = m_0\phi(x), \quad \frac{e^2_0}{4 \pi} = \alpha_0,
    \\
    &u(x) \sqrt{m_0} = r g(r), \quad v(x)\sqrt{m_0} = r f(r),
  \end{aligned}\label{eq:deriv-radi-equat:15}
\end{equation}
which allows one to rewrite the system of equations (\ref{eq:deriv-radi-equat:14}) in compact form
\begin{equation}
  \begin{aligned}
    u^\prime(x) - \frac{u(x)}{x} - (1 - \phi(x)) v(x) &= \lambda v(x),
    \\
    v^\prime(x) + \frac{v(x)}{x} - (1 + \phi(x)) u(x) &= -\lambda u(x).
  \end{aligned}\label{eq:deriv-radi-equat:16}
\end{equation}

In order to complete our derivation we need to add the equation for the potential. For this purpose, we note that the self-consistent potential satisfies the Poisson equation
\begin{align}
  \label{eq:deriv-radi-equat:17}
  \Delta\varphi(\vec x) = -e_{0}\rho(r).
\end{align}
This directly follows from its definition Eq.~(\ref{eq:18}).

Since the density $\rho(r)$ on the right hand side of Eq.~(\ref{eq:deriv-radi-equat:17}) is a spherically symmetric function, the self-consistent potential is also spherically symmetric. Consequently, Eq.~(\ref{eq:deriv-radi-equat:17}) transforms into the following form, which we write in dimensionless variables
\begin{align}
  (x^2 \phi^\prime(x))^\prime &= - \alpha_0 (|u(x)|^2 + |v(x)|^2 - |u_1(x)|^2 - |v_1(x)|^2) = -\alpha_0 \rho(x), \label{eq:deriv-radi-equat:18}
\end{align}
where prime denotes the differentiation with respect to $x$.

We continue further with the solution of Eq.~(\ref{eq:deriv-radi-equat:18}), for which we firstly perform the change of variables
\begin{align}
  \phi(x) = \frac{\tilde \phi}{x}; \quad \phi^\prime(x) = \frac{\tilde \phi^\prime(x)}{x} - \frac{\tilde \phi(x)}{x^2}; \quad \phi^{\prime \prime}(x) = \frac{\tilde \phi^{\prime \prime}(x)}{x} - \frac{2\tilde \phi^\prime(x)}{x^2}+\frac{\tilde \phi(x)}{x^3}, \label{eq:deriv-radi-equat:19}
\end{align}
in which Eq.~(\ref{eq:deriv-radi-equat:18}) looks like
\begin{align}
  \tilde \phi^{\prime \prime}(x) = -\frac{\alpha_0}{x}\rho(x).\label{eq:deriv-radi-equat:20}
\end{align}

Further solution we perform with the help of the Green function
\begin{equation}
  G(x,y) = \left\{\begin{aligned}
      x, \quad x>y
      \\
      y, \quad x<y
    \end{aligned}\right.\label{eq:deriv-radi-equat:21}
\end{equation}
of the free equation $\tilde{\phi}^{\prime \prime}(x) = 0$.

Consequently, the general solution of Eq.~(\ref{eq:deriv-radi-equat:20}) can be written as
\begin{align}
  \phi(x) = -\alpha_0\int_0^x \frac{\rho(y)}{y}dy - \frac{\alpha_0}{x}\int_x^\infty \rho(y)dy + \frac{A}{x} + A_1,\label{eq:deriv-radi-equat:22}
\end{align}
where the integration constants $A$ and $A_{1}$ are to be defined from the conditions that $\phi(x)$ is finite at zero and possesses the correct asymptotic behavior at infinity. Hence,
\begin{align}
  A_1 = \alpha_0 \int_0^\infty \frac{\rho(y)}{y}dy, \quad A = \alpha_0 \int_0^\infty \rho(y)dy.\label{eq:deriv-radi-equat:23}
\end{align}

By plugging Eq.~(\ref{eq:deriv-radi-equat:23}) into Eq.~(\ref{eq:deriv-radi-equat:22}) we come to the final result for the self-consistent potential written in the dimensionless variables
\begin{align}
  \phi(x) &= \frac{\alpha_0}{x}\int_0^x \rho(y)dy + \alpha_0\int_x^\infty \frac{\rho(y)}{y}dy. \label{eq:deriv-radi-equat:24}
\end{align}

Combining all results together, namely the system of Eqs.~(\ref{eq:deriv-radi-equat:16}), with the corresponding system of equations with the subscript $c$, the equation for the self-consistent potential Eq.~(\ref{eq:deriv-radi-equat:24}) and the normalization conditions defined in Eqs.~(\ref{eq:21}) and (\ref{eq:22}), we finally obtain
\begin{align}
  \left\{\begin{aligned}
      &u^\prime(x) - \frac{u(x)}{x} - (1-\phi(x) ) v(x) = \lambda v(x),
      \\
      &v^\prime(x) + \frac{v(x)}{x} - (1+\phi(x)) u(x) = -\lambda u(x),
      \\
      &u_c^\prime(x) - \frac{u_c(x)}{x} - (1-\phi(x) ) v_c(x) = -\lambda_c v_c(x),
      \\
      &v_c^\prime(x) + \frac{v_c(x)}{x} - (1+\phi(x)) u_c(x) = \lambda_c u_c(x),
      \\
      &\phi(x) = \frac{\alpha_0}{x}\int_0^x \rho(y)dy + \alpha_0 \int_x^\infty \frac{\rho(y)}{y}dy,
      \\
      &\int_0^\infty [u^2(x) + v^2(x) + u_c^2(x) + v_c^2(x)]dx = 1.
    \end{aligned}\right. \label{eq:deriv-radi-equat:25}
\end{align}

This system of equations should be complemented with the boundary conditions resulting from the asymptotic behavior of the functions $u(x)$, $v(x)$, $u_{c}(x)$ and $v_{c}(x)$ near zero and infinity respectively:
\begin{equation}
  \begin{aligned}
    &u(x) \sim F_{0}x\left(1+\frac{1-(\lambda - \phi(0))^2}{6}x^2\right),& &v(x) \sim F_{0}\frac{1-(\lambda - \phi(0))}{3}x^2,& &x\rightarrow0,
    \\
    &u(x) \sim F_{\infty} e^{-\sqrt{1- \lambda^2}x},& &v(x) \sim -F_{\infty}\sqrt{\frac{1-\lambda}{1+\lambda}} e^{-\sqrt{1-\lambda^2}x},& &x \rightarrow \infty,
  \end{aligned}\label{eq:deriv-radi-equat:26}
\end{equation}
with the corresponding equations for $u_{c}(x)$ and $v_{c}(x)$.

Concluding, in this Appendix we derived the system of equations which describes the collective excitation of the electron-positron system in quantum electrodynamics in the absence of the photon field.

\section{Continuous analog of Newton method for the numerical solution of the system of equations}
\label{sec:cont-anal-newt}

In this Appendix we present the method of the numerical solution of the radial system of equations of the self-consistent field, which determines the soliton-like solution in quantum electrodynamics
\begin{equation*}
  \left\{\begin{aligned}
      &u_{N}^\prime(x) - \frac{u_N(x)}{x} - (1-\phi(x) ) v_N(x) = \lambda v_N(x),
      \\
      &v_N^\prime(x) + \frac{v_N(x)}{x} - (1+\phi(x)) u_N(x) = -\lambda u_N(x),
      \\
      &u_{cN}^\prime(x) - \frac{u_{cN}(x)}{x} - (1-\phi(x) ) v_{cN}(x) = -\lambda_c v_{cN}(x),
      \\
      &v_{cN}^\prime(x) + \frac{v_{cN}(x)}{x} - (1+\phi(x)) u_{cN}(x) = \lambda_c u_{cN}(x),
      \\
      &\phi(x) = \frac{\alpha_0}{x}\int_0^x \rho(y)dy + \alpha_0 \int_x^\infty \frac{\rho(y)}{y}dy,
      \\
      &\rho(y) = \frac{1}{1+C}(u_{N}^2(y) + v_{N}^2(y)) - \frac{C}{1+C}(u_{cN}^2(y) + v_{cN}^2(y)),
      \\
      &\int_0^\infty dx (u_{N}^2(x) + v_{N}^2(x)) = \int_0^\infty dx (u_{cN}^2(x) + v_{cN}^2(x)) = 1.
    \end{aligned}\right. \tag{\ref{eq:33}}
\end{equation*}
with the corresponding boundary conditions
\begin{equation*}
  \begin{aligned}
    &u_{N}(x) \sim F_{0}x\left(1+\frac{1-(\lambda - \phi(0))^2}{6}x^2\right),& &v_{N}(x) \sim F_{0}\frac{1-(\lambda - \phi(0))}{3}x^2,& &x\rightarrow0,
    \\
    &u_{N}(x) \sim F_{\infty} e^{-\sqrt{1- \lambda^2}x},& &v_{N}(x) \sim -F_{\infty}\sqrt{\frac{1-\lambda}{1+\lambda}} e^{-\sqrt{1-\lambda^2}x},& &x \rightarrow \infty, 
  \end{aligned}\tag{\ref{eq:30}}
\end{equation*}
and the analogous expressions for $u_{cN}(x)$ and $v_{cN}(x)$.

Firstly, we would like to represent Eqs.~(\ref{eq:33}) in symmetric form. For this reason, we make the replacement $\lambda_{c}\to-\lambda_{c}$. Secondly, let us rewrite the boundary conditions in more convenient form, namely excluding the constants $F_{0}$ and $F_{\infty}$ from Eqs.~(\ref{eq:30}). This yields
\begin{align}
  \left\{\begin{aligned}
      &v_{N}(x) - u_{N}(x)x \frac{1 - (\lambda - \phi(0))}{3} = 0,& &v_{N}^\prime(x) - u_{N}^\prime(x) x \frac{2}{3}(1 - (\lambda - \phi(0))),& &x\rightarrow 0
      \\
      &v_{N}(x) + u_{N}(x) \sqrt{\frac{1-\lambda}{1+\lambda}} = 0,& &v_{N}^\prime(x) + u_{N}^\prime(x) \sqrt{\frac{1 - \lambda}{1+ \lambda}} = 0,& &x\rightarrow \infty.
    \end{aligned}\right..  \label{eq:cont-anal-newt:1}
\end{align}

Secondly, we reformulate the system of Eqs.~(\ref{eq:deriv-radi-equat:25}) together with the boundary conditions Eqs.~(\ref{eq:cont-anal-newt:1}) in matrix form
\begin{align}
  \left\{\begin{aligned}
      &\bar{\opa L} \bar{\vec X} = 0
      \\
      &\bar{\opa L}_0 \bar{\vec X} = 0
      \\
      &\bar{\opa L}_\infty \bar{\vec X} = 0
    \end{aligned}\right. \equiv \mathfrak{F}[\bar{\vec X}] = 0, \label{eq:cont-anal-newt:2}
\end{align}
where
\begin{align}\label{eq:cont-anal-newt:3}
  \bar{\opa L} =
  \begin{pmatrix}
    \opa L && 0
    \\
    0 && \opa L^{c}
  \end{pmatrix},
  \quad
  \bar{\opa L}_{0} =
  \begin{pmatrix}
    \opa L_{0} && 0
    \\
    0 && \opa L_{0}^{c}
  \end{pmatrix},
  \quad
  \bar{\opa L}_{\infty} =
  \begin{pmatrix}
    \opa L_{\infty} && 0
    \\
    0 && \opa L_{\infty}^{c}
  \end{pmatrix},
\end{align}
\begin{align}\label{eq:cont-anal-newt:4}
  \opa L &= \begin{pmatrix}
    \frac{\partial}{\partial x} - \frac{1}{x} && -(1+\lambda - \phi(x))
    \\
    -(1 - \lambda + \phi(x)) && \frac{\partial}{\partial x} + \frac{1}{x}
  \end{pmatrix},
  \quad
  \opa L_0 = \begin{pmatrix}
    -x \frac{1 - [\lambda - \phi(0)]}{3} && 1
    \\
    -x \frac{2[1 - \{\lambda -\phi(0)\}]}{3} \frac{\partial}{\partial x} && \frac{\partial}{\partial x}
  \end{pmatrix}, \nonumber
  \\
  \opa L_\infty &= \begin{pmatrix}
    \sqrt{\frac{1 - \lambda}{1 + \lambda}} && 1 
    \\
    \sqrt{\frac{1 - \lambda}{1 + \lambda}}\frac{\partial}{\partial x} && \frac{\partial}{\partial x} 
  \end{pmatrix}.
\end{align}
\begin{align}
  \label{eq:cont-anal-newt:5}
  \bar{\vec X} =
  \begin{pmatrix}
    \vec X
    \\
    \vec X^{c}
  \end{pmatrix},
  \quad
  \vec X = \begin{pmatrix}
    u_{N}(x)
    \\
    v_{N}(x)
  \end{pmatrix},
  \quad
  \vec X^{c} = \begin{pmatrix}
    u_{cN}(x)
    \\
    v_{cN}(x)
  \end{pmatrix}.
\end{align}

In order to obtain the set of operators with the index $c$ in Eq.~(\ref{eq:cont-anal-newt:3}), i.e. $\opa L^{c}$, $\opa L_{0}^{c}$ and $\opa L_{\infty}^{c}$, one needs to add the subscript $c$ to the eigenvalue $\lambda$ in Eq.~(\ref{eq:cont-anal-newt:4}).

Consequently, our task consists in the solution of the nonlinear integro-differential Eq.~(\ref{eq:cont-anal-newt:2}). This will be achieved with the help of some modification of the continuous analog of the Newton method \cite{KOMAROV1978153,GAREEV1977116}, which was applied in a large number of physical problems \cite{ERMAKOV1981235,AirapetyanA1999continuous1,PONOMAREV1978274,MELEZHIK199167,PONOMAREV19731,MELEZHIK1984221,MELEZHIK19861,Hoheisel20121324,Ramm2003}. According to this method the initial problem is substituted with the corresponding evolution equation
\begin{align}
  \mathfrak{F}^{\prime}[\bar{\vec X}(t)] \frac{d\bar{\vec X}}{dt} = - \mathfrak{F}[\bar{\vec X}(t)],\label{eq:cont-anal-newt:6}
\end{align}
where $\mathfrak{F}^{\prime}[\bar{\vec X}]$ is the Fr\'{e}chet derivative of the operator $\mathfrak{F}[\bar{\vec X}]$ and the desired solution $\bar{\vec X}(t)$ is a function of the continuous parameter $t$, $0\leq t < \infty$. Then, under the sufficiently general assumptions \cite{GAREEV1977116,ZhanlavA1992convergence,PuzyninA1999generalized,ZhidkovA1973continuous,GavurinA1958nonlinear,AirapetyanA1999continuous} the evolution Eq.~(\ref{eq:cont-anal-newt:6}) leads to the desired solution $\bar{\vec X}^{*}$, i.e.
\begin{align}
  \label{eq:cont-anal-newt:7}
  \lim_{t\to\infty}\left[\bar{\vec X}(t) - \bar{\vec X}^{*}\right] = 0.
\end{align}

Before proceeding, we recall here that an analogous system of equations appears in the polaron problem \cite{Mitra198791}. However, in that case the two radial Dirac equations are replaced with the Schr\"{o}dinger equation for the wave function $\psi(r)$. Nevertheless, the self-consistent potential is expressed in exactly the same fashion through the density $\rho(r)$ ($\rho(r) = |\psi(r)|^{2}$) as in Eq.~(\ref{eq:18}). Moreover, it was demonstrated in Ref.~\cite{KOMAROV1978153} that the direct application of the evolution Eq.~(\ref{eq:cont-anal-newt:6}) for the polaron problem does not lead to the desired solution for the wave function $\psi(r)$, since it was not possible to prove that the operator $(\mathfrak{F}^{\prime}[\bar{\vec X}(t)])^{-1}$ is bounded from above. Albeit that it is still possible to find the desired solution for which the modification of the Newton method is to be carried out. Namely, during the calculation of the Fr\'{e}chet derivative in Eq.~(\ref{eq:cont-anal-newt:6}) the self-consistent potential should be considered as the $t$-independent function and should be recalculated according to its definition Eq.~(\ref{eq:deriv-radi-equat:24}). Consequently, in this case the operator $(\mathfrak{F}^{\prime}[\bar{\vec X}(t)])^{-1}$ is bounded from above and the evolution does indeed Eq.~(\ref{eq:cont-anal-newt:6}) lead to the desired solution \cite{KOMAROV1978153,GAREEV1977116}. For this reason in what follows we will apply this modified Newton method.

As described in the previous paragraph, the self-consistent potential should be considered as a $t$-independent function. This has a very important implication on the solution of the system of Eqs.~(\ref{eq:cont-anal-newt:2}). Firstly, we notice that the set of operators $\opa L^{c}$, $\opa L_{0}^{c}$, $\opa L_{\infty}^{c}$ is different from the corresponding set $\opa L$, $\opa L_{0}$, $\opa L_{\infty}$ only in terms of the subscript $c$ of the eigenvalue $\lambda$. Secondly, we do not differentiate the self-consistent potential with respect to $t$. As a result, due to the block-diagonal structure of the matrices $\bar{\opa L}$, $\bar{\opa L}_{0}$, $\bar{\opa L}_{\infty}$, the system of equations is split into two equivalent systems, which are coupled only through the self-consistent potential $\phi(x)$. Moreover, during the actual numerical implementation the continuous parameter $t$ is replaced through a set of discrete values $t_{k}$. Consequently, for a given $t_{k}$ the two systems of equations can be solved independently. After this, in the next step, viz. $t_{k+1}$, the self-consistent potential is recalculated according to its definition of Eq.~(\ref{eq:deriv-radi-equat:24}) and the two systems are again solved independently. For this reason, the subsequent relations will be presented only for the expressions without the subscript $c$, as the final result can be simply obtained by adding the corresponding subscript $c$.

In order to continue we insert the system of Eqs.~(\ref{eq:cont-anal-newt:2}) into the evolution Eq.~(\ref{eq:cont-anal-newt:6}). This yields
\begin{equation}
  \label{eq:cont-anal-newt:8}
  \left\{\begin{aligned}
    \opa L \vec V &= - \opa L \vec X + \xi \opa M \vec X,
    \\
    \opa L_{0} \vec V &= - \opa L_{0} \vec X - \xi \opa M_{0} \vec X,
    \\
    \opa L_{\infty} \vec V &= - \opa L_{\infty} \vec X + \frac{\xi}{\varkappa (1 + \lambda)^{2}} \opa M_{\infty} \vec X,
  \end{aligned}\right. \quad \vec V = \frac{d\vec X}{dt},
\end{equation}
where $\xi = d\lambda / dt$ and
\begin{align}
  \frac{d\opa L}{dt} = -\xi \opa M, \quad \opa M =
  \begin{pmatrix}
    0 && 1
    \\
    -1 && 0
  \end{pmatrix},\label{eq:cont-anal-newt:9}
\end{align}
\begin{align}
  \frac{d\opa L_{0}}{dt} = \xi \opa M_{0}, \quad \opa M_{0} =
  \begin{pmatrix}
    \frac{x}{3} && 0
    \\
    \frac{2}{3} x \frac{\partial}{\partial x} && 0
  \end{pmatrix},\label{eq:cont-anal-newt:10}
\end{align}
\begin{align}
  \frac{d\opa L_{\infty}}{dt} = -\frac{\xi}{\varkappa (1+\lambda)^{2}} \opa M_{\infty}, \quad \opa M_{\infty} =
  \begin{pmatrix}
    1 && 0
    \\
    \frac{\partial}{\partial x} && 0
  \end{pmatrix}, \quad \varkappa = \sqrt{\frac{1 - \lambda}{1 + \lambda}}.\label{eq:cont-anal-newt:11}
\end{align}

Further, we perform the discrete approximation of the system of Eqs.~(\ref{eq:cont-anal-newt:8}). For this purpose, we break the semi-infinite interval $0\leq t < \infty$ into sub-intervals with grid points $k = 0, 1, \ldots,$ with the lengths $\tau_{k}$ \cite{ERMAKOV1981235,ZhanlavA1992convergence}. Moreover,
\begin{equation}
  \begin{aligned}
    t_{0} &= 0, \quad t_{k+1} = t_{k} + \tau_{k} ,
    \\
    \vec X_{k+1} &= \vec X_{k} + \tau_{k}\vec V_{k},
    \\
    \lambda_{k + 1} &= \lambda_{k} + \tau_{k} \xi_{k}.
  \end{aligned}\label{eq:cont-anal-newt:12}
\end{equation}
The discretization scheme for the differential equation $\vec V = d\vec X/dt$ is based on the Euler method \cite{PressB2007numerical,SamarskiiB1989numerical} of the solution of differential equations.

Proceeding, we seek the solution for $\vec V$ in the form
\begin{align}
  \vec V_{k} = \vec Z_{k} + \xi \vec Y_{k}.\label{eq:cont-anal-newt:13}
\end{align}

The next step consists of plugging of Eq.~(\ref{eq:cont-anal-newt:13}) into Eq.~(\ref{eq:cont-anal-newt:8}) and equating the terms with the corresponding powers of $\xi$. This leads us to the following result
\begin{equation}
  \begin{aligned}
    &\opa L \vec Z_{k} = - \opa L \vec X_{k},& &\opa L \vec Y_{k} = \opa M \vec X_{k},
    \\
    &\opa L_{0} \vec Z_{k} = - \opa L_{0} \vec X_{k},& &\opa L_{0} \vec Y_{k} = -\opa M_{0} \vec X_{k},
    \\
    &\opa L_{\infty} \vec Z_{k} = - \opa L_{\infty} \vec X_{k},& &\opa L_{\infty} \vec Y_{k} = \frac{1}{\varkappa (1 + \lambda)^{2}}\opa M_{\infty} \vec X_{k}.
  \end{aligned}\label{eq:cont-anal-newt:14}
\end{equation}

As the final step the parameter $\xi$ needs to be determined. For this we utilize the normalization conditions
\begin{align}
  \int_{0}^{\infty}dx \vec X\cdot\vec X = 1, \quad \int_{0}^{\infty}dx \vec X^{c} \cdot\vec X^{c} = 1,\label{eq:cont-anal-newt:15}
\end{align}
which are the direct consequence of Eqs.~(\ref{eq:21}) and (\ref{eq:22}). The differentiation of Eqs.~(\ref{eq:cont-anal-newt:15}) with respect to $t$ and the use of the definition of $\vec V$ yield
\begin{align}
  \xi_{k} = -\frac{1}{2}\frac{\int_{0}^{\infty} dx\vec X_{k}\cdot\vec X_{k} - 1 + 2 \int_{0}^{\infty}dx\vec X_{k}\cdot\vec Z_{k}}{\int_{0}^{\infty}dx \vec X_{k}\cdot\vec Y_{k}},\label{eq:cont-anal-newt:16}
  \\
  \xi^{c}_{k} = -\frac{1}{2}\frac{\int_{0}^{\infty} dx\vec X^{c}_{k}\cdot\vec X^{c}_{k} - 1 + 2 \int_{0}^{\infty}dx\vec X^{c}_{k}\cdot\vec Z^{c}_{k}}{\int_{0}^{\infty}dx \vec X^{c}_{k}\cdot\vec Y^{c}_{k}}.\label{eq:cont-anal-newt:17}
\end{align}

Consequently, we can formulate the algorithm of the numerical solution of the system of equations of the self-consistent field (\ref{eq:deriv-radi-equat:25}):
\begin{enumerate}
\item The initial approximation $\bar{\vec X}_{k}$, $k=0$ for the vector of unknowns $\bar{\vec X}$ is specified.
\item Using $\bar{\vec X}_{k}$, the initial self-consistent potential $\phi_{k}(x)$,  is calculated according to Eq.~(\ref{eq:deriv-radi-equat:24}).
\item The system of boundary value problems, defined by Eq.~(\ref{eq:cont-anal-newt:14}), is solved.
\item With the help of Eqs.~(\ref{eq:cont-anal-newt:16}) and (\ref{eq:cont-anal-newt:17}) the unknown corrections $\xi$ and $\xi^{c}$ to the eigenvalues $\lambda$ and $\lambda_{c}$ are determined.
\item By employing Eq.~(\ref{eq:cont-anal-newt:12}) the new vector of unknowns $\bar{\vec X}_{k + 1}$ is found.
\item The new value of the self-consistent potential $\phi_{k+1}(x)$ is recalculated with the help of $\bar{\vec X}_{k+1}$.
\item Steps 2. - 6. are repeated until either the corrections $\xi_{k}$ and $\xi_{k}^{c}$ become smaller than the given error $\varepsilon$ or $||\bar{\vec X}_{k+1} - \bar{\vec X}_{k}|| < \varepsilon$, for all grid points $x_{i}$.
\end{enumerate}

In addition, we would like to stress that the speed of convergence increases if the state vector $\bar{\vec X}$ is normalized for every iteration.

In actual numerical calculations in order to solve the boundary value problems (\ref{eq:cont-anal-newt:15}) we used a three point template \cite{SamarskiiB1989numerical} for the approximation of derivatives and the corresponding matrix equations were solved by employing the tridiagonal matrix algorithm \cite{PressB2007numerical,SamarskiiB1989numerical}. The spatial grid was logarithmic, i.e. $x_{i} = \exp{(\ln x_{0} + i(\ln x_{f} - \ln x_{0}) / N)}$, $i=\{1,N\}$, while the grid in $t$ was uniform with $\tau_{k} = \tau = 0.7$. The actual number of points $N$ in the spatial grid was $\approx 2\cdot10^{3}$.

In order to evaluate the accuracy of the algorithm we performed a numerical solution of the Dirac and Schr\"{o}dinger equations in the Coulomb field, i.e. the Hydrogen atom and the numerical solution of the polaron problem \cite{Mitra198791}. The accuracy of the calculation of eigenvalues for the Hydrogen atom was greater than 6 decimal digits. Moreover, we were able to reproduce all digits of the well known result for the ground state energy of the polaron problem $E_{\mathrm{p}}=-0.108513$ \cite{MiyakeA1975strong,Mitra198791}, where a similar equation of the self-consistent field arises.

At last we discuss the choice of the initial approximation $\bar{\vec X}_{0}$ for the unknown vector $\bar{\vec X}$. For this purpose, we use the variational estimation for the functional for the energy of the system, which is based on the following wave functions
\begin{align}
  u_{0N}(x) = \mathrm{const}\cdot x e^{-x}, \quad v_{0N}(x) = \mathrm{const}_{1}\cdot x^{2}e^{-x},\label{eq:cont-anal-newt:18}
\end{align}
for $q>0$ and 
\begin{align}
  u_{0N}(x) = \mathrm{const}\cdot x(1 + x^{2}) e^{-x}, \quad v_{0N}(x) = \mathrm{const}_{1}\cdot x^{2}e^{-x}, \label{eq:cont-anal-newt:19}
\end{align}
for $q<0$. The constants $\mathrm{const}$ and $\mathrm{const}_{1}$ are chosen from the normalization condition Eq.~(\ref{eq:cont-anal-newt:15}).

\section{Calculation of the functional (\ref{eq:48})}
\label{sec:calc-expect-kinet}
In this appendix we calculate the expectation value of the functional for the energy of the solution of the second kind Eq.~(\ref{eq:48}). Since the potential part is exactly the same as in the functional for the solution of the first kind Eq.~(\ref{eq:42}) we can write
\begin{align}
  \label{eq:calc-expect-kinet:1}
  &\frac{e_{0}^{2}}{8\pi}\int \frac{d\vec x d\vec y}{|\vec x - \vec y|}\left[\Psi^{\dag}(\vec x)\Psi(\vec x) - \Psi^{c\dag}(\vec x)\Psi^{c}(\vec x)\right]\left[\Psi^{\dag}(\vec y)\Psi(\vec y) - \Psi^{c\dag}(\vec y)\Psi^{c}(\vec y)\right]\nonumber
  \\
  &\mspace{90mu}= \frac{1}{2}\frac{1+C}{1-C} \int d\vec x e_{0}\varphi(\vec x)\frac{g^{2}(r) + f^{2}(r)}{4\pi} \nonumber
  \\
  &\mspace{90mu}= \frac{1}{2} \frac{m_{0}}{\alpha_{0}} q \int_{0}^{\infty}dx \phi(x)[u_{0}^{2}(x) + v_{0}^{2}(x)] = \frac{m_{0}}{\alpha_{0}}\frac{q^{2}}{2}\Pi.
\end{align}
Here we used the definition of the potential part of the total energy of the system Eq.~(\ref{eq:39}) and introduced the dimensionless variables Eq.~(\ref{eq:deriv-radi-equat:15}).

In order to calculate the expectation value of the kinetic part we will employ the properties of the Dirac matrices \cite{LandauQED}, i.e.,
\begin{align}
  \label{eq:calc-expect-kinet:2}
  \alpha_{i}\alpha_{j} + \alpha_{j}\alpha_{i} = 2 \delta_{ij}, \quad \alpha_{i}\beta + \beta \alpha_{i} = 0.
\end{align}

Consequently, the kinetic part of Eq.~(\ref{eq:48}) transforms into
\begin{align}
  \label{eq:calc-expect-kinet:3}
  &\int d\vec x \left[\Psi^{\dag}(\vec x)(\vec \alpha\cdot\vec \nu)(\vec \alpha\cdot\opA{p} +\beta m_0) (\vec \alpha\cdot\vec \nu)\Psi(\vec x) - \Psi^{c\dag}(\vec x)(\vec \alpha\cdot\vec \nu)(\vec \alpha\cdot\opA{p} +\beta m_0)(\vec \alpha\cdot\vec \nu)\Psi^{c}(\vec x)\right] \nonumber
  \\
  &\mspace{90mu}= \int d\vec x \Bigg[\Psi^{\dag}(\vec x)((2 \delta_{ij} - \alpha_{i}\alpha_{j})\opa{p}_{i}n_{j} -\beta m_0 (\vec \alpha\cdot\vec \nu)) (\vec \alpha\cdot\vec \nu)\Psi(\vec x)
  \\
  &\mspace{250mu}- \Psi^{c\dag}(\vec x)((2 \delta_{ij} - \alpha_{i}\alpha_{j}) \opa{p}_{i}n_{j} -\beta m_0 (\vec \alpha\cdot\vec \nu))(\vec \alpha\cdot\vec \nu)\Psi^{c}(\vec x)\Bigg], \nonumber
\end{align}
or after simplification Eq.~(\ref{eq:calc-expect-kinet:3}) reads
\begin{align}
  \label{eq:calc-expect-kinet:4}
  &-\int d\vec x \left[\Psi^{\dag}(\vec x)(\vec \alpha\cdot\opA{p} +\beta m_0) \Psi(\vec x) - \Psi^{c\dag}(\vec x)(\vec \alpha\cdot\opA{p} +\beta m_0)\Psi^{c}(\vec x)\right] \nonumber
  \\
  &\mspace{90mu}+ 2\int d\vec x \left[\Psi^{\dag}(\vec x)(\vec \alpha\cdot\vec \nu)(\opA{p}\cdot\vec \nu) \Psi(\vec x) - \Psi^{c\dag}(\vec x) (\vec \alpha\cdot\vec \nu)(\opA{p}\cdot\vec \nu) \Psi^{c}(\vec x)\right].
\end{align}

The first integral in this equation coincides, up to the minus sign, with the one from the solution of the first kind Eq.~(\ref{eq:38}), i.e. $-m_{0}/\alpha_{0} qT$.

The second integral in Eq.~(\ref{eq:calc-expect-kinet:4}) consists of two parts. Since they are different only with the normalization of the wave functions and the notations for the spherical spinors $\chi_{0}$ and $\chi_{1}$, we perform the calculation only with the first part. The calculation of the second part in the integral is completely analogous to the first one.

In order to calculate the integral in Eq.~(\ref{eq:calc-expect-kinet:4}) we direct the $z$-axis of the coordinate system along the vector $\vec \nu$ and rewrite this expression in the matrix form
\begin{align}
  \label{eq:calc-expect-kinet:5}
  &\frac{1}{1+C}\int d\vec x
  \begin{pmatrix}
    g(r) \chi_{0}^{\dag} && -\ri f(r) \chi_{1}^{\dag}
  \end{pmatrix}
  \begin{pmatrix}
    0 && \sigma_{3}\opa p_{3}
    \\
    \sigma_{3}\opa p_{3} && 0
  \end{pmatrix}
  \begin{pmatrix}
    g(r)\chi_{0}
    \\
    \ri f(r)\chi_{1}
  \end{pmatrix} \nonumber
  \\
  &\mspace{120mu}=\frac{1}{1+C}\int d\vec x [-\ri f(r)\chi_{1}^{\dag} \sigma_{3}\opa p_{3}(g(r)\chi_{0}) + \mathrm{C.C}] \nonumber
  \\
  &\mspace{120mu}=\frac{1}{1+C} \int d\vec x\left[-f(r)\chi_{1}^{\dag}\sigma_{3}\frac{\partial}{\partial z}(g(r)\chi_{0}) + \mathrm{C.C}\right],
\end{align}
where C.C denotes the complex conjugate. The spherical spinor $\chi_{0}$ is independent of the coordinates and consequently, the derivative with respect to $z$ is equal to zero. The derivative $\partial_{z}g(r) = g^{\prime}_{r}(r) z / r = g^{\prime}(r)\sqrt{4\pi/3}Y_{10}$. As a result, Eq.~(\ref{eq:calc-expect-kinet:5}) reads
\begin{align}
  \label{eq:calc-expect-kinet:6}
  &\frac{1}{1+C}\int d\vec x\left[-f(r) \chi_{1}^{\dag} \sigma_{3} g^{\prime}(r) \frac{z}{r} \chi_{0} + \mathrm{C.C}\right] \nonumber
  \\
  &\mspace{120mu} = \frac{1}{1+C}\int d\vec x \Bigg[-f(r)g^{\prime}(r)\sqrt{\frac{1}{3}}Y_{10}\Bigg(-|A_{\frac{1}{2}}|^{2}\sqrt{\frac{1}{3}}Y^{*}_{10} 
  \\
  &\mspace{250mu}- A_{\frac{1}{2}}A_{-\frac{1}{2}}^{*}\sqrt{\frac{2}{3}}Y^{*}_{1-1} - A_{-\frac{1}{2}}A_{\frac{1}{2}}^{*}\sqrt{\frac{2}{3}}Y^{*}_{11} - |A_{-\frac{1}{2}}|^{2}\sqrt{\frac{1}{3}}Y^{*}_{10}\Bigg) + \mathrm{C.C}\Bigg]. \nonumber
\end{align}

By exploiting the orthogonality relation of the spherical harmonics and the condition for the coefficients $|A_{\frac{1}{2}}|^{2} + |A_{-\frac{1}{2}}|^{2} = 1$, one obtains
\begin{align}
  &\frac{1}{1+C}\int_{0}^{\infty} r^{2}dr \frac{2}{3}f(r)g^{\prime}(r) = \frac{1}{1+C} \frac{2}{3} \int_{0}^{\infty}r^{2}dr f(r)\frac{(rg(r))^{\prime} - g(r)}{r} \label{eq:calc-expect-kinet:7} 
  \\
  &\mspace{120mu}=\frac{1}{1+C}m_{0} \frac{1}{3}\int_{0}^{\infty}dx\left[(u_{0}^{\prime}(x)v_{0}(x) - v_{0}^{\prime}(x)u_{0}(x))-\frac{2u_{0}(x)v_{0}(x)}{x}\right]. \nonumber
\end{align}
Here, on the last step we integrated the first term by parts and introduced the dimensionless variables (\ref{eq:deriv-radi-equat:15}).

The calculation of the integral with the functions with the index $c$ in Eq.~(\ref{eq:calc-expect-kinet:4}) is performed in exactly the same fashion. Consequently, combining these two results together we can write
\begin{align}
  \label{eq:calc-expect-kinet:8}
  2\int d\vec x \Big[\Psi^{\dag}(\vec x)\alpha_{3}\opa{p}_{3} \Psi(\vec x) &- \Psi^{c\dag}(\vec x) \alpha_{3}\opa{p}_{3} \Psi^{c}\Big] 
  \\
  &= \frac{m_{0}}{\alpha_{0}}\frac{2}{3}q \int_{0}^{\infty}dx\left[(u_{0}^{\prime}(x)v_{0}(x) - v_{0}^{\prime}(x)u_{0}(x))-\frac{2u_{0}(x)v_{0}(x)}{x}\right]. \nonumber
\end{align}

Finally, the incorporation of all expressions together yields the expectation value of the functional for the solution of the second kind
\begin{align}
  \label{eq:calc-expect-kinet:9}
  \mathfrak{J}^{\prime}[\Psi^{\prime}(\vec  x),\Psi^{\prime c}(\vec x)] &= \frac{m_{0}}{\alpha_{0}}\Bigg(- q T + \frac{q^{2}}{2}\Pi 
  \\
  &\mspace{80mu}+ \frac{2}{3}q \int_{0}^{\infty}dx\left[(u_{0}^{\prime}(x)v_{0}(x) - v_{0}^{\prime}(x)u_{0}(x))-\frac{2u_{0}(x)v_{0}(x)}{x}\right]\Bigg). \nonumber
\end{align}

As the last step we add and subtract $(q^{2}/2)\Pi$ in Eq.~(\ref{eq:calc-expect-kinet:9}). This yields
\begin{align}
  \label{eq:calc-expect-kinet:10}
  \mathfrak{J}^{\prime}&[\Psi^{\prime}(\vec  x),\Psi^{\prime c}(\vec x)] = \frac{m_{0}}{\alpha_{0}}\Bigg(- q T - \frac{q^{2}}{2}\Pi 
  \\
  &\mspace{120mu}+ \frac{2}{3}q \int_{0}^{\infty}dx\left[(u_{0}^{\prime}(x)v_{0}(x) - v_{0}^{\prime}(x)u_{0}(x))-\frac{2u_{0}(x)v_{0}(x)}{x}\right] + q^{2}\Pi\Bigg) \nonumber
  \\
  &= - E_{0} + \frac{m_{0}}{\alpha_{0}}q\left[ \int_{0}^{\infty}dx \left(\frac{2}{3}(u_{0}^{\prime}(x)v_{0}(x) - v_{0}^{\prime}(x)u_{0}(x))-\frac{4}{3}\frac{u_{0}(x)v_{0}(x)}{x}\right) + q\Pi\right]. \nonumber
\end{align}

Here we introduce the energy of the solution of the first kind Eq.~(\ref{eq:42}).

\section{Variation of the functional (\ref{eq:78})}
\label{sec:vari-funct-refeq}
In this appendix we calculate the variation of the functional (\ref{eq:78}). In principle this is a trivial procedure, despite the variation of the self-consistent potential
\begin{align}
  \phi_{0}(x) = \frac{1}{x}\int_{0}^{x}dy (u_{0}^{2} + v_{0}^{2})_{y} + \int_{x}^{\infty}dy\frac{(u_{0}^{2} + v_{0}^{2})_{y}}{y}.\label{eq:vari-funct-refeq:1}
\end{align}

Through this section we will use the notation $()_{y}$, which denotes the dependence of the functions inside the brackets on the variable $y$.

The variational derivative with respect to $u_{0}(x)$ can be written as
\begin{align}
  \label{eq:vari-funct-refeq:2}
  \frac{\delta\mathfrak{I}}{\delta u_{0}}\delta u_{0} &= \delta \left(\int_{0}^{\infty}dx\int_{0}^{x}dy \frac{(u_{0}^{2} + v_{0}^{2})_{x}(u_{0}^{2} + v_{0}^{2})_{y}}{x} + \int_{0}^{\infty}dx\int_{x}^{\infty}dy \frac{(u_{0}^{2} + v_{0}^{2})_{x}(u_{0}^{2} + v_{0}^{2})_{y}}{y}\right) \nonumber
  \\
  &=\int_{0}^{\infty}dx\int_{0}^{x}dy \frac{(2u_{0}\delta u_{0})_{x}(u_{0}^{2} + v_{0}^{2})_{y} + (u_{0}^{2} + v_{0}^{2})_{x}(2u_{0}\delta u_{0})_{y}}{x} \nonumber
  \\
  &+ \int_{0}^{\infty}dx\int_{x}^{\infty}dy \frac{(2u_{0}\delta u_{0})_{x}(u_{0}^{2} + v_{0}^{2})_{y} + (u_{0}^{2} + v_{0}^{2})_{x}(2u_{0}\delta u_{0})_{y}}{y} 
  \\
  &=\int_{0}^{\infty}dx\int_{0}^{x}dy \frac{(2u_{0}\delta u_{0})_{x}(u_{0}^{2} + v_{0}^{2})_{y}}{x} + \int_{0}^{\infty}dx\int_{0}^{x}dy \frac{(u_{0}^{2} + v_{0}^{2})_{x}(2u_{0}\delta u_{0})_{y}}{x} \nonumber
  \\
  &+\int_{0}^{\infty}dx\int_{x}^{\infty}dy \frac{(2u_{0}\delta u_{0})_{x}(u_{0}^{2} + v_{0}^{2})_{y}}{y} + \int_{0}^{\infty}dx\int_{x}^{\infty}dy \frac{(u_{0}^{2} + v_{0}^{2})_{x}(2u_{0}\delta u_{0})_{y}}{y}. \nonumber
\end{align}

The boundary conditions of the radial functions $u_{0}(x)$ and $v_{0}(x)$ are to be satisfied at zero and infinity, respectively. Consequently, in the first and the third integrals the variations are located on the functions, which integration variables have the right limits of integration, while in the second and the last this condition is not satisfied. The integration region of the second integral is the infinitely large triangle located in the first quadrant of the coordinate system $(x,y)$ and lying below the line $x=y$. However, in the fourth integral the integration region is a similar triangle, which is located above the line $x = y$.

Let us change the order of integration in the second and the fourth integrals
\begin{align}
  \int_{0}^{\infty}dx\int_{0}^{x}dy \frac{(u_{0}^{2} + v_{0}^{2})_{x}(2u_{0}\delta u_{0})_{y}}{x} = \int_{0}^{\infty}dy\int_{y}^{\infty}dy \frac{(u_{0}^{2} + v_{0}^{2})_{x}(2u_{0}\delta u_{0})_{y}}{x}, \label{eq:vari-funct-refeq:3}
  \\
  \int_{0}^{\infty}dx\int_{x}^{\infty}dy \frac{(u_{0}^{2} + v_{0}^{2})_{x}(2u_{0}\delta u_{0})_{y}}{y} = \int_{0}^{\infty}dy\int_{0}^{y}dx \frac{(u_{0}^{2} + v_{0}^{2})_{x}(2u_{0}\delta u_{0})_{y}}{y}. \label{eq:vari-funct-refeq:4}
\end{align}

By relabeling $x\leftrightarrow y$ one can observe that the second integral is equal to the third one, while the first integral is equal to the last one. Consequently, we find
\begin{align}
  \label{eq:vari-funct-refeq:5}
  \frac{\delta\mathfrak{I}}{\delta u_{0}}\delta u_{0} = \int_{0}^{\infty}dx 4u_{0}\phi_{0} \delta u_{0}.
\end{align}

As a result we are ready to calculate the full variation of the functional, which yields
\begin{align}
  \label{eq:vari-funct-refeq:6}
  \delta\mathfrak{I} &= \int_{0}^{\infty} dx \Bigg(-v_{0}^{\prime} \delta u_{0} + u_{0}^{\prime} \delta v_{0} - v_{0}^{\prime}\delta u_{0} + u_{0}^{\prime} \delta v_{0} - \frac{2v_{0}}{x}\delta u_{0} - \frac{2u_{0}}{x}\delta v_{0} \nonumber
  \\
  &\mspace{90mu}+ 2u_{0}\delta u_{0} - 2v_{0} \delta v_{0} + 2q \phi_{0}u_{0}\delta u_{0} + 2q \phi_{0}v_{0}\delta v_{0}\Bigg)_{x} - \lambda\int_{0}^{\infty}dx(2u_{0}\delta u_{0} + 2v_{0}\delta v_{0})_{x} \nonumber
  \\
  &\mspace{90mu}-\mu \int_{0}^{\infty}dx\Bigg[\frac{2}{3}\left(-v_{0}^{\prime} \delta u_{0} + u_{0}^{\prime} \delta v_{0} - v_{0}^{\prime} \delta u_{0} + u_{0}^{\prime} \delta v_{0}\right) \nonumber
  \\
  &\mspace{200mu}-\frac{4}{3}\frac{u_{0}}{x}\delta v_{0} - \frac{4}{3}\frac{v_{0}}{x} \delta u_{0} + 4q\phi_{0}u_{0}\delta u_{0} + 4q\phi_{0}v_{0}\delta v_{0}\Bigg]_{x} \nonumber
  \\
  &=\int_{0}^{\infty} dx \delta u_{0}\left[- 2 v_{0}^{\prime} - \frac{2v_{0}}{x} + 2u_{0} + 2u_{0}\phi_{0}q - 2 \lambda u_{0} - \mu \left(-2 \frac{2}{3}v_{0}^{\prime} - 2\frac{2}{3}\frac{v_{0}}{x} + 4q u_{0}\phi_{0}\right)\right]_{x}\nonumber
  \\
  &\mspace{120mu}+\int_{0}^{\infty} dx \delta v_{0}\Bigg[2 u_{0}^{\prime} - \frac{2u_{0}}{x} - 2v_{0} + 2v_{0}\phi_{0}q \nonumber
  \\
  &\mspace{240mu}- 2 \lambda v_{0} - \mu \left(2 \frac{2}{3}u_{0}^{\prime} - 2\frac{2}{3}\frac{u_{0}}{x} + 4q v_{0}\phi_{0}\right)\Bigg]_{x} = 0
\end{align}
and therefore Eqs.~(\ref{eq:52}).

\section{Change of variables in the self-consistent potential, $X$ and $E_{0}$}
\label{sec:change-vari-self}
In this appendix we would like to demonstrate that the change of variables defined by Eq.~(\ref{eq:53}), (\ref{eq:54}) leads to the transformation (\ref{eq:55}) of the self-consistent potential. Indeed
\begin{align}
  \phi_{0}(x) &= \frac{1}{x}\int_{0}^{x}dy (u_{0}^{2} + v_{0}^{2})_{y} + \int_{x}^{\infty}dy\frac{(u_{0}^{2} + v_{0}^{2})_{y}}{y} \nonumber
  \\
              &=a^{2}\Bigg\{\frac{1}{x}\int_{0}^{x}dy\left[\bar{u}_{0}^{2}\left(\frac{y}{1 - \frac{2}{3}\mu}\right) + \bar{v}_{0}^{2}\left(\frac{y}{1 - \frac{2}{3}\mu}\right)\right] \nonumber
  \\
              &\mspace{90mu}+ \int_{x}^{\infty}dy \frac{\left(\bar{u}_{0}^{2}\left(\frac{y}{1 - \frac{2}{3}\mu}\right) + \bar{v}_{0}^{2}\left(\frac{y}{1 - \frac{2}{3}\mu}\right)\right)}{y}\Bigg\} =
    \left[\begin{aligned}
      y = y^{\prime}\left(1 - \frac{2}{3}\mu\right) \nonumber
      \\
      y = x, \quad y^{\prime} = \frac{x}{\left(1 - \frac{2}{3}\mu\right)}
    \end{aligned}\right]\nonumber
  \\
              &=a^{2}\Bigg\{\frac{\left(1 - \frac{2}{3}\mu\right)}{x}\int_{0}^{\frac{x}{\left(1 - \frac{2}{3}\mu\right)}}dy^{\prime}(\bar{u}_{0}^{2}(y^{\prime})+\bar{v}_{0}^{2}(y^{\prime})) \nonumber
  \\
              &\mspace{90mu}+ \int_{\frac{x}{\left(1 - \frac{2}{3}\mu\right)}}^{\infty}\frac{dy^{\prime}}{y^{\prime}}(\bar{u}_{0}^{2}(y^{\prime})+\bar{v}_{0}^{2}(y^{\prime}))\Bigg\} \nonumber
  \\
              &= \frac{1}{\left(1 - \frac{2}{3}\mu\right)}\bar{\phi}_{0}\left(\frac{x}{\left(1 - \frac{2}{3}\mu\right)}\right).\label{eq:change-vari-self:1}
\end{align}

The same procedure for $X$ yields
\begin{align}
  X &= \int_{0}^{\infty} dx \left[\frac{2}{3}(u_{0}^{\prime} v_{0} - v^{\prime}_{0} u_{0})_{x} - \frac{4}{3}\frac{(u_{0}v_{0})_{x}}{x} + q \phi_{0}(x)(u_{0}^{2} + v_{0}^{2})_{x}\right] \nonumber
  \\
  &=a^{2}\int_{0}^{\infty}dx\left[\frac{2}{3}(u_{0}^{\prime} v_{0} - v^{\prime}_{0} u_{0})_{\frac{x}{\left(1 - \frac{2}{3}\mu\right)}} - \frac{4}{3}\frac{(u_{0}v_{0})_{\frac{x}{\left(1 - \frac{2}{3}\mu\right)}}}{x} + q \phi_{0}(x)(u_{0}^{2} + v_{0}^{2})_{\frac{x}{\left(1 - \frac{2}{3}\mu\right)}}\right] \nonumber
  \\
  &=a^{2}\int_{0}^{\infty}dz\left[\frac{2}{3}(u_{0}^{\prime} v_{0} - v^{\prime}_{0} u_{0})_{z} - \frac{4}{3}\frac{(u_{0}v_{0})_{z}}{z} + q \phi_{0}(z)(u_{0}^{2} + v_{0}^{2})_{z}\right] = 0. \label{eq:change-vari-self:2}
\end{align}

For the energy one obtains
\begin{align}
  E_{0} &=\frac{m_{0}}{\alpha_{0}}q \int_{0}^{\infty} dx \left[(u_{0}^{\prime} v_{0} - v^{\prime}_{0} u_{0})_{x} - \frac{2(u_{0}v_{0})_{x}}{x} + (u_{0}^{2} - v_{0}^{2})_{x} + \frac{q}{2} \phi_{0}(x)(u_{0}^{2} + v_{0}^{2})_{x}\right] \nonumber
  \\
        &=\frac{m_{0}}{\alpha_{0}}a^{2}\bar{q}\frac{1-\frac{2}{3}\mu}{1 - 2 \mu}\int_{0}^{\infty}dx\Bigg[(u_{0}^{\prime} v_{0} - v^{\prime}_{0} u_{0})_{\frac{x}{\left(1 - \frac{2}{3}\mu\right)}} - \frac{2(u_{0}v_{0})_{\frac{x}{\left(1 - \frac{2}{3}\mu\right)}}}{x} + (u_{0}^{2} - v_{0}^{2})_{\frac{x}{1 - \frac{2}{3}\mu}} \nonumber
  \\
  &\mspace{240mu}+ \frac{\bar{q}}{2}\frac{1-\frac{2}{3}\mu}{1 - 2 \mu} \phi_{0}(x)(u_{0}^{2} + v_{0}^{2})_{\frac{x}{\left(1 - \frac{2}{3}\mu\right)}}\Bigg] \nonumber
  \\
  &=\frac{m_{0}}{\alpha_{0}}\frac{\bar{q}}{1 - 2 \mu}\int_{0}^{\infty}dz\Bigg[(u_{0}^{\prime} v_{0} - v^{\prime}_{0} u_{0})_{z} - \frac{2(u_{0}v_{0})_{z}}{z} \nonumber
  \\
  &\mspace{180mu}+ \left(1 - \frac{2}{3}\mu\right)(u_{0}^{2} - v_{0}^{2})_{z} + \frac{\bar{q}}{2}\frac{1-\frac{2}{3}\mu}{1 - 2 \mu} \phi_{0}(z)(u_{0}^{2} + v_{0}^{2})_{z}\Bigg].\label{eq:change-vari-self:3}
\end{align}

\section{Calculation of the Jacobian determinant}
\label{sec:calc-jacob-determ}

In this Appendix we will demonstrate that the absolute value of the Jacobian determinant of the variable transformations (\ref{eq:70}), (\ref{eq:71}) is equal to $N^{3}$. We start from showing that the determinant of the transformation of the $x$-component is equal to $N$. Indeed, according to the definition we can write
\begin{align}
  \det J_{x} = \begin{vmatrix}
    \frac{\partial r_{1x}}{\partial R_{1x}} & \frac{\partial r_{1x}}{\partial x_{1x}} & \cdots & \frac{\partial r_{1x}}{\partial x_{1x}}
    \\
    \vdots & \vdots & \cdots & \vdots 
    \\
    \frac{\partial r_{Nx}}{\partial R_{1x}} & \frac{\partial r_{Nx}}{\partial x_{1x}} & \cdots & \frac{\partial r_{Nx}}{\partial x_{1x}}
  \end{vmatrix}, \label{eq:calc-jacob-determ:1}
\end{align}
or by expressing $x_{Nx} = - \sum_{a=1}^{N-1}x_{ax}$ and calculating the derivatives
\begin{align}
  \det J_{x} = \begin{vmatrix}
    1 & 1 & 0 & \cdots & 0
    \\
    1 & 0 & 1 & \cdots & 0
    \\
    \vdots & \vdots & \vdots & \ddots & \vdots
    \\
    1 & 0 & 0 & \cdots & 1
    \\
    1 & -1 & -1 & \cdots & -1
  \end{vmatrix} = 1\underbrace{\begin{vmatrix}
      0 & 1 & 0 & \cdots & 0
      \\
      0 & 0 & 1 & \cdots & 0
      \\
      \vdots & \vdots & \vdots & \ddots & \vdots
      \\
      0 & 0 & 0 & \cdots & 1
      \\
      -1 & -1 & -1 & \cdots & -1
    \end{vmatrix}}_{N-1} + (-1)\underbrace{\begin{vmatrix}
      1 & 1 & 0 & \cdots & 0
      \\
      1 & 0 & 1 & \cdots & 0
      \\
      \vdots & \vdots & \vdots & \ddots & \vdots
      \\
      1 & 0 & 0 & \cdots & 1
      \\
      1 & -1 & -1 & \cdots & -1
    \end{vmatrix}}_{N-1}, \label{eq:calc-jacob-determ:2}
\end{align}
where we have expanded the determinant over the first row. Continuing, it is evident that
\begin{align*}
  |1| = 1, \quad \begin{vmatrix}
    1 & 1
    \\
    1& -1
  \end{vmatrix} = -2.
\end{align*}
Consequently, by using the mathematical induction and expanding Eq.~(\ref{eq:calc-jacob-determ:2}) one obtains
\begin{align*}
	\det J_{x} = (-1)(-1)^{(N-1)+1} + (-1)(-1)^{(N-1)-1}(N-1) = (-1)^{N-1}N,
\end{align*}
that is to be proven.

The overall transformation of variables is expressed through a block diagonal matrix
\begin{equation}
  \label{eq:calc-jacob-determ:3}
  J = \begin{pmatrix}
    J_{x} && 0 && 0
    \\
    0 && J_{y} && 0
    \\
    0 && 0 && J_{z}
  \end{pmatrix},
\end{equation}
and its determinant is equal to the product of the determinants for every coordinate. Consequently, the absolute value of $J$ is equal to $N^{3}$.

\section{Evaluation of the matrix elements in Eq.~(\ref{eq:91})}
\label{sec:eval-matr-elem}
In this appendix we evaluate the remaining two matrix elements in Eq.~(\ref{eq:91}), namely $\langle\psi_{0}^{\prime}|\opa H_{\mathrm{QED}}^{\prime}(\vec P)|\psi_{0}\rangle$ and $\langle\psi_{0}|\opa H_{\mathrm{QED}}(\vec P)|\psi_{0}^{\prime}\rangle$. This requires some care as the expectation value of the quadratic operator needs to be evaluated.

We start the calculation from the term, which is quadratic in density. The basis of the linear combination Eq.~(\ref{eq:83}) consists only of two terms viz. $|\psi_{0}\rangle$ and $|\psi_{0}^{\prime}\rangle$. Consequently, we insert the projection operator between the densities, i.e.
\begin{align}
  \label{eq:eval-matr-elem:1}
  &\langle\psi_{0}|\frac{e_0^2}{8\pi}\int \frac{d\vec x d\vec y}{|\vec x - \vec y|}:\uprho(\vec x,\vec R,\vec P)::\uprho(\vec y,\vec R,\vec P):|\psi_{0}^{\prime}\rangle \nonumber
  \\
  &\mspace{90mu}= \langle\psi_{0}|\frac{e_0^2}{8\pi}\int \frac{d\vec x d\vec y}{|\vec x - \vec y|}:\uprho(\vec x,\vec R,\vec P):(|\psi_{0}\rangle \langle \psi_{0}| + |\psi_{0}^{\prime}\rangle \langle \psi_{0}^{\prime}|):\uprho(\vec y,\vec R,\vec P):|\psi_{0}^{\prime}\rangle \nonumber
  \\
  &\mspace{90mu}= \frac{e_0^2}{8\pi}\int \frac{d\vec x d\vec y}{|\vec x - \vec y|} \Bigg[\langle\psi_{0}|:\uprho(\vec x,\vec R,\vec P):|\psi_{0}\rangle  \langle\psi_{0}|:\uprho(\vec y,\vec R,\vec P):|\psi_{0}^{\prime}\rangle \nonumber
  \\
  &\mspace{250mu}+ \langle\psi_{0}|:\uprho(\vec x,\vec R,\vec P):|\psi_{0}^{\prime}\rangle  \langle\psi_{0}^{\prime}|:\uprho(\vec y,\vec R,\vec P):|\psi_{0}^{\prime}\rangle\Bigg],
\end{align}
and if one introduces the self-consistent potential and calculates the expectation value
\begin{align}
  \label{eq:eval-matr-elem:2}
  e_0\int d\vec x \varphi(r)\left[\Psi^{\dag}(\vec x)\alpha_{3}\Psi(\vec x) - \Psi^{c\dag}(\vec x)\alpha_{3}\Psi^{c}(\vec x)\right].
\end{align}

The self-consistent potential does not depend on the angular variables and, consequently, this integral vanishes due to the orthogonality of the spherical harmonics $Y_{00}$ and $Y_{10}$, $Y_{1-1}$, $Y_{11}$.

The expectation value of the kinetic part, i.e.
\begin{align}
  \label{eq:eval-matr-elem:3}
  &\langle\psi_{0}|\int d\vec x:\uppsi^\dag(\vec x,\vec R,\vec P)(\vec\alpha\cdot \opA p +\beta m_0)\uppsi(\vec x,\vec R,\vec P):|\psi_{0}^{\prime}\rangle \nonumber
  \\
  &\mspace{90mu}= \int d\vec x \Psi^{\dag}(\vec\alpha\cdot \opA p +\beta m_0)\alpha_{3}\Psi(\vec x) - \int d\vec x \Psi^{c\dag}(\vec\alpha\cdot \opA p +\beta m_0)\alpha_{3}\Psi^{c}(\vec x)
\end{align}
is equal to zero. This can be seen if one employs the equations of motion (\ref{eq:25})--(\ref{eq:26}) in Eq.~(\ref{eq:eval-matr-elem:3}). This will yield a similar integral to Eq.~(\ref{eq:eval-matr-elem:2}), which was shown to vanish.

As a result we are left only with the expectation value of the part containing the total momentum viz.
\begin{align}
  \label{eq:eval-matr-elem:4}
  &\langle\psi_{0}|\int d\vec x :\uppsi^\dag(\vec x,\vec R,\vec P)(\vec\alpha\cdot \opA P)\uppsi(\vec x,\vec R,\vec P):|\psi_{0}^{\prime}\rangle \nonumber
  \\
  &\mspace{120mu}= \int d\vec x \Bigg(\Psi^{\dag}(\vec x) (\alpha_{3}P) \alpha_{3}\Psi(\vec x) - \Psi^{c\dag}(\vec x)(-\alpha_{3}P) \alpha_{3}\Psi^{c}(\vec x)\Bigg) = P.
\end{align}

Concluding, we have demonstrated that the matrix elements
\begin{align}
  \label{eq:eval-matr-elem:5}
  \langle\psi_{0}^{\prime}|\opa H_{\mathrm{QED}}^{\prime}(\vec P)|\psi_{0}\rangle = \langle\psi_{0}|\opa H_{\mathrm{QED}}^{\prime}(\vec P)|\psi_{0}^{\prime}\rangle = P.
\end{align}

\bibliography{soliton_like_sol}

\begin{thebibliography}{84}%
\makeatletter
\providecommand \@ifxundefined [1]{%
 \@ifx{#1\undefined}
}%
\providecommand \@ifnum [1]{%
 \ifnum #1\expandafter \@firstoftwo
 \else \expandafter \@secondoftwo
 \fi
}%
\providecommand \@ifx [1]{%
 \ifx #1\expandafter \@firstoftwo
 \else \expandafter \@secondoftwo
 \fi
}%
\providecommand \natexlab [1]{#1}%
\providecommand \enquote  [1]{``#1''}%
\providecommand \bibnamefont  [1]{#1}%
\providecommand \bibfnamefont [1]{#1}%
\providecommand \citenamefont [1]{#1}%
\providecommand \href@noop [0]{\@secondoftwo}%
\providecommand \href [0]{\begingroup \@sanitize@url \@href}%
\providecommand \@href[1]{\@@startlink{#1}\@@href}%
\providecommand \@@href[1]{\endgroup#1\@@endlink}%
\providecommand \@sanitize@url [0]{\catcode `\\12\catcode `\$12\catcode
  `\&12\catcode `\#12\catcode `\^12\catcode `\_12\catcode `\%12\relax}%
\providecommand \@@startlink[1]{}%
\providecommand \@@endlink[0]{}%
\providecommand \url  [0]{\begingroup\@sanitize@url \@url }%
\providecommand \@url [1]{\endgroup\@href {#1}{\urlprefix }}%
\providecommand \urlprefix  [0]{URL }%
\providecommand \Eprint [0]{\href }%
\providecommand \doibase [0]{http://dx.doi.org/}%
\providecommand \selectlanguage [0]{\@gobble}%
\providecommand \bibinfo  [0]{\@secondoftwo}%
\providecommand \bibfield  [0]{\@secondoftwo}%
\providecommand \translation [1]{[#1]}%
\providecommand \BibitemOpen [0]{}%
\providecommand \bibitemStop [0]{}%
\providecommand \bibitemNoStop [0]{.\EOS\space}%
\providecommand \EOS [0]{\spacefactor3000\relax}%
\providecommand \BibitemShut  [1]{\csname bibitem#1\endcsname}%
\let\auto@bib@innerbib\@empty
\bibitem [{\citenamefont {Rajaraman}(1982)}]{RajaramanB1982solitons}%
  \BibitemOpen
  \bibfield  {author} {\bibinfo {author} {\bibfnamefont {R.}~\bibnamefont
  {Rajaraman}},\ }\href {https://books.google.de/books?id=1XucQgAACAAJ} {\emph
  {\bibinfo {title} {Solitons and Instantons: An Introduction to Solitons and
  Instantons in Quantum Field Theory}}},\ North-Holland personal library\
  (\bibinfo  {publisher} {North-Holland Publishing Company},\ \bibinfo {year}
  {1982})\BibitemShut {NoStop}%
\bibitem [{\citenamefont {The Lord~Rayleigh}(1876)}]{RayleighA1876waves}%
  \BibitemOpen
  \bibfield  {author} {\bibinfo {author} {\bibfnamefont {J.~M.}\ \bibnamefont
  {The Lord~Rayleigh}},\ }\href {\doibase 10.1080/14786447608639037} {\bibfield
   {journal} {\bibinfo  {journal} {Philosophical Magazine Series 5}\ }\textbf
  {\bibinfo {volume} {1}},\ \bibinfo {pages} {257} (\bibinfo {year}
  {1876})}\BibitemShut {NoStop}%
\bibitem [{\citenamefont {Korteweg}\ and\ \citenamefont
  {de~Vries}(1895)}]{KortewegA1895change}%
  \BibitemOpen
  \bibfield  {author} {\bibinfo {author} {\bibfnamefont {D.~J.}\ \bibnamefont
  {Korteweg}}\ and\ \bibinfo {author} {\bibfnamefont {G.}~\bibnamefont
  {de~Vries}},\ }\href {\doibase 10.1080/14786449508620739} {\bibfield
  {journal} {\bibinfo  {journal} {Philosophical Magazine Series 5}\ }\textbf
  {\bibinfo {volume} {39}},\ \bibinfo {pages} {422} (\bibinfo {year} {1895})},\
  \Eprint {http://arxiv.org/abs/http://dx.doi.org/10.1080/14786449508620739}
  {http://dx.doi.org/10.1080/14786449508620739} \BibitemShut {NoStop}%
\bibitem [{\citenamefont {Bardeen}\ \emph {et~al.}(1957)\citenamefont
  {Bardeen}, \citenamefont {Cooper},\ and\ \citenamefont
  {Schrieffer}}]{PhysRev.108.1175}%
  \BibitemOpen
  \bibfield  {author} {\bibinfo {author} {\bibfnamefont {J.}~\bibnamefont
  {Bardeen}}, \bibinfo {author} {\bibfnamefont {L.~N.}\ \bibnamefont {Cooper}},
  \ and\ \bibinfo {author} {\bibfnamefont {J.~R.}\ \bibnamefont {Schrieffer}},\
  }\href {\doibase 10.1103/PhysRev.108.1175} {\bibfield  {journal} {\bibinfo
  {journal} {Phys. Rev.}\ }\textbf {\bibinfo {volume} {108}},\ \bibinfo {pages}
  {1175} (\bibinfo {year} {1957})}\BibitemShut {NoStop}%
\bibitem [{\citenamefont {Blatt}(1971)}]{BlattB1971theory}%
  \BibitemOpen
  \bibfield  {author} {\bibinfo {author} {\bibfnamefont {J.~M.}\ \bibnamefont
  {Blatt}},\ }\href@noop {} {\emph {\bibinfo {title} {Theory of
  superconductivity}}}\ (\bibinfo  {publisher} {Academic Press New York},\
  \bibinfo {year} {1971})\BibitemShut {NoStop}%
\bibitem [{\citenamefont {Abrikosov}\ \emph {et~al.}(1975)\citenamefont
  {Abrikosov}, \citenamefont {Gorkov},\ and\ \citenamefont
  {Dzyaloshinski}}]{AbrikosovB1975methods}%
  \BibitemOpen
  \bibfield  {author} {\bibinfo {author} {\bibfnamefont {A.}~\bibnamefont
  {Abrikosov}}, \bibinfo {author} {\bibfnamefont {L.}~\bibnamefont {Gorkov}}, \
  and\ \bibinfo {author} {\bibfnamefont {I.}~\bibnamefont {Dzyaloshinski}},\
  }\href {https://books.google.de/books?id=E\_9NtwNY7UcC} {\emph {\bibinfo
  {title} {Methods of Quantum Field Theory in Statistical Physics}}},\ Dover
  Books on Physics Series\ (\bibinfo  {publisher} {Dover Publications},\
  \bibinfo {year} {1975})\BibitemShut {NoStop}%
\bibitem [{\citenamefont {Weinberg}(2012)}]{WeinbergB2012classical}%
  \BibitemOpen
  \bibfield  {author} {\bibinfo {author} {\bibfnamefont {E.}~\bibnamefont
  {Weinberg}},\ }\href {https://books.google.de/books?id=rfvFNI2C7i8C} {\emph
  {\bibinfo {title} {Classical Solutions in Quantum Field Theory: Solitons and
  Instantons in High Energy Physics}}},\ Cambridge Monographs on Mathematical
  Physics\ (\bibinfo  {publisher} {Cambridge University Press},\ \bibinfo
  {year} {2012})\BibitemShut {NoStop}%
\bibitem [{\citenamefont {Dirac}(1931)}]{DiracA1931quantized}%
  \BibitemOpen
  \bibfield  {author} {\bibinfo {author} {\bibfnamefont {P.~A.~M.}\
  \bibnamefont {Dirac}},\ }\href {\doibase 10.1098/rspa.1931.0130} {\bibfield
  {journal} {\bibinfo  {journal} {Proceedings of the Royal Society of London A:
  Mathematical, Physical and Engineering Sciences}\ }\textbf {\bibinfo {volume}
  {133}},\ \bibinfo {pages} {60} (\bibinfo {year} {1931})},\ \Eprint
  {http://arxiv.org/abs/http://rspa.royalsocietypublishing.org/content/133/821/60.full.pdf}
  {http://rspa.royalsocietypublishing.org/content/133/821/60.full.pdf}
  \BibitemShut {NoStop}%
\bibitem [{\citenamefont {Polyakov}(1974)}]{PolyakovA1974particle}%
  \BibitemOpen
  \bibfield  {author} {\bibinfo {author} {\bibfnamefont {A.~M.}\ \bibnamefont
  {Polyakov}},\ }\href@noop {} {\bibfield  {journal} {\bibinfo  {journal} {JETP
  Lett.}\ }\textbf {\bibinfo {volume} {20}},\ \bibinfo {pages} {194} (\bibinfo
  {year} {1974})}\BibitemShut {NoStop}%
\bibitem [{\citenamefont {Hooft}(1974)}]{HOOFT1974276}%
  \BibitemOpen
  \bibfield  {author} {\bibinfo {author} {\bibfnamefont {G.}~\bibnamefont
  {Hooft}},\ }\href {\doibase 10.1016/0550-3213(74)90486-6} {\bibfield
  {journal} {\bibinfo  {journal} {Nuclear Physics B}\ }\textbf {\bibinfo
  {volume} {79}},\ \bibinfo {pages} {276 } (\bibinfo {year}
  {1974})}\BibitemShut {NoStop}%
\bibitem [{\citenamefont {Bolognesi}\ and\ \citenamefont
  {Konishi}(2002)}]{Bolognesi2002337}%
  \BibitemOpen
  \bibfield  {author} {\bibinfo {author} {\bibfnamefont {S.}~\bibnamefont
  {Bolognesi}}\ and\ \bibinfo {author} {\bibfnamefont {K.}~\bibnamefont
  {Konishi}},\ }\href {\doibase 10.1016/S0550-3213(02)00796-4} {\bibfield
  {journal} {\bibinfo  {journal} {Nuclear Physics B}\ }\textbf {\bibinfo
  {volume} {645}},\ \bibinfo {pages} {337 } (\bibinfo {year}
  {2002})}\BibitemShut {NoStop}%
\bibitem [{\citenamefont {Hu}\ \emph {et~al.}(2016)\citenamefont {Hu},
  \citenamefont {Van~de Graaff}, \citenamefont {Kedar}, \citenamefont {Corson},
  \citenamefont {Cornell},\ and\ \citenamefont {Jin}}]{PhysRevLett.117.055301}%
  \BibitemOpen
  \bibfield  {author} {\bibinfo {author} {\bibfnamefont {M.-G.}\ \bibnamefont
  {Hu}}, \bibinfo {author} {\bibfnamefont {M.~J.}\ \bibnamefont {Van~de
  Graaff}}, \bibinfo {author} {\bibfnamefont {D.}~\bibnamefont {Kedar}},
  \bibinfo {author} {\bibfnamefont {J.~P.}\ \bibnamefont {Corson}}, \bibinfo
  {author} {\bibfnamefont {E.~A.}\ \bibnamefont {Cornell}}, \ and\ \bibinfo
  {author} {\bibfnamefont {D.~S.}\ \bibnamefont {Jin}},\ }\href {\doibase
  10.1103/PhysRevLett.117.055301} {\bibfield  {journal} {\bibinfo  {journal}
  {Phys. Rev. Lett.}\ }\textbf {\bibinfo {volume} {117}},\ \bibinfo {pages}
  {055301} (\bibinfo {year} {2016})}\BibitemShut {NoStop}%
\bibitem [{\citenamefont {J\o{}rgensen}\ \emph {et~al.}(2016)\citenamefont
  {J\o{}rgensen}, \citenamefont {Wacker}, \citenamefont {Skalmstang},
  \citenamefont {Parish}, \citenamefont {Levinsen}, \citenamefont
  {Christensen}, \citenamefont {Bruun},\ and\ \citenamefont
  {Arlt}}]{PhysRevLett.117.055302}%
  \BibitemOpen
  \bibfield  {author} {\bibinfo {author} {\bibfnamefont {N.~B.}\ \bibnamefont
  {J\o{}rgensen}}, \bibinfo {author} {\bibfnamefont {L.}~\bibnamefont
  {Wacker}}, \bibinfo {author} {\bibfnamefont {K.~T.}\ \bibnamefont
  {Skalmstang}}, \bibinfo {author} {\bibfnamefont {M.~M.}\ \bibnamefont
  {Parish}}, \bibinfo {author} {\bibfnamefont {J.}~\bibnamefont {Levinsen}},
  \bibinfo {author} {\bibfnamefont {R.~S.}\ \bibnamefont {Christensen}},
  \bibinfo {author} {\bibfnamefont {G.~M.}\ \bibnamefont {Bruun}}, \ and\
  \bibinfo {author} {\bibfnamefont {J.~J.}\ \bibnamefont {Arlt}},\ }\href
  {\doibase 10.1103/PhysRevLett.117.055302} {\bibfield  {journal} {\bibinfo
  {journal} {Phys. Rev. Lett.}\ }\textbf {\bibinfo {volume} {117}},\ \bibinfo
  {pages} {055302} (\bibinfo {year} {2016})}\BibitemShut {NoStop}%
\bibitem [{\citenamefont {Chen}\ \emph {et~al.}(2015)\citenamefont {Chen},
  \citenamefont {Avila}, \citenamefont {Frantzeskakis}, \citenamefont {Levy},\
  and\ \citenamefont {Asensio}}]{Chen:2015rt}%
  \BibitemOpen
  \bibfield  {author} {\bibinfo {author} {\bibfnamefont {C.}~\bibnamefont
  {Chen}}, \bibinfo {author} {\bibfnamefont {J.}~\bibnamefont {Avila}},
  \bibinfo {author} {\bibfnamefont {E.}~\bibnamefont {Frantzeskakis}}, \bibinfo
  {author} {\bibfnamefont {A.}~\bibnamefont {Levy}}, \ and\ \bibinfo {author}
  {\bibfnamefont {M.~C.}\ \bibnamefont {Asensio}},\ }\href {\doibase
  10.1038/ncomms9585} {\bibfield  {journal} {\bibinfo  {journal} {Nat.
  Commun.}\ }\textbf {\bibinfo {volume} {6}} (\bibinfo {year} {2015}),\
  10.1038/ncomms9585}\BibitemShut {NoStop}%
\bibitem [{\citenamefont {Fröhlich}(1954)}]{Froehlich1954}%
  \BibitemOpen
  \bibfield  {author} {\bibinfo {author} {\bibfnamefont {H.}~\bibnamefont
  {Fröhlich}},\ }\href {\doibase 10.1080/00018735400101213} {\bibfield
  {journal} {\bibinfo  {journal} {Advances in Physics}\ }\textbf {\bibinfo
  {volume} {3}},\ \bibinfo {pages} {325} (\bibinfo {year} {1954})},\ \Eprint
  {http://arxiv.org/abs/http://dx.doi.org/10.1080/00018735400101213}
  {http://dx.doi.org/10.1080/00018735400101213} \BibitemShut {NoStop}%
\bibitem [{\citenamefont {Mitra}\ \emph {et~al.}(1987)\citenamefont {Mitra},
  \citenamefont {Chatterjee},\ and\ \citenamefont
  {Mukhopadhyay}}]{Mitra198791}%
  \BibitemOpen
  \bibfield  {author} {\bibinfo {author} {\bibfnamefont {T.}~\bibnamefont
  {Mitra}}, \bibinfo {author} {\bibfnamefont {A.}~\bibnamefont {Chatterjee}}, \
  and\ \bibinfo {author} {\bibfnamefont {S.}~\bibnamefont {Mukhopadhyay}},\
  }\href {\doibase 10.1016/0370-1573(87)90087-1} {\bibfield  {journal}
  {\bibinfo  {journal} {Physics Reports}\ }\textbf {\bibinfo {volume} {153}},\
  \bibinfo {pages} {91 } (\bibinfo {year} {1987})}\BibitemShut {NoStop}%
\bibitem [{\citenamefont {Gerlach}\ and\ \citenamefont
  {L\"{o}wen}(1991)}]{RevModPhys.63.63}%
  \BibitemOpen
  \bibfield  {author} {\bibinfo {author} {\bibfnamefont {B.}~\bibnamefont
  {Gerlach}}\ and\ \bibinfo {author} {\bibfnamefont {H.}~\bibnamefont
  {L\"{o}wen}},\ }\href {\doibase 10.1103/RevModPhys.63.63} {\bibfield
  {journal} {\bibinfo  {journal} {Rev. Mod. Phys.}\ }\textbf {\bibinfo {volume}
  {63}},\ \bibinfo {pages} {63} (\bibinfo {year} {1991})}\BibitemShut {NoStop}%
\bibitem [{\citenamefont {Spohn}(1987)}]{Spohn1987278}%
  \BibitemOpen
  \bibfield  {author} {\bibinfo {author} {\bibfnamefont {H.}~\bibnamefont
  {Spohn}},\ }\href {\doibase 10.1016/0003-4916(87)90211-9} {\bibfield
  {journal} {\bibinfo  {journal} {Annals of Physics}\ }\textbf {\bibinfo
  {volume} {175}},\ \bibinfo {pages} {278 } (\bibinfo {year}
  {1987})}\BibitemShut {NoStop}%
\bibitem [{\citenamefont {Feranchuk}\ \emph {et~al.}(1984)\citenamefont
  {Feranchuk}, \citenamefont {Fisher},\ and\ \citenamefont
  {Komarov}}]{0022-3719-17-24-012}%
  \BibitemOpen
  \bibfield  {author} {\bibinfo {author} {\bibfnamefont {I.~D.}\ \bibnamefont
  {Feranchuk}}, \bibinfo {author} {\bibfnamefont {S.~I.}\ \bibnamefont
  {Fisher}}, \ and\ \bibinfo {author} {\bibfnamefont {L.~I.}\ \bibnamefont
  {Komarov}},\ }\href {http://stacks.iop.org/0022-3719/17/i=24/a=012}
  {\bibfield  {journal} {\bibinfo  {journal} {Journal of Physics C: Solid State
  Physics}\ }\textbf {\bibinfo {volume} {17}},\ \bibinfo {pages} {4309}
  (\bibinfo {year} {1984})}\BibitemShut {NoStop}%
\bibitem [{\citenamefont {Feynman}(1955)}]{PhysRev.97.660}%
  \BibitemOpen
  \bibfield  {author} {\bibinfo {author} {\bibfnamefont {R.~P.}\ \bibnamefont
  {Feynman}},\ }\href {\doibase 10.1103/PhysRev.97.660} {\bibfield  {journal}
  {\bibinfo  {journal} {Phys. Rev.}\ }\textbf {\bibinfo {volume} {97}},\
  \bibinfo {pages} {660} (\bibinfo {year} {1955})}\BibitemShut {NoStop}%
\bibitem [{\citenamefont {Bardeen}\ \emph {et~al.}(1975)\citenamefont
  {Bardeen}, \citenamefont {Chanowitz}, \citenamefont {Drell}, \citenamefont
  {Weinstein},\ and\ \citenamefont {Yan}}]{PhysRevD.11.1094}%
  \BibitemOpen
  \bibfield  {author} {\bibinfo {author} {\bibfnamefont {W.~A.}\ \bibnamefont
  {Bardeen}}, \bibinfo {author} {\bibfnamefont {M.~S.}\ \bibnamefont
  {Chanowitz}}, \bibinfo {author} {\bibfnamefont {S.~D.}\ \bibnamefont
  {Drell}}, \bibinfo {author} {\bibfnamefont {M.}~\bibnamefont {Weinstein}}, \
  and\ \bibinfo {author} {\bibfnamefont {T.~M.}\ \bibnamefont {Yan}},\ }\href
  {\doibase 10.1103/PhysRevD.11.1094} {\bibfield  {journal} {\bibinfo
  {journal} {Phys. Rev. D}\ }\textbf {\bibinfo {volume} {11}},\ \bibinfo
  {pages} {1094} (\bibinfo {year} {1975})}\BibitemShut {NoStop}%
\bibitem [{\citenamefont {Bolognesi}(2005)}]{Bolognesi2005150}%
  \BibitemOpen
  \bibfield  {author} {\bibinfo {author} {\bibfnamefont {S.}~\bibnamefont
  {Bolognesi}},\ }\href {\doibase 10.1016/j.nuclphysb.2005.09.031} {\bibfield
  {journal} {\bibinfo  {journal} {Nuclear Physics B}\ }\textbf {\bibinfo
  {volume} {730}},\ \bibinfo {pages} {150 } (\bibinfo {year}
  {2005})}\BibitemShut {NoStop}%
\bibitem [{\citenamefont {Bolognesi}(2006)}]{Bolognesi200693}%
  \BibitemOpen
  \bibfield  {author} {\bibinfo {author} {\bibfnamefont {S.}~\bibnamefont
  {Bolognesi}},\ }\href {\doibase 10.1016/j.nuclphysb.2006.06.022} {\bibfield
  {journal} {\bibinfo  {journal} {Nuclear Physics B}\ }\textbf {\bibinfo
  {volume} {752}},\ \bibinfo {pages} {93 } (\bibinfo {year}
  {2006})}\BibitemShut {NoStop}%
\bibitem [{\citenamefont {Skoromnik}\ \emph {et~al.}(2015)\citenamefont
  {Skoromnik}, \citenamefont {Feranchuk}, \citenamefont {Lu},\ and\
  \citenamefont {Keitel}}]{PhysRevD.92.125019}%
  \BibitemOpen
  \bibfield  {author} {\bibinfo {author} {\bibfnamefont {O.~D.}\ \bibnamefont
  {Skoromnik}}, \bibinfo {author} {\bibfnamefont {I.~D.}\ \bibnamefont
  {Feranchuk}}, \bibinfo {author} {\bibfnamefont {D.~V.}\ \bibnamefont {Lu}}, \
  and\ \bibinfo {author} {\bibfnamefont {C.~H.}\ \bibnamefont {Keitel}},\
  }\href {\doibase 10.1103/PhysRevD.92.125019} {\bibfield  {journal} {\bibinfo
  {journal} {Phys. Rev. D}\ }\textbf {\bibinfo {volume} {92}},\ \bibinfo
  {pages} {125019} (\bibinfo {year} {2015})}\BibitemShut {NoStop}%
\bibitem [{\citenamefont {Bjorken}\ and\ \citenamefont
  {Drell}(1965)}]{bjorken1965}%
  \BibitemOpen
  \bibfield  {author} {\bibinfo {author} {\bibfnamefont {J.}~\bibnamefont
  {Bjorken}}\ and\ \bibinfo {author} {\bibfnamefont {S.}~\bibnamefont
  {Drell}},\ }\href {https://books.google.de/books?id=ZczvAAAAMAAJ} {\emph
  {\bibinfo {title} {Relativistic quantum fields}}},\ International series in
  pure and applied physics\ (\bibinfo  {publisher} {McGraw-Hill},\ \bibinfo
  {year} {1965})\BibitemShut {NoStop}%
\bibitem [{\citenamefont {Berestetskii}\ \emph {et~al.}(1982)\citenamefont
  {Berestetskii}, \citenamefont {Lifshitz},\ and\ \citenamefont
  {Pitaevski{\u\i}}}]{LandauQED}%
  \BibitemOpen
  \bibfield  {author} {\bibinfo {author} {\bibfnamefont {V.}~\bibnamefont
  {Berestetskii}}, \bibinfo {author} {\bibfnamefont {E.}~\bibnamefont
  {Lifshitz}}, \ and\ \bibinfo {author} {\bibfnamefont {L.}~\bibnamefont
  {Pitaevski{\u\i}}},\ }\href {https://books.google.de/books?id=URL5NKX8vbAC}
  {\emph {\bibinfo {title} {Quantum Electrodynamics}}},\ Course of theoretical
  physics\ (\bibinfo  {publisher} {Butterworth-Heinemann},\ \bibinfo {year}
  {1982})\BibitemShut {NoStop}%
\bibitem [{\citenamefont {Akhiezer}\ and\ \citenamefont
  {Berestetskii}(1965)}]{AkhiezerB1965quantum}%
  \BibitemOpen
  \bibfield  {author} {\bibinfo {author} {\bibfnamefont {A.}~\bibnamefont
  {Akhiezer}}\ and\ \bibinfo {author} {\bibfnamefont {V.}~\bibnamefont
  {Berestetskii}},\ }\href {https://books.google.de/books?id=3KUgjwEACAAJ}
  {\emph {\bibinfo {title} {Quantum electrodynamics}}},\ Interscience
  monographs and texts in physics and astronomy\ (\bibinfo  {publisher}
  {Interscience},\ \bibinfo {year} {1965})\BibitemShut {NoStop}%
\bibitem [{\citenamefont {Peskin}\ and\ \citenamefont
  {Schroeder}(1995)}]{PeskinB1995introduction}%
  \BibitemOpen
  \bibfield  {author} {\bibinfo {author} {\bibfnamefont {M.}~\bibnamefont
  {Peskin}}\ and\ \bibinfo {author} {\bibfnamefont {D.}~\bibnamefont
  {Schroeder}},\ }\href {https://books.google.de/books?id=i35LALN0GosC} {\emph
  {\bibinfo {title} {An Introduction to Quantum Field Theory}}},\ Advanced book
  classics\ (\bibinfo  {publisher} {Addison-Wesley Publishing Company},\
  \bibinfo {year} {1995})\BibitemShut {NoStop}%
\bibitem [{\citenamefont {Collins}(1984)}]{CollinsB1984renormalization}%
  \BibitemOpen
  \bibfield  {author} {\bibinfo {author} {\bibfnamefont {J.~C.}\ \bibnamefont
  {Collins}},\ }\href {https://books.google.de/books?id=60wX3xfdGRAC} {\emph
  {\bibinfo {title} {Renormalization}}}\ (\bibinfo  {publisher} {Cambridge
  University Press},\ \bibinfo {year} {1984})\BibitemShut {NoStop}%
\bibitem [{\citenamefont {Salmhofer}(1999)}]{SalmhoferB1999renormalization}%
  \BibitemOpen
  \bibfield  {author} {\bibinfo {author} {\bibfnamefont {M.}~\bibnamefont
  {Salmhofer}},\ }\href {\doibase 10.1007/978-3-662-03873-4} {\emph {\bibinfo
  {title} {Renormalization}}}\ (\bibinfo  {publisher} {Springer Science $+$
  Business Media},\ \bibinfo {year} {1999})\BibitemShut {NoStop}%
\bibitem [{\citenamefont {Schwartz}(2013)}]{SchwartzB2014quantum}%
  \BibitemOpen
  \bibfield  {author} {\bibinfo {author} {\bibfnamefont {M.}~\bibnamefont
  {Schwartz}},\ }\href {https://books.google.de/books?id=HbdEAgAAQBAJ} {\emph
  {\bibinfo {title} {Quantum Field Theory and the Standard Model}}}\ (\bibinfo
  {publisher} {Cambridge University Press},\ \bibinfo {year}
  {2013})\BibitemShut {NoStop}%
\bibitem [{\citenamefont {Feranchuk}\ and\ \citenamefont
  {Feranchuk}(2007)}]{FeranchukA2007sigma}%
  \BibitemOpen
  \bibfield  {author} {\bibinfo {author} {\bibfnamefont {I.~D.}\ \bibnamefont
  {Feranchuk}}\ and\ \bibinfo {author} {\bibfnamefont {S.~I.}\ \bibnamefont
  {Feranchuk}},\ }\href {\doibase 10.3842/SIGMA.2007.117} {\bibfield  {journal}
  {\bibinfo  {journal} {SIGMA}\ }\textbf {\bibinfo {volume} {3}},\ \bibinfo
  {pages} {117} (\bibinfo {year} {2007})}\BibitemShut {NoStop}%
\bibitem [{\citenamefont {Hartree}(1928{\natexlab{a}})}]{HartreeA1928wave-1}%
  \BibitemOpen
  \bibfield  {author} {\bibinfo {author} {\bibfnamefont {D.~R.}\ \bibnamefont
  {Hartree}},\ }\href {\doibase 10.1017/S0305004100011920} {\bibfield
  {journal} {\bibinfo  {journal} {Mathematical Proceedings of the Cambridge
  Philosophical Society}\ }\textbf {\bibinfo {volume} {24}},\ \bibinfo {pages}
  {111} (\bibinfo {year} {1928}{\natexlab{a}})}\BibitemShut {NoStop}%
\bibitem [{\citenamefont {Hartree}(1928{\natexlab{b}})}]{HartreeA1928wave-2}%
  \BibitemOpen
  \bibfield  {author} {\bibinfo {author} {\bibfnamefont {D.~R.}\ \bibnamefont
  {Hartree}},\ }\href {\doibase 10.1017/S0305004100011919} {\bibfield
  {journal} {\bibinfo  {journal} {Mathematical Proceedings of the Cambridge
  Philosophical Society}\ }\textbf {\bibinfo {volume} {24}},\ \bibinfo {pages}
  {89} (\bibinfo {year} {1928}{\natexlab{b}})}\BibitemShut {NoStop}%
\bibitem [{\citenamefont {Fock}(1930)}]{FockA1930naeherungs}%
  \BibitemOpen
  \bibfield  {author} {\bibinfo {author} {\bibfnamefont {V.}~\bibnamefont
  {Fock}},\ }\href {\doibase 10.1007/BF01340294} {\bibfield  {journal}
  {\bibinfo  {journal} {Zeitschrift f{\"u}r Physik}\ }\textbf {\bibinfo
  {volume} {61}},\ \bibinfo {pages} {126} (\bibinfo {year} {1930})}\BibitemShut
  {NoStop}%
\bibitem [{\citenamefont {Acikgoz}\ \emph {et~al.}(1995)\citenamefont
  {Acikgoz}, \citenamefont {Barut}, \citenamefont {Kraus},\ and\ \citenamefont
  {Ünal}}]{ACIKGOZ1995126}%
  \BibitemOpen
  \bibfield  {author} {\bibinfo {author} {\bibfnamefont {I.}~\bibnamefont
  {Acikgoz}}, \bibinfo {author} {\bibfnamefont {A.}~\bibnamefont {Barut}},
  \bibinfo {author} {\bibfnamefont {J.}~\bibnamefont {Kraus}}, \ and\ \bibinfo
  {author} {\bibfnamefont {N.}~\bibnamefont {Ünal}},\ }\href {\doibase
  10.1016/0375-9601(94)00979-Y} {\bibfield  {journal} {\bibinfo  {journal}
  {Physics Letters A}\ }\textbf {\bibinfo {volume} {198}},\ \bibinfo {pages}
  {126 } (\bibinfo {year} {1995})}\BibitemShut {NoStop}%
\bibitem [{\citenamefont {Barut}\ \emph {et~al.}(1992)\citenamefont {Barut},
  \citenamefont {Kraus}, \citenamefont {Salamin},\ and\ \citenamefont
  {\"Unal}}]{PhysRevA.45.7740}%
  \BibitemOpen
  \bibfield  {author} {\bibinfo {author} {\bibfnamefont {A.~O.}\ \bibnamefont
  {Barut}}, \bibinfo {author} {\bibfnamefont {J.}~\bibnamefont {Kraus}},
  \bibinfo {author} {\bibfnamefont {Y.}~\bibnamefont {Salamin}}, \ and\
  \bibinfo {author} {\bibfnamefont {N.}~\bibnamefont {\"Unal}},\ }\href
  {\doibase 10.1103/PhysRevA.45.7740} {\bibfield  {journal} {\bibinfo
  {journal} {Phys. Rev. A}\ }\textbf {\bibinfo {volume} {45}},\ \bibinfo
  {pages} {7740} (\bibinfo {year} {1992})}\BibitemShut {NoStop}%
\bibitem [{\citenamefont {Barut}(1992)}]{BARUT19921469}%
  \BibitemOpen
  \bibfield  {author} {\bibinfo {author} {\bibfnamefont {A.}~\bibnamefont
  {Barut}},\ }\href {\doibase 10.1016/0020-7225(92)90157-C} {\bibfield
  {journal} {\bibinfo  {journal} {International Journal of Engineering
  Science}\ }\textbf {\bibinfo {volume} {30}},\ \bibinfo {pages} {1469 }
  (\bibinfo {year} {1992})}\BibitemShut {NoStop}%
\bibitem [{\citenamefont {Barut}\ and\ \citenamefont
  {Dowling}(1989)}]{PhysRevA.39.2796}%
  \BibitemOpen
  \bibfield  {author} {\bibinfo {author} {\bibfnamefont {A.~O.}\ \bibnamefont
  {Barut}}\ and\ \bibinfo {author} {\bibfnamefont {J.~P.}\ \bibnamefont
  {Dowling}},\ }\href {\doibase 10.1103/PhysRevA.39.2796} {\bibfield  {journal}
  {\bibinfo  {journal} {Phys. Rev. A}\ }\textbf {\bibinfo {volume} {39}},\
  \bibinfo {pages} {2796} (\bibinfo {year} {1989})}\BibitemShut {NoStop}%
\bibitem [{\citenamefont {Barut}\ and\ \citenamefont
  {Van~Huele}(1985)}]{PhysRevA.32.3187}%
  \BibitemOpen
  \bibfield  {author} {\bibinfo {author} {\bibfnamefont {A.~O.}\ \bibnamefont
  {Barut}}\ and\ \bibinfo {author} {\bibfnamefont {J.~F.}\ \bibnamefont
  {Van~Huele}},\ }\href {\doibase 10.1103/PhysRevA.32.3187} {\bibfield
  {journal} {\bibinfo  {journal} {Phys. Rev. A}\ }\textbf {\bibinfo {volume}
  {32}},\ \bibinfo {pages} {3187} (\bibinfo {year} {1985})}\BibitemShut
  {NoStop}%
\bibitem [{\citenamefont {Barut}\ and\ \citenamefont
  {Kraus}(1983)}]{Barut1983}%
  \BibitemOpen
  \bibfield  {author} {\bibinfo {author} {\bibfnamefont {A.~O.}\ \bibnamefont
  {Barut}}\ and\ \bibinfo {author} {\bibfnamefont {J.}~\bibnamefont {Kraus}},\
  }\href {\doibase 10.1007/BF01889480} {\bibfield  {journal} {\bibinfo
  {journal} {Foundations of Physics}\ }\textbf {\bibinfo {volume} {13}},\
  \bibinfo {pages} {189} (\bibinfo {year} {1983})}\BibitemShut {NoStop}%
\bibitem [{\citenamefont {Barut}\ and\ \citenamefont
  {Kraus}(1977)}]{PhysRevD.16.161}%
  \BibitemOpen
  \bibfield  {author} {\bibinfo {author} {\bibfnamefont {A.~O.}\ \bibnamefont
  {Barut}}\ and\ \bibinfo {author} {\bibfnamefont {J.}~\bibnamefont {Kraus}},\
  }\href {\doibase 10.1103/PhysRevD.16.161} {\bibfield  {journal} {\bibinfo
  {journal} {Phys. Rev. D}\ }\textbf {\bibinfo {volume} {16}},\ \bibinfo
  {pages} {161} (\bibinfo {year} {1977})}\BibitemShut {NoStop}%
\bibitem [{\citenamefont {Heitler}(1954)}]{HeitlerB1954quantum}%
  \BibitemOpen
  \bibfield  {author} {\bibinfo {author} {\bibfnamefont {W.}~\bibnamefont
  {Heitler}},\ }\href {https://books.google.de/books?id=L7w7UpecbKYC} {\emph
  {\bibinfo {title} {The Quantum Theory of Radiation}}},\ Dover Books on
  Physics\ (\bibinfo  {publisher} {Dover Publications},\ \bibinfo {year}
  {1954})\BibitemShut {NoStop}%
\bibitem [{\citenamefont {Kivshar}\ and\ \citenamefont
  {Malomed}(1989)}]{RevModPhys.61.763}%
  \BibitemOpen
  \bibfield  {author} {\bibinfo {author} {\bibfnamefont {Y.~S.}\ \bibnamefont
  {Kivshar}}\ and\ \bibinfo {author} {\bibfnamefont {B.~A.}\ \bibnamefont
  {Malomed}},\ }\href {\doibase 10.1103/RevModPhys.61.763} {\bibfield
  {journal} {\bibinfo  {journal} {Rev. Mod. Phys.}\ }\textbf {\bibinfo {volume}
  {61}},\ \bibinfo {pages} {763} (\bibinfo {year} {1989})}\BibitemShut
  {NoStop}%
\bibitem [{\citenamefont {Shifman}\ and\ \citenamefont
  {Yung}(2007)}]{RevModPhys.79.1139}%
  \BibitemOpen
  \bibfield  {author} {\bibinfo {author} {\bibfnamefont {M.}~\bibnamefont
  {Shifman}}\ and\ \bibinfo {author} {\bibfnamefont {A.}~\bibnamefont {Yung}},\
  }\href {\doibase 10.1103/RevModPhys.79.1139} {\bibfield  {journal} {\bibinfo
  {journal} {Rev. Mod. Phys.}\ }\textbf {\bibinfo {volume} {79}},\ \bibinfo
  {pages} {1139} (\bibinfo {year} {2007})}\BibitemShut {NoStop}%
\bibitem [{\citenamefont {Landau}\ and\ \citenamefont
  {Lifshitz}(1977)}]{LandauQM}%
  \BibitemOpen
  \bibfield  {author} {\bibinfo {author} {\bibfnamefont {L.}~\bibnamefont
  {Landau}}\ and\ \bibinfo {author} {\bibfnamefont {E.}~\bibnamefont
  {Lifshitz}},\ }\href {https://books.google.de/books?id=J9ui6KwC4mMC} {\emph
  {\bibinfo {title} {Quantum Mechanics: Non-relativistic Theory}}},\
  Butterworth Heinemann\ (\bibinfo  {publisher} {Butterworth-Heinemann},\
  \bibinfo {year} {1977})\BibitemShut {NoStop}%
\bibitem [{\citenamefont {Negele}(1982)}]{RevModPhys.54.913}%
  \BibitemOpen
  \bibfield  {author} {\bibinfo {author} {\bibfnamefont {J.~W.}\ \bibnamefont
  {Negele}},\ }\href {\doibase 10.1103/RevModPhys.54.913} {\bibfield  {journal}
  {\bibinfo  {journal} {Rev. Mod. Phys.}\ }\textbf {\bibinfo {volume} {54}},\
  \bibinfo {pages} {913} (\bibinfo {year} {1982})}\BibitemShut {NoStop}%
\bibitem [{\citenamefont {Dunne}\ and\ \citenamefont
  {Thies}(2014)}]{PhysRevD.89.025008}%
  \BibitemOpen
  \bibfield  {author} {\bibinfo {author} {\bibfnamefont {G.~V.}\ \bibnamefont
  {Dunne}}\ and\ \bibinfo {author} {\bibfnamefont {M.}~\bibnamefont {Thies}},\
  }\href {\doibase 10.1103/PhysRevD.89.025008} {\bibfield  {journal} {\bibinfo
  {journal} {Phys. Rev. D}\ }\textbf {\bibinfo {volume} {89}},\ \bibinfo
  {pages} {025008} (\bibinfo {year} {2014})}\BibitemShut {NoStop}%
\bibitem [{\citenamefont {Dunne}\ and\ \citenamefont
  {Thies}(2013)}]{PhysRevLett.111.121602}%
  \BibitemOpen
  \bibfield  {author} {\bibinfo {author} {\bibfnamefont {G.~V.}\ \bibnamefont
  {Dunne}}\ and\ \bibinfo {author} {\bibfnamefont {M.}~\bibnamefont {Thies}},\
  }\href {\doibase 10.1103/PhysRevLett.111.121602} {\bibfield  {journal}
  {\bibinfo  {journal} {Phys. Rev. Lett.}\ }\textbf {\bibinfo {volume} {111}},\
  \bibinfo {pages} {121602} (\bibinfo {year} {2013})}\BibitemShut {NoStop}%
\bibitem [{\citenamefont {Boehmer}\ \emph {et~al.}(2008)\citenamefont
  {Boehmer}, \citenamefont {Fritsch}, \citenamefont {Kraus},\ and\
  \citenamefont {Thies}}]{PhysRevD.78.065043}%
  \BibitemOpen
  \bibfield  {author} {\bibinfo {author} {\bibfnamefont {C.}~\bibnamefont
  {Boehmer}}, \bibinfo {author} {\bibfnamefont {U.}~\bibnamefont {Fritsch}},
  \bibinfo {author} {\bibfnamefont {S.}~\bibnamefont {Kraus}}, \ and\ \bibinfo
  {author} {\bibfnamefont {M.}~\bibnamefont {Thies}},\ }\href {\doibase
  10.1103/PhysRevD.78.065043} {\bibfield  {journal} {\bibinfo  {journal} {Phys.
  Rev. D}\ }\textbf {\bibinfo {volume} {78}},\ \bibinfo {pages} {065043}
  (\bibinfo {year} {2008})}\BibitemShut {NoStop}%
\bibitem [{\citenamefont {Feranchuk}\ \emph {et~al.}(2015)\citenamefont
  {Feranchuk}, \citenamefont {Ivanov}, \citenamefont {Le},\ and\ \citenamefont
  {Ulyanenkov}}]{FeranchukB2015nonperturbative}%
  \BibitemOpen
  \bibfield  {author} {\bibinfo {author} {\bibfnamefont {I.~D.}\ \bibnamefont
  {Feranchuk}}, \bibinfo {author} {\bibfnamefont {A.}~\bibnamefont {Ivanov}},
  \bibinfo {author} {\bibfnamefont {V.~H.}\ \bibnamefont {Le}}, \ and\ \bibinfo
  {author} {\bibfnamefont {A.~P.}\ \bibnamefont {Ulyanenkov}},\ }\href
  {https://books.google.de/books?id=uvWuoQEACAAJ} {\emph {\bibinfo {title}
  {Non-perturbative Description of Quantum Systems}}},\ Lecture Notes in
  Physics\ (\bibinfo  {publisher} {Springer International Publishing},\
  \bibinfo {year} {2015})\BibitemShut {NoStop}%
\bibitem [{\citenamefont {Alkofer}\ \emph {et~al.}(1996)\citenamefont
  {Alkofer}, \citenamefont {Reinhardt},\ and\ \citenamefont
  {Weigel}}]{ALKOFER1996139}%
  \BibitemOpen
  \bibfield  {author} {\bibinfo {author} {\bibfnamefont {R.}~\bibnamefont
  {Alkofer}}, \bibinfo {author} {\bibfnamefont {H.}~\bibnamefont {Reinhardt}},
  \ and\ \bibinfo {author} {\bibfnamefont {H.}~\bibnamefont {Weigel}},\ }\href
  {\doibase 10.1016/0370-1573(95)00018-6} {\bibfield  {journal} {\bibinfo
  {journal} {Physics Reports}\ }\textbf {\bibinfo {volume} {265}},\ \bibinfo
  {pages} {139 } (\bibinfo {year} {1996})}\BibitemShut {NoStop}%
\bibitem [{\citenamefont {Christov}\ \emph {et~al.}(1996)\citenamefont
  {Christov}, \citenamefont {Blotz}, \citenamefont {Kim}, \citenamefont
  {Pobylitsa}, \citenamefont {Watabe}, \citenamefont {Meissner}, \citenamefont
  {Arriola},\ and\ \citenamefont {Goeke}}]{CHRISTOV199691}%
  \BibitemOpen
  \bibfield  {author} {\bibinfo {author} {\bibfnamefont {C.}~\bibnamefont
  {Christov}}, \bibinfo {author} {\bibfnamefont {A.}~\bibnamefont {Blotz}},
  \bibinfo {author} {\bibfnamefont {H.-C.}\ \bibnamefont {Kim}}, \bibinfo
  {author} {\bibfnamefont {P.}~\bibnamefont {Pobylitsa}}, \bibinfo {author}
  {\bibfnamefont {T.}~\bibnamefont {Watabe}}, \bibinfo {author} {\bibfnamefont
  {T.}~\bibnamefont {Meissner}}, \bibinfo {author} {\bibfnamefont {E.~R.}\
  \bibnamefont {Arriola}}, \ and\ \bibinfo {author} {\bibfnamefont
  {K.}~\bibnamefont {Goeke}},\ }\href {\doibase 10.1016/0146-6410(96)00057-9}
  {\bibfield  {journal} {\bibinfo  {journal} {Progress in Particle and Nuclear
  Physics}\ }\textbf {\bibinfo {volume} {37}},\ \bibinfo {pages} {91 }
  (\bibinfo {year} {1996})}\BibitemShut {NoStop}%
\bibitem [{\citenamefont {Dashen}\ \emph
  {et~al.}(1974{\natexlab{a}})\citenamefont {Dashen}, \citenamefont
  {Hasslacher},\ and\ \citenamefont {Neveu}}]{PhysRevD.10.4130}%
  \BibitemOpen
  \bibfield  {author} {\bibinfo {author} {\bibfnamefont {R.~F.}\ \bibnamefont
  {Dashen}}, \bibinfo {author} {\bibfnamefont {B.}~\bibnamefont {Hasslacher}},
  \ and\ \bibinfo {author} {\bibfnamefont {A.}~\bibnamefont {Neveu}},\ }\href
  {\doibase 10.1103/PhysRevD.10.4130} {\bibfield  {journal} {\bibinfo
  {journal} {Phys. Rev. D}\ }\textbf {\bibinfo {volume} {10}},\ \bibinfo
  {pages} {4130} (\bibinfo {year} {1974}{\natexlab{a}})}\BibitemShut {NoStop}%
\bibitem [{\citenamefont {Jauch}\ and\ \citenamefont
  {Rohrlich}(2012)}]{jauch2012theory}%
  \BibitemOpen
  \bibfield  {author} {\bibinfo {author} {\bibfnamefont {J.}~\bibnamefont
  {Jauch}}\ and\ \bibinfo {author} {\bibfnamefont {F.}~\bibnamefont
  {Rohrlich}},\ }\href {https://books.google.de/books?id=Epn6CAAAQBAJ} {\emph
  {\bibinfo {title} {The Theory of Photons and Electrons: The Relativistic
  Quantum Field Theory of Charged Particles with Spin One-half}}},\ Theoretical
  and Mathematical Physics\ (\bibinfo  {publisher} {Springer Berlin
  Heidelberg},\ \bibinfo {year} {2012})\BibitemShut {NoStop}%
\bibitem [{\citenamefont {Messiah}(1981)}]{MessiahB1981quantum}%
  \BibitemOpen
  \bibfield  {author} {\bibinfo {author} {\bibfnamefont {A.}~\bibnamefont
  {Messiah}},\ }\href {https://books.google.de/books?id=VR93vUk8d\_8C} {\emph
  {\bibinfo {title} {Quantum Mechanics}}},\ \bibinfo {series} {Quantum
  Mechanics}\ No.\ \bibinfo {number} {v. 2}\ (\bibinfo  {publisher}
  {North-Holland},\ \bibinfo {year} {1981})\BibitemShut {NoStop}%
\bibitem [{\citenamefont {Dashen}\ \emph
  {et~al.}(1974{\natexlab{b}})\citenamefont {Dashen}, \citenamefont
  {Hasslacher},\ and\ \citenamefont {Neveu}}]{PhysRevD.10.4114}%
  \BibitemOpen
  \bibfield  {author} {\bibinfo {author} {\bibfnamefont {R.~F.}\ \bibnamefont
  {Dashen}}, \bibinfo {author} {\bibfnamefont {B.}~\bibnamefont {Hasslacher}},
  \ and\ \bibinfo {author} {\bibfnamefont {A.}~\bibnamefont {Neveu}},\ }\href
  {\doibase 10.1103/PhysRevD.10.4114} {\bibfield  {journal} {\bibinfo
  {journal} {Phys. Rev. D}\ }\textbf {\bibinfo {volume} {10}},\ \bibinfo
  {pages} {4114} (\bibinfo {year} {1974}{\natexlab{b}})}\BibitemShut {NoStop}%
\bibitem [{\citenamefont {Dashen}\ \emph
  {et~al.}(1974{\natexlab{c}})\citenamefont {Dashen}, \citenamefont
  {Hasslacher},\ and\ \citenamefont {Neveu}}]{PhysRevD.10.4138}%
  \BibitemOpen
  \bibfield  {author} {\bibinfo {author} {\bibfnamefont {R.~F.}\ \bibnamefont
  {Dashen}}, \bibinfo {author} {\bibfnamefont {B.}~\bibnamefont {Hasslacher}},
  \ and\ \bibinfo {author} {\bibfnamefont {A.}~\bibnamefont {Neveu}},\ }\href
  {\doibase 10.1103/PhysRevD.10.4138} {\bibfield  {journal} {\bibinfo
  {journal} {Phys. Rev. D}\ }\textbf {\bibinfo {volume} {10}},\ \bibinfo
  {pages} {4138} (\bibinfo {year} {1974}{\natexlab{c}})}\BibitemShut {NoStop}%
\bibitem [{\citenamefont {Fl{\"u}gge}(1994)}]{FlueggeB1994practical}%
  \BibitemOpen
  \bibfield  {author} {\bibinfo {author} {\bibfnamefont {S.}~\bibnamefont
  {Fl{\"u}gge}},\ }\href {https://books.google.de/books?id=VpggN9qIFUcC} {\emph
  {\bibinfo {title} {Practical Quantum Mechanics}}},\ Classics in Mathematics\
  (\bibinfo  {publisher} {Springer Berlin Heidelberg},\ \bibinfo {year}
  {1994})\BibitemShut {NoStop}%
\bibitem [{\citenamefont {Komarov}\ \emph {et~al.}(1978)\citenamefont
  {Komarov}, \citenamefont {Krylov},\ and\ \citenamefont
  {Feranchuk}}]{KOMAROV1978153}%
  \BibitemOpen
  \bibfield  {author} {\bibinfo {author} {\bibfnamefont {L.}~\bibnamefont
  {Komarov}}, \bibinfo {author} {\bibfnamefont {E.}~\bibnamefont {Krylov}}, \
  and\ \bibinfo {author} {\bibfnamefont {I.}~\bibnamefont {Feranchuk}},\ }\href
  {\doibase 10.1016/0041-5553(78)90176-3} {\bibfield  {journal} {\bibinfo
  {journal} {USSR Computational Mathematics and Mathematical Physics}\ }\textbf
  {\bibinfo {volume} {18}},\ \bibinfo {pages} {153 } (\bibinfo {year}
  {1978})}\BibitemShut {NoStop}%
\bibitem [{\citenamefont {Gareev}\ \emph {et~al.}(1977)\citenamefont {Gareev},
  \citenamefont {Goncharov}, \citenamefont {Zhidkov}, \citenamefont {Puzynin},
  \citenamefont {Khoromskii},\ and\ \citenamefont {Yamaleev}}]{GAREEV1977116}%
  \BibitemOpen
  \bibfield  {author} {\bibinfo {author} {\bibfnamefont {F.}~\bibnamefont
  {Gareev}}, \bibinfo {author} {\bibfnamefont {S.}~\bibnamefont {Goncharov}},
  \bibinfo {author} {\bibfnamefont {E.}~\bibnamefont {Zhidkov}}, \bibinfo
  {author} {\bibfnamefont {I.}~\bibnamefont {Puzynin}}, \bibinfo {author}
  {\bibfnamefont {B.}~\bibnamefont {Khoromskii}}, \ and\ \bibinfo {author}
  {\bibfnamefont {R.}~\bibnamefont {Yamaleev}},\ }\href {\doibase
  10.1016/0041-5553(77)90041-6} {\bibfield  {journal} {\bibinfo  {journal}
  {USSR Computational Mathematics and Mathematical Physics}\ }\textbf {\bibinfo
  {volume} {17}},\ \bibinfo {pages} {116 } (\bibinfo {year}
  {1977})}\BibitemShut {NoStop}%
\bibitem [{\citenamefont {Airapetyan}(1999)}]{AirapetyanA1999continuous}%
  \BibitemOpen
  \bibfield  {author} {\bibinfo {author} {\bibfnamefont {R.}~\bibnamefont
  {Airapetyan}},\ }\href {\doibase 10.1080/00036819908840791} {\bibfield
  {journal} {\bibinfo  {journal} {Applicable Analysis}\ }\textbf {\bibinfo
  {volume} {73}},\ \bibinfo {pages} {463} (\bibinfo {year} {1999})},\ \Eprint
  {http://arxiv.org/abs/http://dx.doi.org/10.1080/00036819908840791}
  {http://dx.doi.org/10.1080/00036819908840791} \BibitemShut {NoStop}%
\bibitem [{\citenamefont {Ermakov}\ and\ \citenamefont
  {Kalitkin}(1981)}]{ERMAKOV1981235}%
  \BibitemOpen
  \bibfield  {author} {\bibinfo {author} {\bibfnamefont {V.}~\bibnamefont
  {Ermakov}}\ and\ \bibinfo {author} {\bibfnamefont {N.}~\bibnamefont
  {Kalitkin}},\ }\href {\doibase 10.1016/0041-5553(81)90022-7} {\bibfield
  {journal} {\bibinfo  {journal} {USSR Computational Mathematics and
  Mathematical Physics}\ }\textbf {\bibinfo {volume} {21}},\ \bibinfo {pages}
  {235 } (\bibinfo {year} {1981})}\BibitemShut {NoStop}%
\bibitem [{\citenamefont {Gavurin}(1958)}]{GavurinA1958nonlinear}%
  \BibitemOpen
  \bibfield  {author} {\bibinfo {author} {\bibfnamefont {M.~K.}\ \bibnamefont
  {Gavurin}},\ }\href@noop {} {\bibfield  {journal} {\bibinfo  {journal} {Izv.
  Vyssh. Uchebn. Zaved. Mat.}\ ,\ \bibinfo {pages} {18}} (\bibinfo {year}
  {1958})}\BibitemShut {NoStop}%
\bibitem [{\citenamefont {Puzynin}\ \emph {et~al.}(1999)\citenamefont
  {Puzynin}, \citenamefont {Amirkhanov}, \citenamefont {Zemlyanaya},
  \citenamefont {Pervushin}, \citenamefont {Puzynina}, \citenamefont {Strizh},\
  and\ \citenamefont {Lakhno}}]{PuzyninA1999generalized}%
  \BibitemOpen
  \bibfield  {author} {\bibinfo {author} {\bibfnamefont {I.~V.}\ \bibnamefont
  {Puzynin}}, \bibinfo {author} {\bibfnamefont {I.~V.}\ \bibnamefont
  {Amirkhanov}}, \bibinfo {author} {\bibfnamefont {E.~V.}\ \bibnamefont
  {Zemlyanaya}}, \bibinfo {author} {\bibfnamefont {V.~N.}\ \bibnamefont
  {Pervushin}}, \bibinfo {author} {\bibfnamefont {T.~P.}\ \bibnamefont
  {Puzynina}}, \bibinfo {author} {\bibfnamefont {T.~A.}\ \bibnamefont
  {Strizh}}, \ and\ \bibinfo {author} {\bibfnamefont {V.~D.}\ \bibnamefont
  {Lakhno}},\ }\href {\doibase 10.1134/1.953099} {\bibfield  {journal}
  {\bibinfo  {journal} {Physics of Particles and Nuclei}\ }\textbf {\bibinfo
  {volume} {30}},\ \bibinfo {pages} {87} (\bibinfo {year} {1999})}\BibitemShut
  {NoStop}%
\bibitem [{\citenamefont {Zhidkov}\ \emph {et~al.}(1973)\citenamefont
  {Zhidkov}, \citenamefont {Makarenko},\ and\ \citenamefont
  {Puzynin}}]{ZhidkovA1973continuous}%
  \BibitemOpen
  \bibfield  {author} {\bibinfo {author} {\bibfnamefont {E.}~\bibnamefont
  {Zhidkov}}, \bibinfo {author} {\bibfnamefont {G.}~\bibnamefont {Makarenko}},
  \ and\ \bibinfo {author} {\bibfnamefont {I.}~\bibnamefont {Puzynin}},\
  }\href@noop {} {\bibfield  {journal} {\bibinfo  {journal} {Sov. J. Part.
  Nucl}\ ,\ \bibinfo {pages} {127}} (\bibinfo {year} {1973})}\BibitemShut
  {NoStop}%
\bibitem [{\citenamefont {Gross}(1976)}]{GROSS19761}%
  \BibitemOpen
  \bibfield  {author} {\bibinfo {author} {\bibfnamefont {E.~P.}\ \bibnamefont
  {Gross}},\ }\href {\doibase 10.1016/0003-4916(76)90082-8} {\bibfield
  {journal} {\bibinfo  {journal} {Annals of Physics}\ }\textbf {\bibinfo
  {volume} {99}},\ \bibinfo {pages} {1 } (\bibinfo {year} {1976})}\BibitemShut
  {NoStop}%
\bibitem [{\citenamefont {Tyablikov}(1951)}]{TyablikovA1951adyabatic}%
  \BibitemOpen
  \bibfield  {author} {\bibinfo {author} {\bibfnamefont {S.~V.}\ \bibnamefont
  {Tyablikov}},\ }\href@noop {} {\bibfield  {journal} {\bibinfo  {journal}
  {Sov. Phys. JETP}\ }\textbf {\bibinfo {volume} {3}},\ \bibinfo {pages} {377}
  (\bibinfo {year} {1951})}\BibitemShut {NoStop}%
\bibitem [{\citenamefont {Bogolyubov}(1950)}]{BogolyubovA1950about}%
  \BibitemOpen
  \bibfield  {author} {\bibinfo {author} {\bibfnamefont {N.~N.}\ \bibnamefont
  {Bogolyubov}},\ }\href@noop {} {\bibfield  {journal} {\bibinfo  {journal}
  {Ukrainian Mathematical Journal}\ }\textbf {\bibinfo {volume} {2}},\ \bibinfo
  {pages} {3} (\bibinfo {year} {1950})}\BibitemShut {NoStop}%
\bibitem [{\citenamefont {Pobylitsa}\ \emph {et~al.}(1992)\citenamefont
  {Pobylitsa}, \citenamefont {Arriola}, \citenamefont {Meissner}, \citenamefont
  {Grummer}, \citenamefont {Goeke},\ and\ \citenamefont
  {Broniowski}}]{0954-3899-18-9-008}%
  \BibitemOpen
  \bibfield  {author} {\bibinfo {author} {\bibfnamefont {P.~V.}\ \bibnamefont
  {Pobylitsa}}, \bibinfo {author} {\bibfnamefont {E.~R.}\ \bibnamefont
  {Arriola}}, \bibinfo {author} {\bibfnamefont {T.}~\bibnamefont {Meissner}},
  \bibinfo {author} {\bibfnamefont {F.}~\bibnamefont {Grummer}}, \bibinfo
  {author} {\bibfnamefont {K.}~\bibnamefont {Goeke}}, \ and\ \bibinfo {author}
  {\bibfnamefont {W.}~\bibnamefont {Broniowski}},\ }\href
  {http://stacks.iop.org/0954-3899/18/i=9/a=008} {\bibfield  {journal}
  {\bibinfo  {journal} {Journal of Physics G: Nuclear and Particle Physics}\
  }\textbf {\bibinfo {volume} {18}},\ \bibinfo {pages} {1455} (\bibinfo {year}
  {1992})}\BibitemShut {NoStop}%
\bibitem [{\citenamefont {Feynman}(1998)}]{FeynmanB1998statistical}%
  \BibitemOpen
  \bibfield  {author} {\bibinfo {author} {\bibfnamefont {R.}~\bibnamefont
  {Feynman}},\ }\href {https://books.google.de/books?id=Ou4ltPYiXPgC} {\emph
  {\bibinfo {title} {Statistical Mechanics: A Set Of Lectures}}},\ Advanced
  Books Classics Series\ (\bibinfo  {publisher} {Westview Press},\ \bibinfo
  {year} {1998})\BibitemShut {NoStop}%
\bibitem [{\citenamefont {Bjorken}\ and\ \citenamefont
  {Drell}(1964)}]{bjorkenB1964}%
  \BibitemOpen
  \bibfield  {author} {\bibinfo {author} {\bibfnamefont {J.}~\bibnamefont
  {Bjorken}}\ and\ \bibinfo {author} {\bibfnamefont {S.}~\bibnamefont
  {Drell}},\ }\href {https://books.google.de/books?id=pAdRAAAAMAAJ} {\emph
  {\bibinfo {title} {Relativistic quantum mechanics}}},\ International series
  in pure and applied physics\ (\bibinfo  {publisher} {McGraw-Hill},\ \bibinfo
  {year} {1964})\BibitemShut {NoStop}%
\bibitem [{\citenamefont {Airapetyan}\ \emph {et~al.}(1999)\citenamefont
  {Airapetyan}, \citenamefont {Ramm},\ and\ \citenamefont
  {Smirnova}}]{AirapetyanA1999continuous1}%
  \BibitemOpen
  \bibfield  {author} {\bibinfo {author} {\bibfnamefont {R.~G.}\ \bibnamefont
  {Airapetyan}}, \bibinfo {author} {\bibfnamefont {A.~G.}\ \bibnamefont
  {Ramm}}, \ and\ \bibinfo {author} {\bibfnamefont {A.~B.}\ \bibnamefont
  {Smirnova}},\ }\href {\doibase 10.1142/S0218202599000233} {\bibfield
  {journal} {\bibinfo  {journal} {Mathematical Models and Methods in Applied
  Sciences}\ }\textbf {\bibinfo {volume} {09}},\ \bibinfo {pages} {463}
  (\bibinfo {year} {1999})},\ \Eprint
  {http://arxiv.org/abs/http://www.worldscientific.com/doi/pdf/10.1142/S0218202599000233}
  {http://www.worldscientific.com/doi/pdf/10.1142/S0218202599000233}
  \BibitemShut {NoStop}%
\bibitem [{\citenamefont {Ponomarev}\ \emph {et~al.}(1978)\citenamefont
  {Ponomarev}, \citenamefont {Puzynin}, \citenamefont {Puzynina},\ and\
  \citenamefont {Somov}}]{PONOMAREV1978274}%
  \BibitemOpen
  \bibfield  {author} {\bibinfo {author} {\bibfnamefont {L.}~\bibnamefont
  {Ponomarev}}, \bibinfo {author} {\bibfnamefont {I.}~\bibnamefont {Puzynin}},
  \bibinfo {author} {\bibfnamefont {T.}~\bibnamefont {Puzynina}}, \ and\
  \bibinfo {author} {\bibfnamefont {L.}~\bibnamefont {Somov}},\ }\href
  {\doibase 10.1016/0003-4916(78)90033-7} {\bibfield  {journal} {\bibinfo
  {journal} {Annals of Physics}\ }\textbf {\bibinfo {volume} {110}},\ \bibinfo
  {pages} {274 } (\bibinfo {year} {1978})}\BibitemShut {NoStop}%
\bibitem [{\citenamefont {Melezhik}(1991)}]{MELEZHIK199167}%
  \BibitemOpen
  \bibfield  {author} {\bibinfo {author} {\bibfnamefont {V.}~\bibnamefont
  {Melezhik}},\ }\href {\doibase 10.1016/0021-9991(91)90292-S} {\bibfield
  {journal} {\bibinfo  {journal} {Journal of Computational Physics}\ }\textbf
  {\bibinfo {volume} {92}},\ \bibinfo {pages} {67 } (\bibinfo {year}
  {1991})}\BibitemShut {NoStop}%
\bibitem [{\citenamefont {Ponomarev}\ \emph {et~al.}(1973)\citenamefont
  {Ponomarev}, \citenamefont {Puzynin},\ and\ \citenamefont
  {Puzynina}}]{PONOMAREV19731}%
  \BibitemOpen
  \bibfield  {author} {\bibinfo {author} {\bibfnamefont {L.}~\bibnamefont
  {Ponomarev}}, \bibinfo {author} {\bibfnamefont {I.}~\bibnamefont {Puzynin}},
  \ and\ \bibinfo {author} {\bibfnamefont {T.}~\bibnamefont {Puzynina}},\
  }\href {\doibase 10.1016/0021-9991(73)90121-6} {\bibfield  {journal}
  {\bibinfo  {journal} {Journal of Computational Physics}\ }\textbf {\bibinfo
  {volume} {13}},\ \bibinfo {pages} {1 } (\bibinfo {year} {1973})}\BibitemShut
  {NoStop}%
\bibitem [{\citenamefont {Melezhik}\ \emph {et~al.}(1984)\citenamefont
  {Melezhik}, \citenamefont {Puzynin}, \citenamefont {Puzynina},\ and\
  \citenamefont {Somov}}]{MELEZHIK1984221}%
  \BibitemOpen
  \bibfield  {author} {\bibinfo {author} {\bibfnamefont {V.}~\bibnamefont
  {Melezhik}}, \bibinfo {author} {\bibfnamefont {I.}~\bibnamefont {Puzynin}},
  \bibinfo {author} {\bibfnamefont {T.}~\bibnamefont {Puzynina}}, \ and\
  \bibinfo {author} {\bibfnamefont {L.}~\bibnamefont {Somov}},\ }\href
  {\doibase 10.1016/0021-9991(84)90115-3} {\bibfield  {journal} {\bibinfo
  {journal} {Journal of Computational Physics}\ }\textbf {\bibinfo {volume}
  {54}},\ \bibinfo {pages} {221 } (\bibinfo {year} {1984})}\BibitemShut
  {NoStop}%
\bibitem [{\citenamefont {Melezhik}(1986)}]{MELEZHIK19861}%
  \BibitemOpen
  \bibfield  {author} {\bibinfo {author} {\bibfnamefont {V.}~\bibnamefont
  {Melezhik}},\ }\href {\doibase 10.1016/0021-9991(86)90001-X} {\bibfield
  {journal} {\bibinfo  {journal} {Journal of Computational Physics}\ }\textbf
  {\bibinfo {volume} {65}},\ \bibinfo {pages} {1 } (\bibinfo {year}
  {1986})}\BibitemShut {NoStop}%
\bibitem [{\citenamefont {Hoheisel}\ \emph {et~al.}(2012)\citenamefont
  {Hoheisel}, \citenamefont {Kanzow}, \citenamefont {Mordukhovich},\ and\
  \citenamefont {Phan}}]{Hoheisel20121324}%
  \BibitemOpen
  \bibfield  {author} {\bibinfo {author} {\bibfnamefont {T.}~\bibnamefont
  {Hoheisel}}, \bibinfo {author} {\bibfnamefont {C.}~\bibnamefont {Kanzow}},
  \bibinfo {author} {\bibfnamefont {B.}~\bibnamefont {Mordukhovich}}, \ and\
  \bibinfo {author} {\bibfnamefont {H.}~\bibnamefont {Phan}},\ }\href {\doibase
  10.1016/j.na.2011.06.039} {\bibfield  {journal} {\bibinfo  {journal}
  {Nonlinear Analysis: Theory, Methods \& Applications}\ }\textbf {\bibinfo
  {volume} {75}},\ \bibinfo {pages} {1324 } (\bibinfo {year} {2012})},\
  \bibinfo {note} {variational Analysis and Its Applications}\BibitemShut
  {NoStop}%
\bibitem [{\citenamefont {Ramm}\ \emph {et~al.}(2003)\citenamefont {Ramm},
  \citenamefont {Smirnova},\ and\ \citenamefont {Favini}}]{Ramm2003}%
  \BibitemOpen
  \bibfield  {author} {\bibinfo {author} {\bibfnamefont {G.~A.}\ \bibnamefont
  {Ramm}}, \bibinfo {author} {\bibfnamefont {B.~A.}\ \bibnamefont {Smirnova}},
  \ and\ \bibinfo {author} {\bibfnamefont {A.}~\bibnamefont {Favini}},\ }\href
  {\doibase 10.1007/s10231-002-0054-0} {\bibfield  {journal} {\bibinfo
  {journal} {Annali di Matematica Pura ed Applicata}\ }\textbf {\bibinfo
  {volume} {182}},\ \bibinfo {pages} {37} (\bibinfo {year} {2003})}\BibitemShut
  {NoStop}%
\bibitem [{\citenamefont {Zhanlav}\ and\ \citenamefont
  {Puzynin}(1992)}]{ZhanlavA1992convergence}%
  \BibitemOpen
  \bibfield  {author} {\bibinfo {author} {\bibfnamefont {T.}~\bibnamefont
  {Zhanlav}}\ and\ \bibinfo {author} {\bibfnamefont {I.~V.}\ \bibnamefont
  {Puzynin}},\ }\href@noop {} {\bibfield  {journal} {\bibinfo  {journal}
  {Computational Mathematics and Mathematical Physics}\ }\textbf {\bibinfo
  {volume} {32}},\ \bibinfo {pages} {729} (\bibinfo {year} {1992})}\BibitemShut
  {NoStop}%
\bibitem [{\citenamefont {Press}(2007)}]{PressB2007numerical}%
  \BibitemOpen
  \bibfield  {author} {\bibinfo {author} {\bibfnamefont {W.}~\bibnamefont
  {Press}},\ }\href {https://books.google.de/books?id=1aAOdzK3FegC} {\emph
  {\bibinfo {title} {Numerical Recipes 3rd Edition: The Art of Scientific
  Computing}}}\ (\bibinfo  {publisher} {Cambridge University Press},\ \bibinfo
  {year} {2007})\BibitemShut {NoStop}%
\bibitem [{\citenamefont {Samarski{\u\i}}\ and\ \citenamefont
  {Nikolaev}(1989)}]{SamarskiiB1989numerical}%
  \BibitemOpen
  \bibfield  {author} {\bibinfo {author} {\bibfnamefont {A.}~\bibnamefont
  {Samarski{\u\i}}}\ and\ \bibinfo {author} {\bibfnamefont {E.}~\bibnamefont
  {Nikolaev}},\ }\href {https://books.google.de/books?id=t4uWtKG6NooC} {\emph
  {\bibinfo {title} {Numerical Methods for Grid Equations}}},\ \bibinfo
  {series} {Numerical Methods for Grid Equations}\ No.\ \bibinfo {number} {v.
  2}\ (\bibinfo  {publisher} {Birkh{\"a}user},\ \bibinfo {year}
  {1989})\BibitemShut {NoStop}%
\bibitem [{\citenamefont {Miyake}(1975)}]{MiyakeA1975strong}%
  \BibitemOpen
  \bibfield  {author} {\bibinfo {author} {\bibfnamefont {S.~J.}\ \bibnamefont
  {Miyake}},\ }\href {\doibase 10.1143/JPSJ.38.181} {\bibfield  {journal}
  {\bibinfo  {journal} {Journal of the Physical Society of Japan}\ }\textbf
  {\bibinfo {volume} {38}},\ \bibinfo {pages} {181} (\bibinfo {year} {1975})},\
  \Eprint {http://arxiv.org/abs/http://dx.doi.org/10.1143/JPSJ.38.181}
  {http://dx.doi.org/10.1143/JPSJ.38.181} \BibitemShut {NoStop}%
\end{thebibliography}%

\end{document}